\documentclass[
    reprint,
    amsmath,
    amssymb,
    aps,
    pra,
    floatfix,
    natbib
]{revtex4-2}

\usepackage{graphicx}      
\usepackage{dcolumn}       
\usepackage{bm}            
\usepackage{color,xcolor}  
\usepackage{times}         
\usepackage{url}           
\usepackage{booktabs}      
\usepackage{paralist}      
\usepackage{comment}       

\usepackage{mathtools}     
\usepackage{physics}       
\usepackage{qcircuit}      
\usepackage{tikz}          
\usetikzlibrary{external}
\usetikzlibrary{decorations.pathmorphing,math}
\tikzset{snake it/.style={decorate, decoration=snake}}

\usepackage{silence}
\usepackage{hyperref}      
\hypersetup{
    colorlinks=true,
    linkcolor=blue,
    citecolor=blue,
    urlcolor=blue,
}
\usepackage[capitalise,noabbrev]{cleveref}  

\makeatletter
\let\frontmatter@abstractwidth\linewidth
\makeatother

\usepackage{appendix}

\definecolor{ibmOrange}{HTML}{FFB000}
\definecolor{ibmRed}{HTML}{FE6100}
\definecolor{ibmMagenta}{HTML}{DC267F}
\definecolor{ibmViolet}{HTML}{785EF0}
\definecolor{ibmBlue}{HTML}{648FFF}

\newcommand{\nshots}{N_{s}}

\newcommand{\ntwirling}{N_{t}}

\newcommand{\mpsims}{Majorana propagation simulations }

\makeatletter
\def\l@subsubsection#1#2{}
\makeatother

\let\doaddcontentsline\addcontentsline
\newcommand{\noaddcontentsline}[3]{}

\newcommand{\tocoff}{\let\addcontentsline\noaddcontentsline}
\newcommand{\tocon}{\let\addcontentsline\doaddcontentsline}

\newcommand{\notocline}[2]{%
  \let\addcontentsline\noaddcontentsline%
  #1{#2}%
  \let\addcontentsline\doaddcontentsline%
}

\tocoff 

\begin{document}

\title{Fermionic dynamics on a trapped-ion quantum computer beyond exact classical simulation}

\renewcommand{\andname}{with}

\author{Phasecraft}\email{info@phasecraft.io}
\author{Quantinuum collaborators{\hyperlink{author_dagger}{$^\dagger$}}}
\noaffiliation
\date{\today}

\begin{abstract}
{\bf
Simulation of the time-dynamics of fermionic many-body systems has long been predicted to be one of the key applications of quantum computers~\cite{georgescu14}.
Such simulations -- for which classical methods are often inaccurate -- are critical to advancing our knowledge and understanding of quantum chemistry and materials, underpinning a wide range of fields, from biochemistry to clean-energy technologies and chemical synthesis. However, the performance of all previous digital quantum simulations of fermions has been matched by classical methods, and it has thus far remained unclear whether near-term, intermediate-scale quantum hardware could offer any computational advantage in this area. Here, we implement an efficient quantum simulation algorithm on Quantinuum's System Model H2 trapped-ion quantum computer for the time dynamics of a 56-qubit system that is too complex for exact classical simulation. We focus on the periodic spinful 2D Fermi-Hubbard model and present evidence of spin-charge separation \cite{Anderson1995}, where the elementary electron's charge and spin decouple. In the limited cases where ground truth is available through exact classical simulation, we find that it agrees with the results we obtain from the quantum device. Employing long-range Wilson operators \cite{Wilson} to study deconfinement of the effective gauge field between spinons and the effective potential between charge carriers \cite{Lee_2006,Lee_2008}, we find behaviour that differs from predictions made by classical tensor network methods. Our results herald the use of quantum computing for simulating strongly correlated electronic systems beyond the capacity of classical computing.
}
\end{abstract}

\maketitle


The Fermi-Hubbard model exemplifies the key challenge in many-body physics: accurately modelling systems of interacting particles.
Although a highly simplified model of interacting electrons, it contains a rich phase diagram \cite{Dagotto_1994}, making it an ideal laboratory in which to study phenomena such as spin-charge separation~\cite{Anderson1995,arute20,Vijayan2020}, the metal-insulator transition~\cite{Akiyama_2022}, and magnetic ordering~\cite{Hart2015,Parsons_2016,Mazurenko_2017}. However, the straightforward definition of the model hides an inherent complexity. The most complex instance of the model whose ground state has been exactly computed numerically is just 17 electrons on 22 sites~\cite{Yamada_2005}. On the other hand, several state-of-the-art approximate methods can simulate low-energy states on hundreds of sites at different filling fractions \cite{LeBlanc2015}, while special-purpose analogue simulators can address larger instances still (between hundreds and thousands of sites)~\cite{bakr2025,Xu2025}, serving as significant tools to probe equilibrium properties of interacting systems.

Simulating dynamical properties of the Fermi-Hubbard model appears to be significantly harder than equilibrium low-energy properties for classical computers. Indeed, the largest reported instances where exact time-dynamics has been simulated classically have 14 sites \cite{Innerberger_2020}; although approximate methods reaching $6\times 5$ sites have been demonstrated for certain observables~\cite{thompson2025}, the level of accuracy of these is unknown.
Quantum computers are predicted to be able to simulate the Fermi-Hubbard model efficiently, yet all previous digital simulations of the dynamics of the Fermi-Hubbard model on quantum computing hardware are either 1D instances \cite{arute20,Vilchez_Estevez_2025,chowdhury2025}, or relatively small 2D instances~\cite{evered25, H_mery_2024}, and can be simulated accurately using tensor network methods. Analogue quantum simulators can simulate dynamical properties for systems on thousands of sites~\cite{Xu2025,Guardado2020Subdiffusion, Brown2019Bad,Nichols2019Spin},
but face significant limitations on the initial states they can prepare and the quantities that can be measured at the end of the simulation.

\begin{figure}[tpb!]
    \centering
    \includegraphics[width=\columnwidth]
    {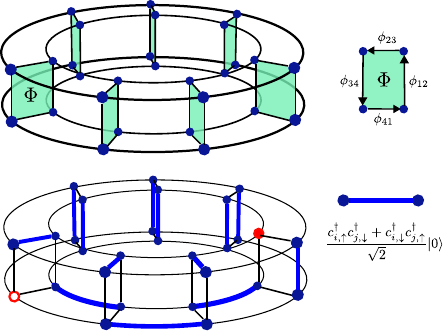}
    \includegraphics[width=\columnwidth]{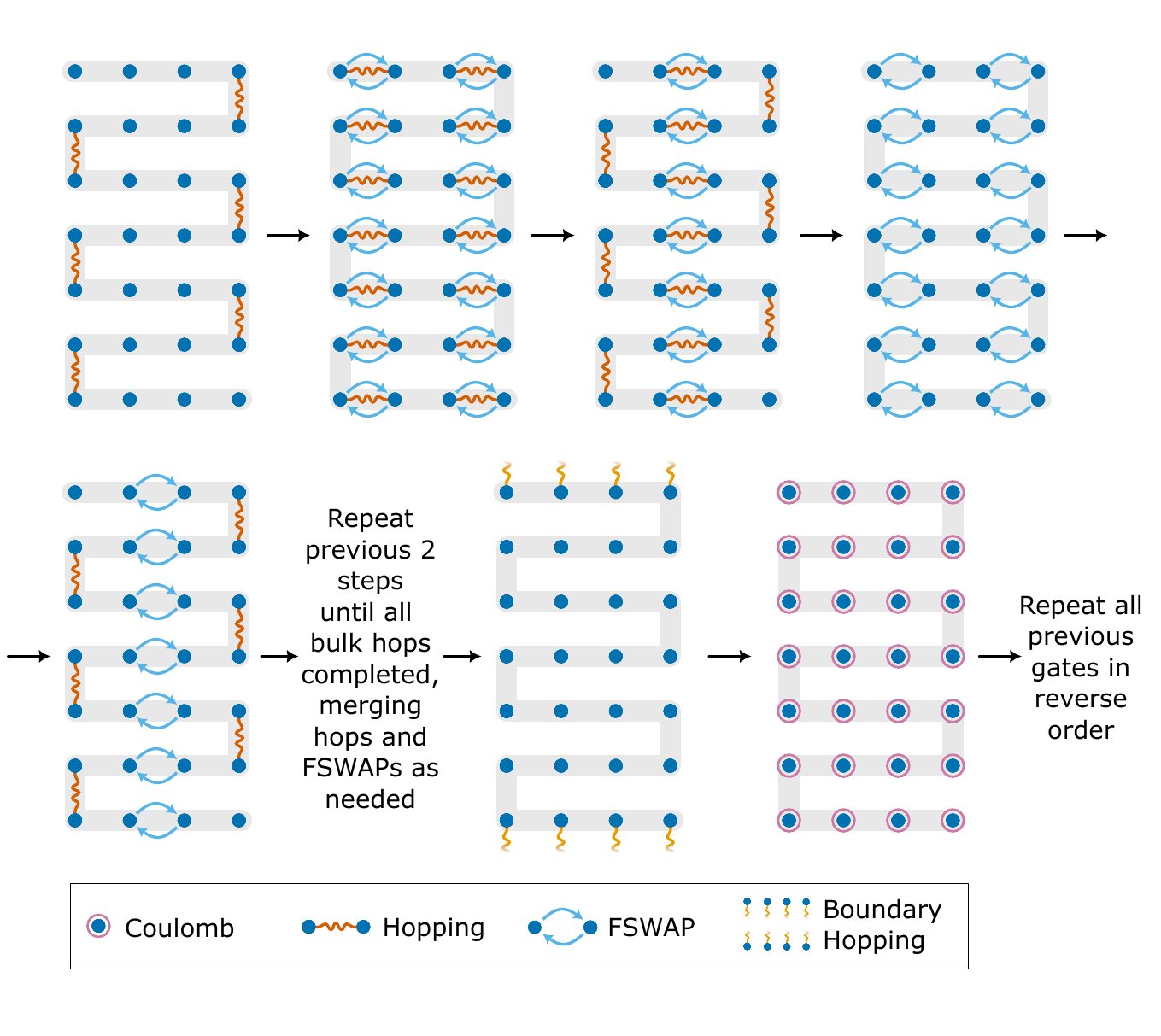}
    \caption{
    \textbf{Fermi-Hubbard lattice, initial state and map of a spin sector on the quantum circuit.}
    \textbf{Top,} The Fermi-Hubbard model instance studied in this work is a double-periodic system of size $|\mathcal{L}|=L_x\times L_y=7\times 4$ (56 qubits), with $\Phi=\pi$ flux on the long direction. Around each highlighted plaquette $\phi_{ij}=\pi/4$, with $i$, $j$ ordered such that they form a directed loop (example on the right) . Each site (blue circles) can accommodate spin-up and spin-down electrons.
    \textbf{Middle,} The initial state is a dimerised configuration where we place a maximally entangled state, shown as a blue link. Each link corresponds to the $S_z^{\rm total}=0$ triplet state. The empty and full red circles correspond to a holon and a doublon, respectively.
    \textbf{Bottom,} Structure of the quantum circuit for a single Trotter step on qubits (blue circles) representing a single spin-sector. The grey line shows the structure of the Jordan-Wigner encoding. ``Boundary hopping'' means a hopping term across the vertical boundary, these are not caught by the swap network and must be implemented with a separate circuit, see \cref{sec:circuits}. }
    \label{fig:model}
\end{figure}

Here we implement an efficient quantum algorithm for simulating the time-dynamics of the 2D Fermi-Hubbard model with periodic boundary conditions on the Quantinuum H2-2 trapped-ion quantum computer. Our system is described by the Hamiltonian
\begin{align}\label{eq:FH_Ham_main}
H = -J\sum_{\langle i,j\rangle\in \mathcal{L},\sigma}(e^{i\phi_{ij}}c_{i,\sigma}^\dagger c_{j,\sigma} + {\rm h.c.}) +U\sum_{i\in \mathcal{L}}n_{i,\uparrow}n_{i,\downarrow},
\end{align}
where $c_{i,\sigma}^\dagger$, $c_{j,\sigma}$ are fermionic creation and annihilation operators at sites $i$, $j$, $n_{i,\sigma} = c^\dagger_{i,\sigma}c_{i,\sigma}$ and spin $\sigma=\{\uparrow,\downarrow\}$. The first term describes the hopping of electrons in the lattice $\mathcal{L}$, characterised by the hopping integral $-J$. The pattern of phases $\phi_{ij}$ corresponds to the insertion of magnetic flux in the system.
We study this model on a doubly-periodic lattice (torus) with $|\mathcal{L}|=7\times 4$ sites and a $\pi$ phase flux in the long direction -- i.e.\ the Peierls phase in the short direction (see \cref{fig:model} top). While the $\pi$-flux phase appears naturally in the study of Dirac semi-metals, here we add them to lower the average energy of the initial state \cite{Lieb1994}. This lattice size is beyond the reach of exact classical simulation in practice: the Hilbert space explored by the system is $\ge 2^{43}$-dimensional, even taking symmetries into account, and exactly simulating the dynamics of a quantum system is more complex than simply writing down the state of that system.

Periodic boundary conditions reduce spurious boundary effects that do not contribute in the thermodynamic limit.
This is necessary for mitigating the need to use large model sizes to approximately capture translational invariance, and we anticipate that this will remain relevant in the quantum simulation of such systems. Trapped-ion quantum computers are particularly well-suited to modelling this periodicity because of their connectivity model. The qubits (ions) are freely reconfigurable with a comparatively low cost in error and run-time compared to gate operations, yielding an effective all-to-all connectivity model. This affords the freedom of non-local interactions with minimal cost overhead, allowing for efficient implementation of periodic boundaries.

We choose to start with a dimerised state at half-filling (see \cref{fig:model} middle). The initial state corresponds to a dimer covering that is broken at two points, where a holon (i.e.\ a fully unoccupied site) and a doublon (a doubly occupied site) are located. Our experiment begins with this state, time-evolves for times $t \in [0.1,2]$ in $0.1$ increments (all times are in units of inverse hopping $J^{-1}$), and measures in the real-space occupation basis (computational basis in the qubit representation), enabling any property constructed from spin-resolved densities to be determined.

We consider the non-interacting ($U/J=0$) and interacting ($U/J=4$) regimes. As is conventional, we set $J=1$ and measure all energies in units of the hopping strength $J$.
The non-interacting Fermi-Hubbard model is well-known to generally be exactly solvable classically, both for simulating time-evolution and for ground-state properties. However, in our case, straightforward classical simulability does not apply, even for the non-interacting model.
This is because the initial state we use is not a fermionic Gaussian state, i.e.\ it is not a single Slater determinant in some basis. Instead, it is an example of a so-called fermionic magic state~\cite{hebenstreit2019all,fermionsampling}: that is, one of a family of states which promote fermionic linear optics to universal quantum computation. This means that, as one scales up the family of instances considered in this work, we should not expect even non-interacting time-dynamics starting with these states to continue to be classically simulable. Nevertheless, for small operator weights, we are able to show that the $U=0$ regime remains simulable in this setting (\cref{sec:nearflo}).

In this work, we study several signatures of spin-charge separation, including local and global charge and spin correlations, together with extended many-body correlators (Wilson loops and open lines) that diagnose the potential between spin carriers (spinons) and the potential between charge carriers (holons and doublons). It is believed that spin-charge separation plays a fundamental role in the appearance of high temperature superconductivity \cite{Senthil2000,bennemann2008superconductivity}.

\section*{Results}
\begin{figure*}[!htbp]
    \centering
    \includegraphics[width=\textwidth]{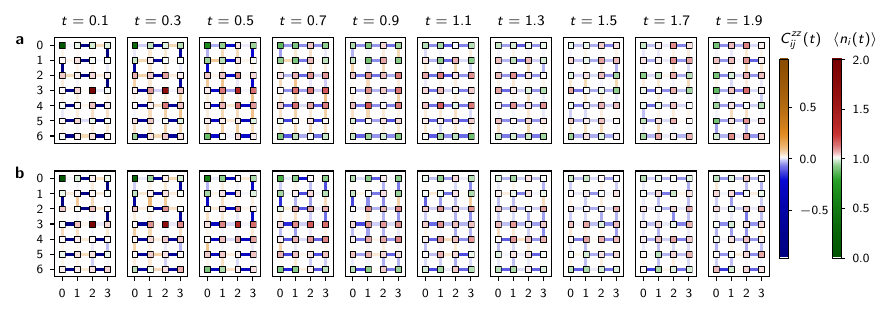}
    \caption{
    \textbf{Evolution of local charge, and spin correlations.}
    Local charge density $\langle n_{i}(t)\rangle $ (squares as sites) and spin (connected) correlation function $C^{zz}_{ij}(t)$ between nearest-neighbours (represented by links) as a function of time for \textbf{a,} $U=0$ and \textbf{b,} $U=4$ for the error mitigated (TFLO + GPR) experimental data.
    In the charge sector, we see diffusion from the initial holon-doublon configuration towards the uniform state. In the free case \textbf{(a)}, the charges develop a charge-density profile oscillating in the y direction, while in the interacting case \textbf{(b)}, the charge profile is more disordered. In the spin sector, the initial triplet configuration takes $\sim t=0.7$ to melt, leaving behind a residual antiferromagnetic correlation, greater for the interacting case (see also \cref{fig:doublons_and_triplets}).}
     \label{fig:densities_and_spin_correlation}
\end{figure*}

We begin by using our simulation algorithm to obtain an overall picture of the time evolution of our Fermi-Hubbard model instance. \cref{fig:densities_and_spin_correlation} shows the evolution of charge densities $\langle n_{i}(t)\rangle = \langle n_{i\uparrow}(t)\rangle  + \langle n_{i\downarrow}(t)\rangle$, and the neighbouring spin (connected) correlations
\begin{equation}\label{eq:neighbour_SZSZ}
    C^{zz}_{ij}(t) = 4\left(\langle S^z_i(t) S^z_j(t)\rangle-\langle S^z_i(t) \rangle \langle S^z_j(t)\rangle\right),
\end{equation}
as a function of time, for $U=0$ (panel a) and $U=4$ (panel b). Here $S_i^z=(n_{i,\uparrow} - n_{i,\downarrow})/2$. We observe that the initial charge configuration diffuses radially, while the initial spin correlations vanish. While in both cases the charge tends towards the homogeneous state, several differences emerge.
For $U=0$ at large times ($t=1.9$), the charge arranges into charge-density waves in the short direction of the lattice (see also \cref{app:additional_results}).
Note as well a slight antiferromagnetic tendency in the $y$ direction, which can be attributed to the initial triplet configuration (see \cref{fig:model}), while no such charge-density wave appears for $U=4$.
The presence of interactions also slightly favours overall antiferromagnetic order at late times, in a pattern that is completely disordered with respect to the initial template of correlations.

\begin{figure*}[!htbp]
    \includegraphics[width=\textwidth]{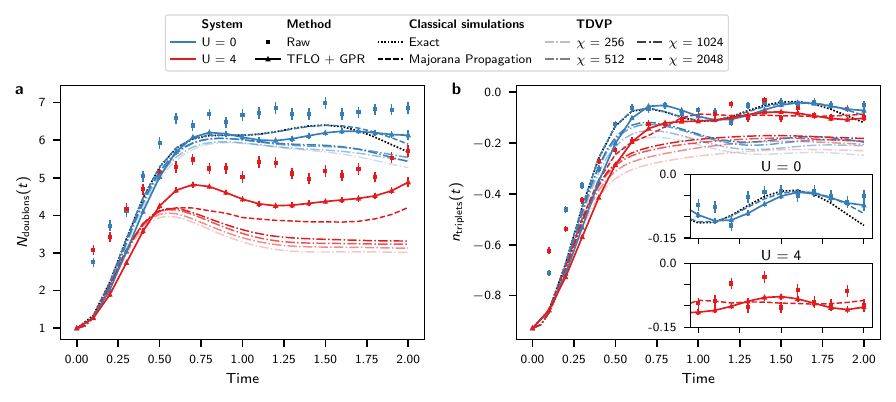}
    \caption{
    \textbf{Evolution of global doublon charge and magnetic correlations.}
    \textbf{a,} Total number of doublons and \textbf{b,} triplet density as a function of time. While the initial state has a fixed energy density, it is not an equilibrium state.
    In particular, the number of doublons evolves nontrivially with time (panel \textbf{a}). By mitigating the raw data with TFLO + GPR, we recover a signal that achieves better agreement with the ground truth, here represented by Majorana propagation. For $U = 0$, we verified that the latter overlaps exactly with the FLO simulation of the Trotterised circuit (not shown). In both panels, the dotted line represents the exact, untrotterised FLO simulation of the time evolution for $U = 0$.
    The TFLO + GPR curve has been obtained by symmetrising over doublons and holons (since $N_{\mathrm{doublons}}(t) = N_{\mathrm{holons}}(t)$) via $N^{\mathrm{sym}}_{\mathrm{doublons}}  =(N_{\mathrm{doublons}} + N_{\mathrm{holons}})/2$.
    Since $N_{\mathrm{doublons}}$ and $ N_{\mathrm{holons}}$ are not independent, in computing the error bars of $N^{\mathrm{sym}}_{\mathrm{doublons}}$, we assumed perfectly correlated errors (i.e., $\sigma^{\mathrm{sym}}_{\mathrm{doublon}} = \sigma_{\mathrm{doublon}}+\sigma_{\mathrm{holon}}$).
    Note that for $N_{\rm doublons}$, the results from Majorana propagation are equally distant from the TDVP simulations as from the experimental data, but the experiment is capturing the late increase in doublon number.
    From panel \textbf{b}, we can see that while the initial melting of the antiferromagnetic order is slower for the interacting system, the spin ordering is lost essentially at $t\gtrsim 0.5$. In $n_{\rm triplets}$, the agreement between the mitigated experimental signal and the results from Majorana propagation is within the error bars, while the results from tensor network TDVP simulations converge to a smaller negative value, indicating more antiferromagnetic order in the tensor network simulations than expected for this state. The insets isolate the late-time dynamics of $n_{\rm triplets}$ for ease of comparison. In both panels, error bars indicate one standard deviation from the mean.}
    \label{fig:doublons_and_triplets}
\end{figure*}

At half-filling, the charge carriers are doublons and holons, which can emerge without constraints when the interaction vanishes, as the spin sectors are totally decoupled in this regime. In contrast, for large interactions, the doublon creation has to overcome an energy $\sim U$. We can directly inspect the proliferation of doublons (and holons), captured by the observable
\begin{align}
    N_{\rm doublons} \coloneqq \sum_{i\in \mathcal L} n_{i,\uparrow}n_{i,\downarrow}
    \label{eq:doublon_sum}
\end{align}
in \cref{fig:doublons_and_triplets}a, which equilibrates to a value that decreases as the interaction increases. Note that for $U=4$, we observe that the creation of holon-doublon pairs is not accurately modeled by the $t$-$J$ model \cite{Spalek_1978}, which assumes a constant number of pairs, in contrast to the proliferation seen in  \cref{fig:doublons_and_triplets}.
We also include the results from the Trotterised circuit (for $U=0$) to show the accumulation of Trotter error at large times. Note that this error is smaller than the statistical uncertainty due to sampling measurement results.

We assess the magnetic ordering through the nearest-neighbour triplet density
\begin{equation}
   n_{\mathrm{triplets}}(t) \coloneqq \frac{2}{L_x L_y} \sum_{\langle i,j\rangle }C^{zz}_{ij}(t),
   \label{eq:triplet_density}
\end{equation}
shown in \cref{fig:doublons_and_triplets}b. Initially, the order melts, rapidly approaching zero. The late-time behaviour of the signal shows equilibration with slightly stronger antiferromagnetic tendency in the presence of interactions compared with the non-interacting case.
For both the number of doublons and the triplet density, we note that the experimental results differ from the tensor network simulations at times $t\gtrsim 0.5$ for $U=4$.
For the doublon number, we observe that, while Majorana propagation predicts a smaller number of doublons than the experiment, the trends match, including the late-time increase in doublon population.
Likewise, the number of triplets exhibits agreement between Majorana propagation and the experimental data, within small experimental errors.
For $U=4$, it is not possible to exactly determine the effect of Trotter error, and it is possible that the late time increase in the doublon number is an artifact of this. However, small-scale simulations (\cref{sec:trotter-error}) suggest that Trotter error may be substantially lower than worst-case commutator bounds would indicate.

\begin{figure}[!htbp]
    \centering
    \includegraphics[width=0.95\columnwidth]{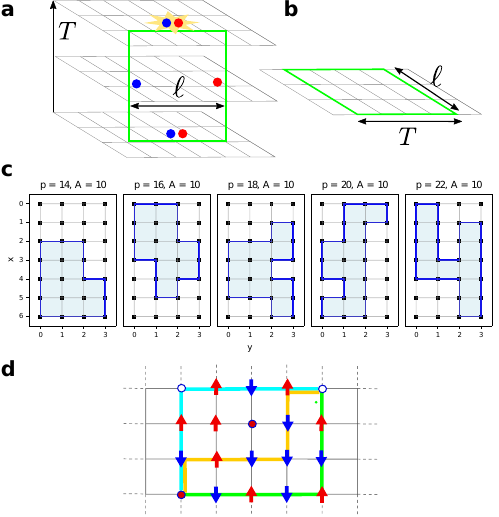}
    \caption{
    \textbf{Wilson open lines. }
    \textbf{a,} In (lattice) gauge theory, the Wilson loops that serve as order parameters of the confinement/deconfinement transition are usually taken in space-time, here sketched as a green rectangle spanning a region of size $\ell$ in space and (Euclidean) time $T$.
    \textbf{b,} In our analysis, we fix the time slice and analyse the expectation of Wilson lines as a function of perimeter or area for different real times.
    \textbf{c,} Examples of some of the closed Wilson loops considered in the computation of $\mathcal{W}_d(\mathcal{C})$ in \cref{fig:wilson_time_averaged}a below for area $A = 10$ and varying perimeters $p$.
    \textbf{d,} Different paths in the computation of holon-doublon Wilson lines, located in opposite corners and represented by a white and red circle, respectively. (Green) This path has a full Néel order of the spinons in between. (Yellow) where the Néel order is broken. (Blue) Path where holons are present.}
    \label{fig:Wilson_space_time}
\end{figure}

\subsection*{Spin-charge separation and deconfinement}
The average energy of the initial state is comparable to an equilibrium state with temperature $T\sim3J$ (see \cref{app:physics}). A Fermi-Hubbard system at half-filling with that temperature is expected to host fractionalised quasiparticles \cite{Lee_2006, Lee_2008}. We study the fractionalisation of the constituent electron into a fermionic particle carrying spin 1/2 (spinon) and a boson carrying the charge (holon/doublon).
We investigate this by expressing the Fermi-Hubbard model in terms of a modified Kotliar-Ruckenstein representation \cite{Kotliar_1986} of the fermion operator, which ultimately leads to a dual $U(1)$ gauge theory description of the Fermi-Hubbard Hamiltonian (see \cref{app:Fractionalization} for details). In this dual picture, spinons interact with an emergent gauge field described by the motion of the doublons/holons.
 Constructing gauge-invariant operators in terms of this gauge field, we find the following Wilson loop operators
\begin{align}\label{eq:Wilson_loops}
  \mathcal{W}_{d}(\mathcal{C})\coloneqq \prod_{j\in\mathcal{C}}(1-n_{j,\uparrow}n_{j,\downarrow}),
\end{align}
where $\mathcal{C}$ is a closed loop of sites in the lattice. The expectation value of these Wilson loops can be used to diagnose the confinement/deconfinement transition of the gauge field mediating the interaction between spinons. Usually, the Wilson loop that detects confinement/deconfinement is taken in a space-time loop (see also \cref{fig:Wilson_space_time}a), which can be directly understood as the Euclidean action of a potential between two charges \cite{POLYAKOV1978477}.
In our case, we measure the Wilson loop at fixed space-like regions (i.e.\ for a fixed time): see e.g.\ \cref{fig:Wilson_space_time}b.
These two descriptions are related, but not identical. It has been shown~\cite{BORGS1985455} that the fixed time (or horizontal) Wilson lines are not a faithful indicator of confinement, as they can show an area law in the deconfined phase. As discussed in \cite{BORGS1985455}, if the system in spatial dimension $d-1$ is confined, then the horizontal Wilson lines will exhibit an area law, regardless of the true nature of the system in $d$ dimensions.
However, in our setting, we can still use the horizontal Wilson lines as indicators of the confinement/deconfinement transition, because in one dimension the Fermi-Hubbard model is deconfined for any interaction parameter, thus avoiding this constraint.

\begin{figure*}[t]
    \centering
    \includegraphics[width=\textwidth]{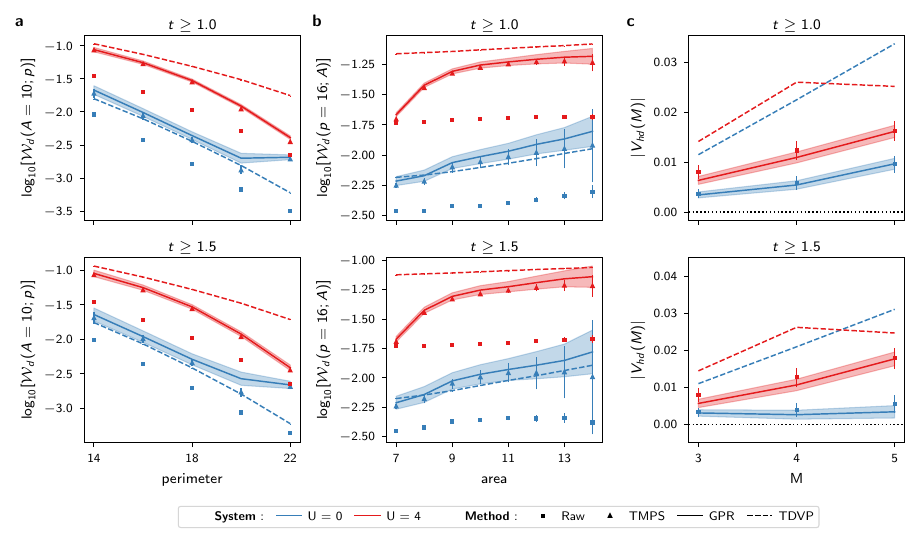}
    \caption{
    \textbf{Time averaged Wilson loops and lines.}
    Time averaged Wilson loops as defined in \cref{eq:Wilson_loops} with:
    \textbf{a,} fixed area, varying perimeter; \textbf{b,} fixed perimeter, varying area. Here, we observe the scaling of space-like Wilson loops with the perimeter.
    This operator is the order parameter of the $U(1)$ gauge field mediating the interactions between spinons. The nontrivial scaling with perimeter signals deconfinement.
    \textbf{c,} Time average of the expectation of the (absolute value of the) open Wilson line $V_{hd}(M)$ as a function of the Manhattan distances $M$.The expectation value of this operator measures the effective interaction potential between a doublon and a holon. Increasing interaction with distance signals confinement.
    Both experimental and TDVP data have been smoothed using GPR.
    In all panels, square points (Raw) indicate the raw device signal, triangular points (TMPS) indicate device signal corrected by the training with MPS (TMPS) procedure described in \cref{app:subsec:tflo}, and the solid lines (GPR) are the curves produced by GPR applied to the triangular (\textbf{a}, \textbf{b}) or square (\textbf{c}) points, in both time and perimeter/area directions. TMPS data have been obtained by training with TDVP data up to $t=0.3$. Dashed lines are produced by sampling $10,000$ shots from the MPS obtained by TDVP with $\chi = 2048$.
    }
    \label{fig:wilson_time_averaged}
\end{figure*}

In Figures~\ref{fig:wilson_time_averaged}a and~\ref{fig:wilson_time_averaged}b, we plot the behaviour of area vs.\ perimeter in Wilson loops. Here, we fix the area (respectively, perimeter) and study the expectation value of this observable as a function of the perimeter (area). An illustration of the type of loops chosen for fixed area and varying perimeter is shown in \cref{fig:Wilson_space_time}c.
To decouple the transient effects of the initial state, and obtain a signal defined by the initial energy density, we perform a time average over times $t$ larger than 1.0 (1.5).

We show both raw data and data mitigated using the short-time results from MPS-based TDVP simulation.
The latter is discussed in more detail in \cref{app:subsec:tflo}.
The values reported by the error-mitigated experiment diverge from tensor network results for the largest available times $t$.
The observables corresponding to the largest Wilson loops we consider have over a million terms spanning all Majorana weight sectors up to weight~44. Given the failure of Majorana propagation to accurately simulate the open Wilson lines of \cref{eq:open_wilson_lines} below, which involve a significantly smaller number of terms, we do not expect Majorana propagation methods to be effective for the closed Wilson loops of \cref{eq:Wilson_loops}.
This suggests that the confinement/deconfinement transition witnessed by $\mathcal{W}_d$ cannot be captured by Majorana propagation.

We observe that $\mathcal{W}_d$ scales with the perimeter of the loop (for fixed area), for both the non-interacting and the interacting cases.
While the quantitative predictions made by the classical and quantum methods disagree, both predict a perimeter scaling of the Wilson loops, with the quantum computation showing a stronger dependence on the perimeter for small loops, and also a nontrivial area scaling in the presence of interactions (\cref{fig:wilson_time_averaged} b), saturating to a value in agreement with the classical computations.
 As $\mathcal{W}_d$ is a Wilson loop constructed from \textit{link} operators mediating the interaction between spinons, this indicates their deconfinement.
 To probe spin-charge separation, we need to compare the effective potential between charge carriers with the potential between spinons.
 The operator that measures the potential between charge carriers is an open Wilson line motivated by the dual $U(1)$ lattice gauge theory description of the model (see \cref{app:Fractionalization}) and takes the form
 \begin{align}
     V_{\alpha\beta}(M)
     \coloneqq \sum_{\substack{\textnormal{sites }  i, j:\\ |i-j| = M}}  \frac{n^\alpha_{i} n^\beta_{j}}{N_{\rm pairs}}
    \sum_{\substack{\textnormal{paths } \gamma:\\ \textnormal{ from } i \textnormal{ to } j }}\frac{\sum_{(m, n) \in \gamma} S_{m}^{z}S_{n}^{z}}{N_{\rm paths}(i,j)}
    \label{eq:open_wilson_lines}
 \end{align}
 where $\alpha, \beta = \{h, d\}$ and $n_i^d\coloneqq n_{i\uparrow}n_{i\downarrow}$, $n_{i}^h\coloneqq (1-n_{i\uparrow})(1-n_{i\downarrow})$. The outermost sum runs over all pairs of sites with distance $M$, while the innermost sum runs over all pairs of sites $(m, n)$ along a path $\gamma$ from $i$ to $j$. Here $N_{\rm paths}(i,j)$ is the number of paths between a given pair $(i,j)$. An illustration of the type of paths that enter in the computation of the potential $V_{\alpha\beta}$ is shown in \cref{fig:Wilson_space_time}d.
 We extract this from the measured shots in the computational basis. In \cref{fig:wilson_time_averaged}c we show the value of this observable as a function of the (Manhattan) distance between a doublon and a holon, obtained from raw experimental data and from sampling the tensor network state at bond dimension $\chi =  2048$.
 We do not include Majorana propagation results because these failed to converge as we decreased the truncation parameters, even for low times, as detailed in \cref{app:sec:majorana_propagation_simulations}.

 In \cref{fig:wilson_time_averaged}c we observe some striking differences between different simulation methods, as well as between interacting and non-interacting models.  As time progresses, the experimental values show a clear separation between the $U=0$ and the $U=4$ regimes, where $U=0$ shows no signal of a growing confining potential between a holon and a doublon with distance, as opposed to the interacting case $U=4$. In contrast, the classical simulation predicts a confining potential even for zero interaction. It is possible that the difference between the experimental signal and the TDVP results could be attributed to Trotter error, although it appears at times where Trotter error is expected to be small (see \cref{sec:trotter-error}).

Another plausible source of difference between computational methods could be the effective energy density of the states produced by the different methods. The lower production of doublons and higher antiferromagnetic order for mid-to-late times in the classical tensor network simulations observed in \cref{fig:doublons_and_triplets} suggest the state produced through TDVP has a smaller energy density than the corresponding one explored by the quantum computer. To illuminate this, we also explore the behaviour of the closed and open Wilson lines as a function of their size for the ground-state of the system, obtained through Density Matrix Renormalisation Group (see \cref{app:subsec:dmrg} for implementation details).
This result is shown in \cref{fig:wilson_panel_dmrg} in \cref{app:additional_results}, for both the closed and open Wilson lines. While we observe similar perimeter scaling of $\mathcal{W}_d$ for the ground-state, TDVP and quantum results, the expectation of the open line operators for the Wilson loops in the ground-state differ from the ones obtained experimentally for $U=4$, while being similar to the results obtained by TDVP. We take this as another signature of the different energy density that the states can achieve.
Taking the experimental results at face value, the closed Wilson lines show deconfinement of the gauge field acting between spinons, while the open Wilson lines between holons and doublons show a signature of a growing confining potential between them, which are the charge carriers. This signals spin-charge separation for the state at this energy density.

\section*{Outlook}


We have shown that it is possible to analyse the time-dynamics of a periodic materials model that is beyond the capacity of exact classical simulation, using an algorithm running on a digital quantum computer and executing over 2000 two-qubit quantum gates.
In some cases, the results we obtain diverge both quantitatively and qualitatively from those obtained through all the advanced approximate classical simulation techniques available to us.
In particular, we provide evidence that -- at the energy densities explored -- the Fermi-Hubbard model exhibits spin-charge separation by examining the deconfinement of the field mediating the spinon interactions and the potential between charge carriers.
We have validated our results by comparing them against exact classical simulations in the interaction-free case, and approximate classical simulations in the regimes where they are expected to be accurate. Beyond the observables we consider, we believe that many additional physical properties of the Fermi-Hubbard model can be obtained from the data we collected.

To determine whether a time-dynamics simulation outperforms the capabilities of classical computers, we require that the following desiderata are satisfied: (i)~the results obtained from the quantum algorithm agree with exact classical methods (here, FLO) in a classically tractable regime (here, $U=0$); (ii)~the feasible approximate classical methods (here, MPS-based TDVP simulations) do not agree with exact classical methods in a classically tractable regime; (iii)~the results from the quantum algorithm are not reproducible by approximate classical methods in a regime that is intractable for exact classical methods (here, $U=4$).
Criterion~(iii) provides evidence that at least one of the device and classical simulation methods produces an inaccurate result for $U=4$. Criteria~(i) and~(ii) together provide evidence that, of the two methods, the device is the one more likely to be producing an accurate result.
The Wilson lines/loops discussed here fulfill these criteria, with further evidence provided by cross-entropy benchmarking. This heralds the use of quantum computing for simulating strongly-correlated electronic systems beyond the capacity of classical computing.
The boundary of classical tractability is not final and changes with the development of more specialised and efficient classical methods. We have made all experimental data publicly available, and invite the community to attempt to replicate these findings using classical techniques.

We used a quantum simulation algorithm based on a second-order Trotter formula, which was found to be effective in a recent study~\cite{thrift}. Future work could explore more efficient algorithms. Ongoing improvements in quantum gate fidelities will enable access to longer evolution times and larger system sizes, elucidating physics such as the role of charge fluctuations or external time-dependent pulses in the development of superconducting order~\cite{Fava2024-lo}. We expect that the path from here to full fault-tolerance will be full of discoveries, with error correction playing an increasingly fundamental role as quantum gate fidelities improve and the sizes of simulated systems increase.

\section*{Data availability}
Raw data were taken from Quantinuum's System Model H2-2 trapped-ion quantum computer in June and July 2025.
Those data are available at~\cite{zenodo_dataset}, along with expectation values of observables presented in figures throughout, computed from shot data as well as using other numerical techniques.

\section*{Acknowledgments}
We wish to acknowledge the Quantinuum team for their guidance in developing this experiment.
We would also like to thank Vanya Eccles, Callum MacPherson, Sam White and the TKET support team for technical assistance throughout the experiment.
We thank Andrew Childs for helpful comments on an earlier version.
This work received funding from the European Research Council (ERC) under the European Union's Horizon 2020 research and innovation programme (grant agreement No.\ 817581).

{
\setlength{\parindent}{0pt}

\newpage 
\hypertarget{author_dagger}{
\noindent {$^\dagger$} \textbf{Authors}\\
}

\noindent \textbf{Phasecraft} \\ 
Faisal Alam$^1$, 
Jan Lukas Bosse$^1$, 
Ieva \v{C}epait\.{e}$^1$, 
Adrian Chapman$^2$, 
Laura Clinton$^1$, 
Marcos Crichigno$^2$, 
Elizabeth Crosson$^2$, 
Toby Cubitt$^{1,3}$, 
Charles Derby$^1$, 
Oliver Dowinton$^1$, 
Norhan Eassa$^1$,
Paul K. Faehrmann$^{1,4}$, 
Steve Flammia$^{2,5}$, 
Brian Flynn$^1$,
Filippo Maria Gambetta$^1$, 
Ra\'ul Garc\'ia-Patr\'on$^{1,6}$, 
Max Hunter Gordon$^1$, 
Glenn Jones$^1$, 
Abhishek Khedkar$^1$, 
Joel Klassen$^1$, 
Michael Kreshchuk$^2$, 
Edward Harry McMullan$^1$, 
Lana Mineh$^1$, 
Ashley Montanaro$^{1,7}$,
Caterina Mora$^1$,
John J. L. Morton$^{1,8}$, 
Alberto Nocera$^1$,
Dhrumil Patel$^{2,5}$, 
Pete Rolph$^1$, 
Raul A. Santos$^1$, 
James R. Seddon$^1$, 
Evan Sheridan$^1$, 
Wilfrid Somogyi$^1$, 
Marika Svensson$^1$, 
Niam Vaishnav$^1$, 
Sabrina Yue Wang$^1$,
Gethin Wright$^1$\\
\\
\textbf{Quantinuum collaborators}\\
Eli Chertkov$^9$,
Henrik Dreyer$^{10}$,
Michael Foss-Feig$^9$\\

\noindent\textbf{Author contributions}\newline
JK, RAS, AM conceived of the project and provided leadership and oversight. 
JK, CD, RAS, BF, SYW designed the quantum circuits. 
BF, SYW, JLB, EHM implemented and executed the circuits on hardware. 
MH, JK, JLB, AM, SYW, EC, FMG developed and applied the error mitigation techniques. 
BF, SYW, RAS, FMG, AK, PR, LM performed the tensor network simulations. 
SYW, JLB, BF, AM performed the FLO simulations. 
JLB implemented the Majorana propagation. 
LC, JK performed the Trotter error analysis. 
EC, TC, SYW, MHG, MS, BF, NV performed the XEB analysis. 
RAS, SYW, FMG, MK, AM undertook the physics analysis. 
BF, MHG, EHM, JLB, SYW, FMG developed the underpinning algorithmic infrastructure used for data analysis.
EC, HD and MF provided scientific and project consultation and liaised between the Phasecraft and Quantinuum teams.
All authors wrote and revised the manuscript and the Supplementary Information.

\vfill

\begin{center}
\rule{0.25\textwidth}{0.4pt}  
\end{center}

$^1$ Phasecraft Ltd, London, UK\\
$^2$ Phasecraft Inc, Washington DC, USA\\
$^3$ Department of Computer Science, University College London, UK\\
$^4$ Dahlem Center for Complex Quantum Systems, Freie Universit\"{a}t Berlin, 14195 Berlin, Germany\\
$^5$ Department of Computer Science, Virginia Tech, USA \\
$^6$ School of Informatics, QSL, University of Edinburgh, UK \\
$^7$ University of Bristol, UK\\
$^{8}$ Department of Electrical and Electronic Engineering, UCL, London, UK\\
$^9$ Quantinuum, 303 S Technology Ct, Broomfield, CO 80021, USA\\
$^{10}$ Quantinuum, Leopoldstrasse 180, 80804 Munich, Germany
\\

}

\clearpage
\twocolumngrid
\section*{Methods}

The first step in representing a fermionic model on a quantum computer is to choose a fermionic encoding, which maps fermionic modes to qubits while preserving fermionic anti-symmetry. Here we use the well-known Jordan-Wigner transform, which represents the Fermi-Hubbard model space-optimally, using $2$ qubits per site and hence $56$ qubits to represent a $7\times 4$ system. The price paid is that certain hopping terms in \cref{eq:FH_Ham_main} are mapped to high-weight Pauli strings of the form $(X_i X_j + Y_i Y_j)Z_{i+1}\dots Z_{j-1}$. However, we can mitigate this cost using fermionic swap networks~\cite{kivlichan18,cade20}, which enable sequences of long-range -- and hence high-weight -- operations in the Jordan-Wigner transform to be implemented efficiently.

We simulate time-dynamics using a second-order Trotter formula,
\begin{equation}
    e^{-iHt} \approx \left(\prod_{h\in \mathcal H}^\rightarrow e^{-ih\frac{t}{2k}}\prod_{o\in\mathcal O}e^{-iot/k}\prod_{h\in\mathcal H}^\leftarrow e^{-ih\frac{t}{2k}}\right)^k,
\end{equation}
where $\mathcal H$ denotes the set of hopping terms, $\mathcal O$ denotes the set of onsite terms, and the ordering of the terms in $\mathcal H$ is determined by the swap network. We execute $k=4$ second-order Trotter steps (with a similar structure to 8 first-order steps, but with better accuracy than directly applying first-order bounds would suggest) and time-evolve the initial state up to time $t=2$. Exact classical simulation of low-weight observables for $U=0$ shows that, in this instance, Trotterised time-dynamics approximates the true dynamics up to small errors -- far beyond the regime where theoretical bounds hold (see \cref{sec:errormitigation}). As we cannot exactly simulate the interacting case or the dynamics of high-weight observables, we resort to smaller-scale experiments to estimate the level of accuracy of Trotterisation in these cases. Based on this (see \cref{sec:trotter-error}), and on comparison against the ground truth in the $U=0$ setting for low-weight observables, we estimate that (noiseless) Trotterised dynamics are accurate up to time $t\approx 1.5$.

The circuit that we implement for initial state preparation and for time evolution is illustrated in \cref{fig:model}.
We obtain additional gate savings beyond a standard swap network by observing that there is no need to reorder the qubits at the end of the swap network to return to the initial ordering, given that the symmetric structure of a second-order Trotter step naturally undoes any shuffling of the ordering. In addition, we use the structure of the second-order Trotter formula to merge each layer's final time evolution by hopping terms (which are executed in reverse order) with the next layer's first time evolution by hopping terms in forward order. Further details may be found in \Cref{sec:circuits}.

Previous uses of fermionic swap networks for the Fermi-Hubbard model have focused on open boundary conditions~\cite{kivlichan18,cade20,google-fhvqe}. Periodic boundary conditions in the short direction of the lattice can be implemented without any increase in circuit complexity, because all modes pass each other at some point during the fermionic swap network, and hopping terms can be merged with fermionic swap operations without any additional cost. Furthermore, periodic boundary conditions in the long direction can also be implemented at little additional cost. This is because for states of fixed parity in each spin sector, the parity operator -- a $Z$ string across the entire spin sector -- is constant, implying that we can replace the long $Z$ strings that would extend between two distant qubits with $Z$ strings on the complementary set of qubits. Some remaining parity corrections need to be made, which can be implemented with a small number of additional controlled-$Z$ gates.

The quantum circuits we execute contain at most 2,415 two-qubit gates and 4,627 one-qubit gates.
The two-qubit gates are CPHASE gates with varying angles, which are native gates on the device, up to one-qubit gates.
See \cref{sec:circuits} for a further description of gate decomposition, breakdown of gate counts and details of optimisations applied.

\subsection*{Error mitigation}

We implement three main techniques to improve the quality of our experimental results (see \cref{sec:errormitigation} for further details and validation).

The first of these is an error suppression technique known as \emph{Pauli pseudo-twirling} \cite{pseudo-twirling,quantinuum}. Pseudo-twirling is a variant of the well-known twirling technique, which is based on the idea that the primary source of error on many quantum circuit platforms -- and in particular on ion traps -- is two-qubit gates. Systematic errors on these gates can be reduced and converted into incoherent errors by conjugating each gate with Paulis and actively flipping the sign of the gate generator when it anti-commutes with the conjugate Pauli.

The second technique is an error mitigation method known as \emph{Training with Fermionic Linear Optics (TFLO)}~\cite{tflo}. This method uses the fact that quantum circuits consisting solely of so-called fermionic linear optics (FLO) operations can be efficiently simulated classically. This allows data sets of noisy and exact observable values to be prepared, enabling the inference of a map between exact and noisy data, thus allowing the effect of noise on a given observable to be inverted.
This map can then be applied to experimental data, which is not classically simulable, with the expectation that the error behaviour will be similar.
Time-evolution of the Fermi-Hubbard model is particularly well suited to the TFLO technique, since the quantum circuit for simulating time dynamics is FLO in the case $U=0$.
As the initial state we consider is not a Gaussian state, standard classical FLO computational techniques~\cite{terhal02} do not apply. We are nevertheless able to develop efficient classical algorithms for computing low-weight observables for the output of our experiment (see \cref{sec:nearflo}), allowing these to be computed exactly in seconds. We expect that high-weight observables (and also sampling from measurement outcomes) will require exponential cost, albeit lower than in the $U=4$ case. Thus, to mitigate errors in high-weight observables, we use an alternative technique where training data is produced using tensor network techniques for short times, which are expected to be accurate in that regime.

The final post-processing technique we use is \emph{Gaussian process regression (GPR)} \cite{Rasmussen_2004,prml}.
This technique enables us to obtain meaningful results from a very small number of shots per data point computed: only 160 shots per point, made up of 16 pseudo-twirled instances with 10 shots for each instance.
GPR produces estimates based on the assumption that each experimental value is sampled from a Gaussian distribution and nearby points along the time parameterized curves are correlated, with the level of correlation depending on the distance between the parameters. Additionally, this naturally allows the inclusion of prior information such as the function value or its derivative at special known points.
Since the Gaussian assumption is not strictly satisfied, we cross-validate the GPR method using a particle filter (see~\cref{subsec:ParticleFilter}).

\subsection*{Classical simulation}
Our quantum circuits on 56 qubits are beyond the capacity of direct state-vector simulation on classical computers. However, it is also necessary to consider more advanced classical simulation techniques, as these can sometimes simulate surprisingly large-scale and complex quantum computations. Here we considered multiple such techniques: direct tensor contraction; time-evolving a matrix product state (MPS) via the time-dependent variational principle (TDVP) method~\cite{haegeman2011time}, as well as alternative time-evolution techniques~\cite{provazza_fast_2024}; and Majorana propagation~\cite{majorana_propagation}.
We found that the computational resources required by direct tensor contraction scale poorly with the lattice size, with direct contraction already ruled out for a $6\times4$ lattice (see \cref{app:sec:quimb}).
We evaluated several approximate tensor network techniques for simulating the Fermi-Hubbard model and found that using TDVP on an MPS ansatz performed most reliably, so we report its results below alongside the experimental results, together with those of Majorana propagation.
See \cref{app:sec:tensor_network_simulations}
for a description of our simulations and an overview of alternative implementations considered.

As an overall test of how well our error-mitigated experiment fared against TDVP, we use cross-entropy benchmarking to compare the outputs of each of the experiments and TDVP against the exactly simulable ground truth distribution for $U=0$.
We find (see \cref{sec:xeb}) that the experiment achieves higher accuracy than simulation through TDVP with respect to this metric.
More concretely, we observe linear cross-entropy benchmarking fidelities between one and five percent after error mitigation, as compared with the ideal, continuous-time FLO calculations for $U=0$. Note that this comparison also accounts for the effect of Trotter error, as well as hardware error.


\clearpage
\appendix
\newpage
\onecolumngrid

\setcounter{page}{1}  
\appendix

\tocon
\tableofcontents
\clearpage

\section{The Fermi-Hubbard model}\label{app:physics}

The Fermi-Hubbard (FH) model represents a paradigmatic system in the study of strongly correlated materials. It is described by the single-band Hamiltonian
\begin{align}
H = -J\sum_{\langle i,j\rangle\in \mathcal{L},\sigma}(e^{i\phi_{ij}}c_{i,\sigma}^\dagger c_{j,\sigma} + {\rm h.c.}) +U\sum_{i\in \mathcal{L}}n_{i,\uparrow}n_{i,\downarrow},
\label{eqn:single_band_ham}
\end{align}
where $c_{j,\sigma}$ $(c_{j,\sigma}^\dagger)$ is a fermionic destruction (creation) operator at site $j$ and spin $\sigma=\{\uparrow,\downarrow\}$ satisfying the canonical anti-commutation rules $\{c_{j,\sigma},c^\dagger_{k,\sigma'}\}=\delta_{ij}\delta_{\sigma\sigma'}$. The density operator is given by $n_{j,\sigma}=c^\dagger_{j,\sigma}c_{j,\sigma}$. The first term describes the hopping of electrons in the lattice $\mathcal{L}$, characterized by the hopping integral $-J$. The pattern of phases $\phi_{ij}$ corresponds to the insertion of magnetic flux in the system.
In this model, interactions appear whenever a site is doubly occupied. This is modeled by the second term in \cref{eqn:single_band_ham}. In our experimental setup, we considered a rectangular $L_x\times L_y=7\times4$ lattice with double periodic boundary conditions and nearest neighbour hopping. 
As shown schematically in \cref{fig:model} in the main text (top), we added a small $\pi$ flux in the smaller direction (shown in light green), corresponding to magnetic field inside the tours. In bigger systems, the effect of this flux depends strongly on how the system is enlarged. Fixing $L_y$ and increasing $L_x\rightarrow \infty$, the system corresponds to an $L_y$-legged cylinder with flux. Using DMRG and field theory techniques, the low-energy description of this model can be understood qualitatively \cite{White_1994,Sierra_1996, Dagotto_1996, Cabra_1998}. At half filling, Umklapp terms open a charge gap in the presence of interactions. A spin gap also opens for even $L_y$, while the spin sector remains gapless for odd $L_y$. This is a direct manifestation of the Haldane conjecture \cite{Haldane_1983,Haldane_1983b}. In contrast, extending the system by repeating its unit cell in three dimensions corresponds to studying a stack of Dirac semimetals, a system that is radically different to the standard Fermi-Hubbard model \cite{Otsuka_2002,Otsuka_2016}.
In our particular scenario, adding a $\pi$ magnetic flux changes the eigenstates and eigenenergies of \cref{eqn:single_band_ham} compared to the case of zero flux, without breaking time-reversal symmetry. At zero interaction and half-filling, the ground-state goes from being unique ($\Phi=0$) to being 36-fold degenerate $(\Phi=\pi)$. More importantly for the following analysis, the Aharonov-Bohm flux substantially changes the time-dynamics evolution of the electrons. At $\pi$-flux, the paths connecting opposite sites on a plaquette pierced by the magnetic field interfere destructively, as opposed to constructively in the zero flux case. 

While at finite temperatures the Mermin-Wagner theorem \cite{Mermin_1966} prevents the existence of antiferromagnetic long-range order (AFLRO) in the half-filled FH model, it is widely accepted that at exactly zero temperature and high interaction strengths the half-filled FH system possesses AFLRO \cite{Anderson_1952,Takahashi_1989,Huse_1988,Okabe_1988,Reger_1988}. 
 
The low-energy description of the half-filled FH model maps to a 2D Heisenberg antiferromagnet, with gapless spin excitations. On the other hand, neutron scattering \cite{Lee_2003} and ARPES \cite{Hashimoto2014-hm} experiments reveal a systematic tendency of spin-gap formation above the critical superconducting temperature $T_c$ in a wide range of cuprates. The formation of a spin-gap and the emergence of superconductivity are expected to be related. A spin-gap implies that low-energy spin fluctuations are suppressed, thus reducing the effect of scattering channels that can break Cooper pairs, strengthening $d-$wave unconventional superconductivity \cite{Li_2018}.

This opens the question of how this physics can appear in the FH model, where the spin excitations are gapless. One possibility is that the system spontaneously dimerizes. In this scenario, the reduction in magnetic exchange energy is larger than the increase in energy from an elastic distortion in the lattice.

We study the melting of this dimerized state by performing a quench from the dimerized state \cref{fig:model} (bottom) by time evolving it with the Hamiltonian defined in \cref{eqn:single_band_ham}. The dimerization pattern that we chose allows us to reduce the circuit complexity of preparing the initial state, as it follows the Jordan-Wigner line discussed in \cref{sec:fermi-qubit}. This dimerization is broken at two sites to put a holon and a doublon, shown as an open and full circle in \cref{fig:model} (bottom).

The average energy of the initial state is $\langle \Psi|H|\Psi\rangle=U$, and its average density $n=N_{el}/L_xL_y=1$, where $N_{el}$ is the total number of electrons. Using the equation of state for the FH model \cite{LeBlanc_2013}, this energy density corresponds to that of a system with approximate temperature $T\sim 3J$. In the phase diagram, a system with this temperature and filling lies above the dome where antiferromagnetic order is dominant 
and right in the intersection of the strange metal and pseudogap regions.

At half filling, the charge carriers are doublons and holons, which can emerge without constraints when the interaction vanishes, as the spin sectors are totally decoupled in this regime. In contrast, for large interactions, the doublons have to overcome an energy $\sim U$ to appear. As we will further discuss in the next section, introducing a basis where the doublons, holons and spinons (the carriers of spin but no charge) are explicit, we map the FH model into an equivalent $U(1)$ lattice gauge theory. 

\newpage
\section{Additional results} \label{app:additional_results}

\begin{figure}[htpb]
    \centering
    \includegraphics[width=\textwidth]{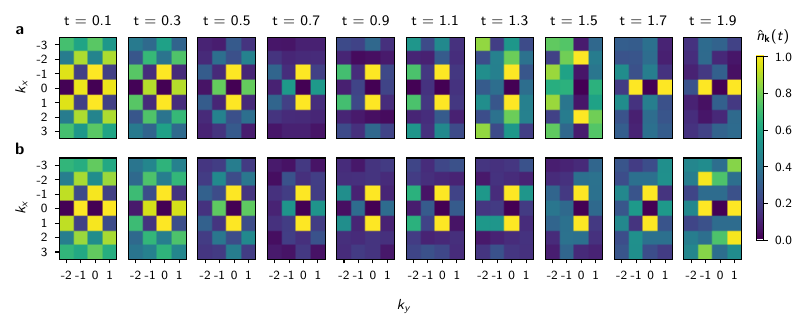}
    \caption{Normalised space Fourier transform of the mitigated experimental density profiles shown in \cref{fig:densities_and_spin_correlation} for a) $U = 0$ and b) $U = 4$. We set the central peak at $(k_x, k_y) = 0$. Note that the stripe charge density manifests here as a sharp peak at $k_x=0$ for long times in the non-interacting case.
    }
   \label{fig:spatial_FT_c}
\end{figure}

The initial state and the Hamiltonian are invariant under several symmetries. A particularly important one is spin-reflection $R c_{i,\uparrow}R^\dagger = c_{i,\downarrow}$. The presence of this symmetry immediately separates the spin and charge responses. The spin operator $S_i^z = \frac{1}{2}(n_{i,\uparrow}-n_{i,\downarrow})$ is odd under the symmetry action as $RS^z_i(t)R^\dagger= -S^z_i(t)$, while the initial state is even under this transformation. This implies that the expectation value of any operator consisting of an odd number of spin operators vanishes identically for all times. We discuss the use of this symmetry for error mitigation of observables in 
\cref{app:symmetry_averaging}.
On the other hand the charge density $n_i = n_{i,\uparrow}+n_{i,\downarrow}$ evolves non-trivially. As shown in \cref{fig:densities_and_spin_correlation}, the initial inhomogeneous charge configuration dilutes towards the uniform density state. To 
quantitatively capture this melting, and the spatial movement of the charges, in \cref{fig:spatial_FT_c} we include the results of the spatial Fourier transform of the charge density $\tilde{n}( \bm{k})\coloneqq \sum_r e^{i \bm{k} \cdot \bm{r}} n_{\bm{r}}-L_xL_y$ at different time steps, where we have substracted the zero momentum component. Note that initially (up to $t\sim 0.8$), the evolution of the charge in the interacting and non-interacting cases is very similar, while later times show a divergence in the signal, with late-time dynamics being of charge-density wave type (in the short direction)  for $U=0$  and disordered for $U/J=4$.

\begin{figure}[htpb]
    \centering
        \includegraphics[width=0.9\textwidth]{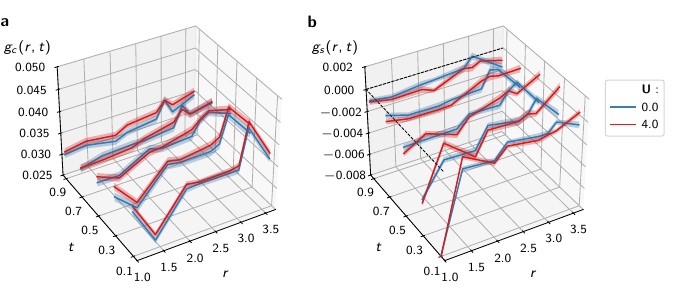}
    \caption{a) Pair correlation function for charges. (b) Pair correlation function for spins. In both panels, data for TFLO + GPR are shown, with shaded regions corresponding to errors.}
    \label{fig:pair_correlation}
\end{figure}

  The space-time evolution of the charges shows the expected equilibration towards the uniform density state in a way that resembles a fluid. 
  We can gain insight about the melting of the original order, and the effect of interactions by analysing the pair correlation function for charges
(see \cref{fig:pair_correlation} (left))
      \begin{align}
          g_c(r,t)=\frac{1}{L_xL_yZ(r)}\sum_{\bm{x}}\langle n_{\bm{x}}(t)n_{\bm{x}+\bm{r}}(t)\rangle, 
      \end{align}
      and spins (see \cref{fig:pair_correlation} (right))
      \begin{align}
          g_s(r,t)=\frac{1}{L_xL_yZ(r)}\sum_{\bm{x}}\langle S^z_{\bm{x}}(t)S^z_{\bm{x}+\bm{r}}(t)\rangle, 
      \end{align}
      where $Z(r)$ is the number of sites at a particular (Euclidean) distance $r$, so in the square lattice $Z(r=1)=4, Z(\sqrt{2})=4, Z(2)=4$ and so on. As seen from \cref{fig:pair_correlation}, the system displays a pairs correlations in the spin and charge sector that resemble a liquid at short times. As time progresses, the $g_{s,c}$ flatten, indicating a gas-like behavior. The peaks appearing at the largest Euclidean distances are a result of the radius $r$ wrapping around the system.

At zero interaction, the movement of holon-doublons should distort the antiferromagnetic order of the initial state. As the interaction increases, the exchange interaction between local spin moments penalizes a disruption of the antiferromagnetic order. We see this effect in \cref{fig:densities_and_spin_correlation} (right) where the connected spin-correlation function between nearest neighbor sites $i,j$ $C_{ij}^{zz}(t)$, defined in \cref{eq:neighbour_SZSZ}
is shown as a function of time. Note that the initial triplet ($m_z^{\rm total}=0$) state is an eigenstate with eigenvalue zero of the dimer Hamiltonian for any value of the interaction parameter. Since the total Hamiltonian can be viewed as the union of dimer Hamiltonians in all possible nearest-neighbour bonds, the initial order takes a considerable amount of time $t=0.4J^{-1}$ to melt. Once the melting has happened, we observe that higher interaction leads to a higher antiferromagnetic order at final times.
Summing over all possible links between nearest neighbors, we define the (triplet) density $ n_{\mathrm{triplets}}(t)$ in \cref{eq:triplet_density}. Its behavior as a function of time is shown in \cref{fig:doublons_and_triplets}.

The expectation of a considerably higher antiferromagnetic order in the interacting regime contrasts with the death of triplet correlations shown in \cref{fig:doublons_and_triplets} (right). This seemingly puzzling behavior can be understood from a microscopic point of view. In all the results above, we observe an interplay between the spin and the charge degree of freedom, for non-zero interaction. In particular, the spin order is dampened by the presence of a non-zero density of mobile charge carriers, which can move and distort the spin alignment of the background.
In order to disentangle the effect of the mobile charges from the spin background, we study the fractionalisation of the electron into spin and charge quasiparticles.

\subsection{Fractionalisation}\label{app:Fractionalization}

In the one-dimensional FH model, it is possible to show \cite{Ogata_1990} that the fundamental electronic degree of freedom fractionalises into spin and charge modes that propagate independently. The situation in two dimensions is not so sharp. It is believed \cite{Lee_2006} that the electron fractionalises into spin and charge degrees of freedom above some temperature, while the ground-state is confining, but there is no conclusive evidence of this.
One manifestation of this physics appears in the behaviour of extended Wilson operators. In this subsection, we explain this connection and present numerical and experimental results on the expectation values of these operators, which support the idea that for the state studied, the electronic degrees of freedom are indeed fractionalised.

We study this problem from the point of view of a dual-lattice gauge theory description of the FH model, where the electron is fractionalised into the (modified) Kotliar-Ruckenstein representation \cite{Kotliar_1986}
\begin{align}
c_{i\sigma}^{\dagger}=f_{i\sigma}^{\dagger}h_{i}+\sum_{\sigma'}\epsilon_{\sigma\sigma'}f_{i\sigma'}d_{i}^{\dagger},
\end{align}
where $\epsilon_{\sigma\sigma'}=-\epsilon_{\sigma'\sigma}$ and $\epsilon_{\uparrow\downarrow}=1.$
Here $f_{i\sigma}$ are fermionic fields satisfying the usual anticommutation
relation $\{f_{i\sigma},f_{j\sigma}^{\dagger}\}=0$, $\{f_{i\sigma},f_{j\sigma}^{\dagger}\}=\delta_{ij}\delta_{\sigma'\sigma}$
while $h_{i},d_{i}$ are bosonic fields satisfying the $\mathfrak{sl}(2, \mathbb{C}) $ algebra
\begin{align}
[h_{i},h_{j}^{\dagger}] & =-\delta_{ij}Z_{j}^{h},\quad[Z_{j}^{h},h_{i}^{\dagger}]=\delta_{ij}h_{j}^{\dagger},\\
[d_{i},d_{j}^{\dagger}] & =-\delta_{ij}Z_{j}^{d},\quad[Z_{j}^{d},d_{i}^{\dagger}]=\delta_{ij}d_{j}^{\dagger},
\end{align}
and $[d_{i},f_{j\sigma}]=[h_{j},f_{j\sigma}]=0$. The boson operators
$h_{i}^{\dagger},d_{i}^{\dagger}$ create a holon and a doublon, respectively.
Since we do not want more than a doublon or a holon per site, we fix the
representation of the $h_{j}$ and $d_{j}$ by the constraint $d_{j}^{\dagger2}=h_{j}^{\dagger2}=0.$
This fixes the operators $h_{i},d_{i}$ to be in the fundamental representation of $\mathfrak{sl}(2, \mathbb{C}) $ and can be interpreted as the usual $\sigma^{-}=\frac{1}{2}(X-iY)$
matrices with $\{h_{i},h_{i}^{\dagger}\}=\{d_{i},d_{i}^{\dagger}\}=1.$ Representing the bosonic degrees of freedom as operators in a finite-dimensional Hilbert space is the main difference between that representation \cite{Kotliar_1986} and the one that we use in this work.
In this new enlarged basis, the anticommutation relation of the physical fermion operators becomes
\begin{align}
\{c_{i\sigma},c_{i\sigma}^{\dagger}\}
 & =f_{i\sigma}^{\dagger}f_{i\sigma}+f_{i\bar{\sigma}}^{\dagger}f_{i\bar{\sigma}}+h_{i}^{\dagger}h_{i}+d_{i}^{\dagger}d_{i}-2f_{i\sigma}^{\dagger}f_{i\sigma}h_{i}^{\dagger}h_{i}-2f_{i\bar{\sigma}}^{\dagger}f_{i\bar{\sigma}}d_{i}^{\dagger}d_{i},
\end{align}
where $\bar{\uparrow}= \downarrow$ ($\bar{\downarrow}= \uparrow$).  This means that to recover the physical states, we have to impose a
constraint. The simplest one is
\begin{align}
C_{i}=\sum_{\sigma=\uparrow,\downarrow}f_{i\sigma}^{\dagger}f_{i\sigma}+h_{i}^{\dagger}h_{i}+d_{i}^{\dagger}d_{i}=1,
\end{align}
such that the physical states $|\rm phys\rangle$ satisfy 
$C_{j}|{\rm phys\rangle}=|{\rm phys\rangle}$
and in the physical states we have $\{c_{i\sigma},c_{i\sigma}^{\dagger}\}=1$.
The operator $C_{j}$ has integer eigenvalues $\lambda(C_{j})=(0,1,2,3,4)$
with multiplicities $m(C_{j})=(1,4,6,4,1)$. The projector onto the
physical states can be written as $\mathcal{P\coloneqq \prod}_{j}\mathcal{P}_{j}$
with the local projector
\begin{align}
\mathcal{P}_{j}=\frac{C_{j}(2-C_{j})(3-C_{j})(4-C_{j})}{6}.
\end{align}
Lastly, the fermion operator $c_{j,\sigma}=f_{i\sigma}^{\dagger}h_{i}+\epsilon_{\sigma\sigma'}f_{i\sigma'}d_{i}^{\dagger}$
commutes with the projector as $\left[c_{j\sigma},C_{j}\right]=0$ implies $\left[c_{j\sigma},\mathcal{P}_{j}\right]=0$. As a consequence, the physical operator does not create transitions between the physical states and the non-physical ones.

The physical Hamiltonian (as the enlarged Hamiltonian
acting on the physical states) is given by $H=H_{n}+H_{s}+H_{{\rm int}}$ with
\begin{align}\label{eq:H_normal}
H_{n} & =-J\sum_{\sigma \langle ij\rangle }e^{i\phi_{ij}}(f_{i\sigma}^{\dagger}(h_{i}h_{j}^{\dagger})f_{j\sigma}+f_{i{\sigma}}(d_{i}^{\dagger}d_{j})f_{j{\sigma}}^{\dagger})+{\rm h.c.},\\
H_{s} & =-J\sum_{\langle ij\rangle}e^{i\phi_{ij}}[(f_{i\uparrow}^{\dagger}f_{j\downarrow}^{\dagger}-f_{i\downarrow}^{\dagger}f_{j\uparrow}^{\dagger})h_{i}d_{j}+(f_{i\downarrow}f_{j\uparrow}-f_{i\uparrow}f_{j\downarrow})d_{i}^{\dagger}h_{j}^{\dagger}]+{\rm h.c.},\\
H_{{\rm int}} & =U\sum_{i}d_{i}^{\dagger}d_{i}.
\end{align}
This Hamiltonian has a local $U(1)$ gauge symmetry generated by the
vertex operators $F_{j\sigma}^{\theta}\coloneqq e^{i\theta f_{j\sigma}^{\dagger}f_{j\sigma}},$
$H_{j}^{\theta}\coloneqq e^{i\theta h_{j}^{\dagger}h_{j}}$ and $D_{j}^{\theta}\coloneqq e^{i\theta d_{j}^{\dagger}d_{j}}$ with
$F_{j\sigma}^{\theta}f_{j\sigma}F_{j\sigma}^{\dagger\theta} =e^{i\theta}f_{j\sigma}^{\dagger},$
(with similar relations for the other operators). The local transformations $g_{j}^{\theta}\coloneqq H_{j}^{\theta}F_{j\uparrow}^{\theta}D_{j}^{\theta}F_{j\downarrow}^{\theta}$
generates a $U(1)$ transformation that leaves the Hamiltonian invariant
as
\begin{align}
g_{j}^{\theta}c_{j\sigma}g_{j}^{\dagger\theta} & =f_{j\sigma}^{\dagger}h_{j}+\epsilon_{\sigma\sigma'}f_{j\sigma'}d_{j}^{\dagger}.
\end{align}
Motivated by the Hamiltonian $H_n$ in \cref{eq:H_normal}, we can define the following gauge field mediating the interaction between the spinon
 $\mathcal{G}_{ij}\coloneqq (d^{\dagger}_{i}d_{j})$ which transforms as
a gauge link operator $
g_{i}^{\theta_{j}}g_{j}^{\theta_{i}}\mathcal{G}_{ij}g_{j}^{\dagger\theta_{i}}g_{i}^{\dagger\theta_{j}}  =e^{-i(\theta_{j}-\theta_{i})}\mathcal{G}_{ij}.$
From this link operator, we can construct Wilson lines that are gauge
invariants. For example, over a plaquette with sites from 1 to 4, we
have
\begin{align}
\mathcal{W}_{d}(\square) & \coloneqq {\rm Tr}\left[\prod_{j\in\square}(d^\dagger_{j-1}d_{j})\right]={\rm Tr}[(d^\dagger_{1}d_{2})(d^\dagger_{2}d_{3})(d^\dagger_{3}d_{4})(d^\dagger_{4}d_{1})]=\prod_{j\in\square}d_jd_j^\dagger.
\end{align}
More generally, for any space-like loop $\mathcal{C}$ we define
\begin{align}
\mathcal{W}_{d}(\mathcal{C})\coloneqq \prod_{j\in\mathcal{C}}d_{j}d^\dagger_{j}=\prod_{j\in\mathcal{C}}(1-n_{j,\uparrow}n_{j,\downarrow}),
\end{align}
where, in the last step, we used the map to the physical degrees of freedom.

\begin{figure}[htpb]
    \centering
    \includegraphics[width=\textwidth]{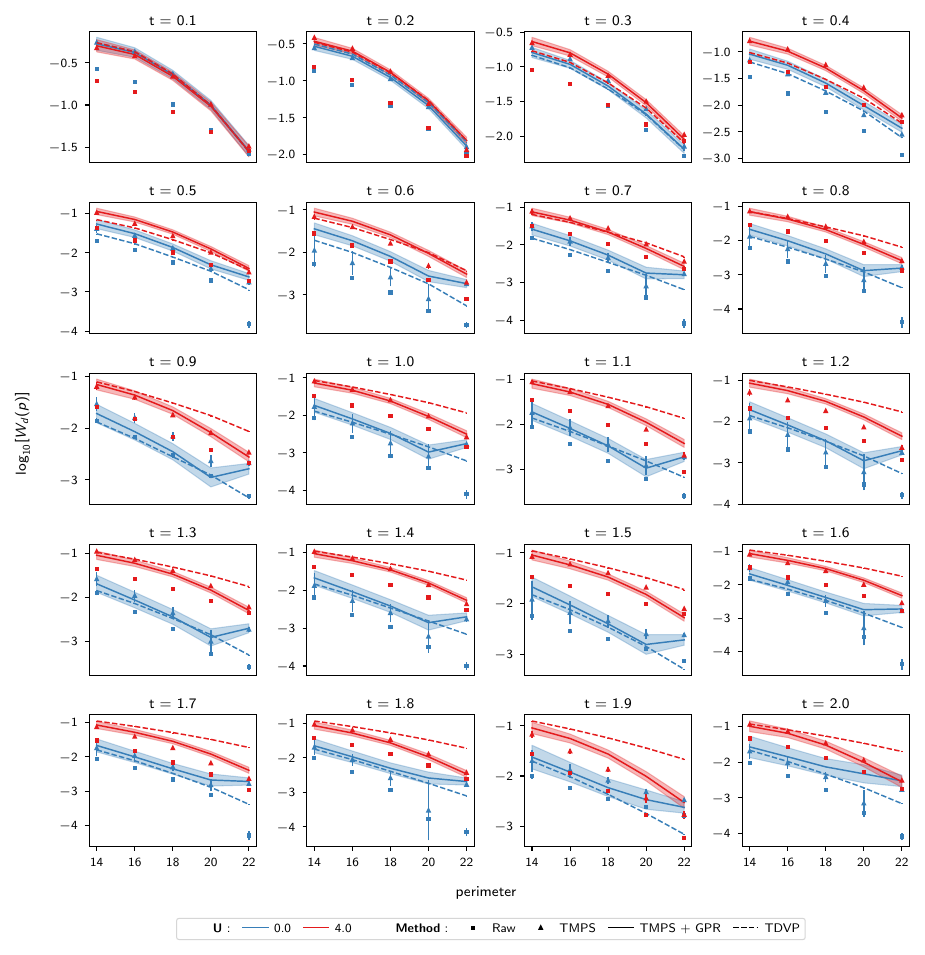} 
    \caption{Wilson loop with fixed area ($A = 10$) and varying perimeter. Same as in \cref{fig:wilson_time_averaged}a but all times shown.}
    \label{fig:all_wilson_loops_fixed_area}
\end{figure}

\begin{figure}[htpb]
    \centering
    \includegraphics[width=\textwidth]{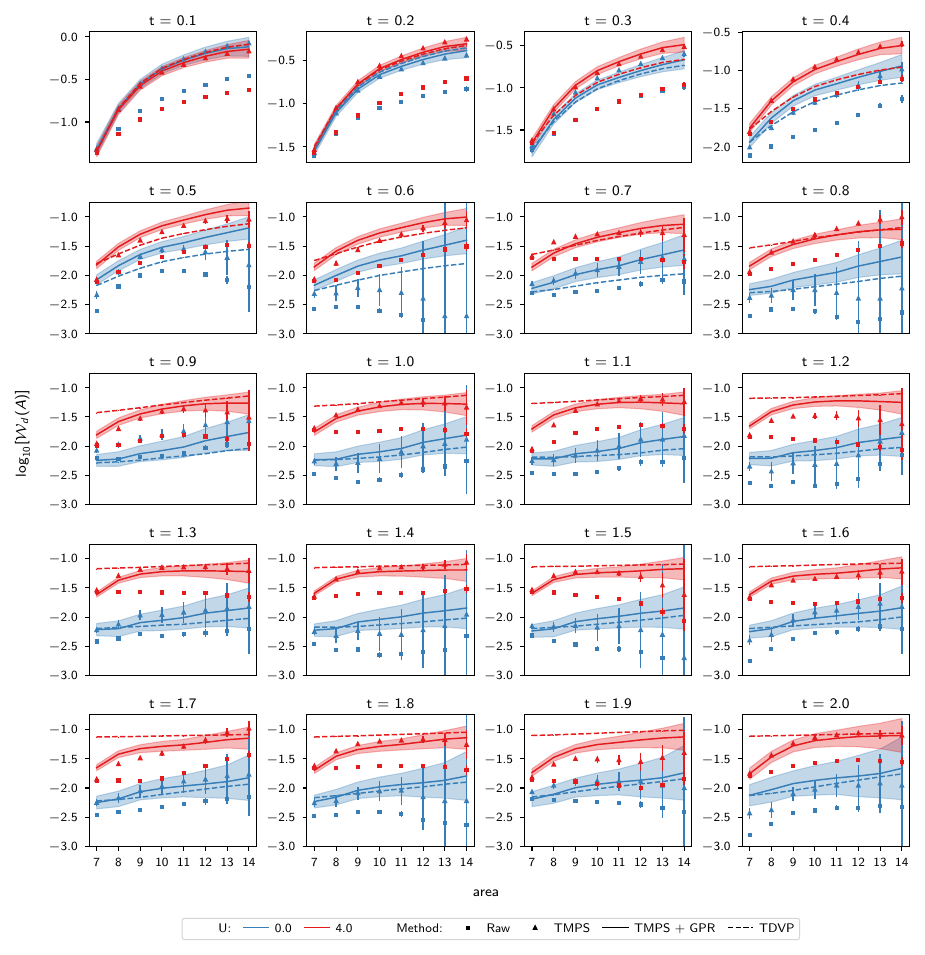} 
    \caption{Wilson loop with fixed perimeter ($p=16$) and varying area. Same as in \cref{fig:wilson_time_averaged}b, but all times shown.}
    \label{fig:all_wilson_loops_fixed_perimeter}
\end{figure}

In \cref{fig:all_wilson_loops_fixed_area,fig:all_wilson_loops_fixed_perimeter} we observe the behaviour of the expectation of horizontal (doublon) Wilson loops $\langle \mathcal{W}_d(t)\rangle$ for different times, as we fix the area and vary the perimeter (\cref{fig:all_wilson_loops_fixed_area}), or for fixed perimeter and varying the area (\cref{fig:all_wilson_loops_fixed_perimeter}).  We observe that the $\mathcal{W}_d$ has a perimeter scaling, for any of the values of interaction $U$ considered. Although these Wilson loops are horizontal (i.e, taken at a fixed time), they can still serve as order parameters for deconfinement (of spinons). Note that even at $U=0$, we observe a perimeter law, with similar scaling but weaker strength for larger times. This is not surprising. As $\mathcal{W}_d$ only probes the interaction between spinons, we cannot claim spin-charge separation from this signature alone because the charge carriers could be deconfined in the same way without a clear separation between them, as would happen in the case of normal non-interacting fermions.

To really compare the confinement/deconfinement between charge and spin carriers, we can define, inspired by $H_n$ in \cref{eq:H_normal}, the complementary link operator between doublons given by
\begin{align}
    d_{i}^{\dagger}\left(\sum_{\sigma}f_{i\sigma}f_{j\sigma}^{\dagger}\right)d_{j}=d_{i}^{\dagger}\left(f_{i\uparrow}f_{j\uparrow}^{\dagger}+f_{i\downarrow}f_{j\downarrow}^{\dagger}\right)d_{j}=d_{i}^{\dagger}\begin{bmatrix}f_{i\uparrow} & f_{i\downarrow}\end{bmatrix}\begin{bmatrix}f_{j\uparrow}^{\dagger}\\
f_{j\downarrow}^{\dagger}
\end{bmatrix}d_{j}
\end{align}
The local gauge-invariant operator is then 
\begin{align}
    W_{h,j}\coloneqq \begin{bmatrix}f_{j\uparrow}^{\dagger}\\
f_{j\downarrow}^{\dagger}
\end{bmatrix}\begin{bmatrix}f_{j\uparrow} & f_{j\downarrow}\end{bmatrix}=\begin{bmatrix}f_{j\uparrow}^{\dagger}f_{j\uparrow} & f_{j\uparrow}^{\dagger}f_{j\downarrow}\\
f_{j\downarrow}^{\dagger}f_{j\uparrow} & f_{j\downarrow}^{\dagger}f_{j\downarrow}
\end{bmatrix}=\begin{bmatrix}Q_{j}+S_{j}^{3} & S_{j}^{+}\\
S_{j}^{-} & Q_{j}-S_{j}^{3}
\end{bmatrix},  
\end{align}
where we have introduced the spin and charge operators $S_{j}^{3}\coloneqq \frac{1}{2}(f_{j\uparrow}^{\dagger}f_{j\uparrow}-f_{j\downarrow}^{\dagger}f_{j\downarrow})$, $S_{j}^{+}\coloneqq f_{j\uparrow}^{\dagger}f_{j\downarrow}$ and 
$Q_{j}\coloneqq \frac{1}{2}(f_{j\uparrow}^{\dagger}f_{j\uparrow}+f_{j\downarrow}^{\dagger}f_{j\downarrow})$.
In the subspace of the physical states where the charge operator is fixed to $Q_j=1/2$, $W_h,j$ corresponds exactly to the Lax operator associated with the quantum inverse scattering method \cite{faddeev1996} for the 1D Heisenberg chain.

The operator that is gauge invariant is given by the open Wilson line between doublons, and it takes the form
\begin{align}
d_i^\dagger\mathcal{S}_{ij}d_j\coloneqq d_i^\dagger\begin{bmatrix}f_{i\uparrow} & f_{i\downarrow}\end{bmatrix}\left (\prod_{k \in \mathcal C_{ij}}W_{h,k}\right)\begin{bmatrix}f_{j\uparrow}^{\dagger}\\
f_{j\downarrow}^{\dagger}
\end{bmatrix}d_{j},
\end{align}
where $\mathcal{C}_{ij}$ is a path connecting the sites $i$ and $j$.
Remarkably, the product of gauge-invariant operators $\prod_{k\in \mathcal{C}}W_{k,h}$ along a closed path satisfying  $Q_k=1/2$ $\forall k \in \mathcal{C}$ corresponds to the monodromy matrix of the Heisenberg model \cite{faddeev1996}. This monodromy matrix and the intertwining relation of the Lax operator can be used to prove the existence of $N$ conserved quantities on the $N$-site Heisenberg model, where one of those is the Heisenberg Hamiltonian. We will discuss this connection in depth in a separate publication. 
We use this connection to isolate a meaningful operation associated with the quantum operator $\prod_{k \in \mathcal C_{ij}}W_{h,k}$ that can be extracted from measurements in the computational basis. The simplest non-trivial operator in the family generated by the product $\prod_{k \in \mathcal C_{ij}}W_{h,k}$ acting on the paths containing only spinons is the $z$ component of the Heisenberg Hamiltonian. Using this, we {\it define} the line operator $ H_{line}= \sum_{j\in line}S_j^zS^z_{j+1},$ that we measure in a line connecting two doublons or a doublon and a hole. Then we consider the expectation value of the operator
 \begin{align}
     V_{\alpha\beta}(M)
     \coloneqq \sum_{\substack{\textnormal{sites }  i, j:\\ |i-j| = M}}  \frac{n^\alpha_{i} n^\beta_{j}}{N_{\rm pairs}}
    \sum_{\substack{\textnormal{paths } \gamma:\\ \textnormal{ from } i \textnormal{ to } j }}\frac{\sum_{(m, n) \in \gamma} S_{m}^{z}S_{n}^{z}}{N_{\rm paths}(i,j)}
 \end{align}
as a measure of the potential between the particles $\alpha, \beta = h, d$. Here $n_i^d\coloneqq n_{i\uparrow}n_{i\downarrow}$, $n_{i}^h\coloneqq (1-n_{i\uparrow})(1-n_{i\downarrow})$.
We extract $V_{hd}(M)$ from the measured shots in the computational basis using the following algorithm
\begin{itemize}
\item Define a Manhattan distance $M$.
\item Loop over the shots to find one doublon and one hole at distance $M$. This is a configuration, if no shots are found return zero, otherwise
\item For path $\in$ the paths with Manhattan distance $M$ that connect the two doublons, compute $R_{\rm path}=\sum_{k\in \rm {path}}S_{k}^{z}S_{k+1}^{z}$. 
\item Sum $R_{\rm path}$ over all paths, return this as $S(\text{config})$.
\item Finally, return the sum over configurations and divide by the total number of paths. This is $V_{hd}(M)$.
\end{itemize}

In \cref{fig:Wilson_space_time} we sketch an example of this procedure for configurations starting with a doublon and ending in a holon.

\begin{figure*}[t]
    \centering
    \includegraphics[width=\textwidth]{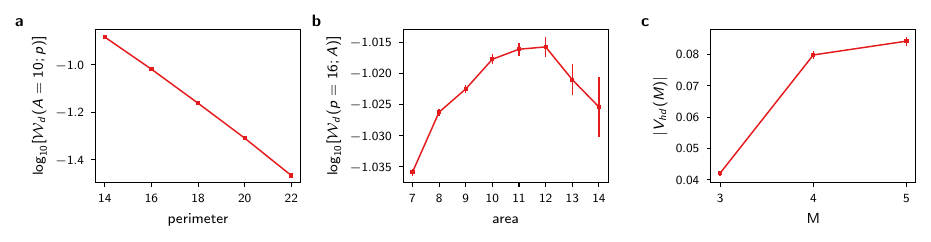}
    \caption{Wilson loop with a) fixed area, varying perimeter; b) fixed perimeter, varying area; c) open Wilson line $V_{hd}(M)$ computed using 10000 shots from a DMRG simulation that approximates the ground-state of the FH model with $U=4$.}
    \label{fig:wilson_panel_dmrg}
\end{figure*}

\begin{figure}[htpb]
    \centering
    \includegraphics[width=\textwidth]{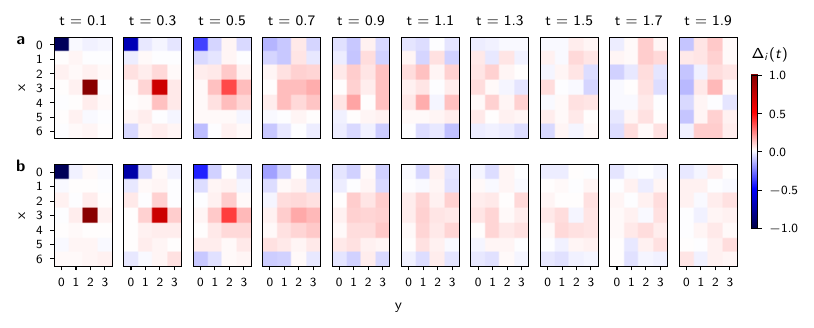}
    \caption{Evolution of the charge difference $\Delta_i=\langle n^d_i\rangle -\langle n_i^h\rangle$ between doublons and holons. Note the similarities in the spread up to times $t\sim 0.5$ and the periodic nature of the system apparent from the wrapping around of the holon density. The total density $\Delta_i$ decays to zero faster in the interacting case compared with the non-interacting case.}
    \label{fig:holon-doublon}
\end{figure}

As discussed in the main text, the growth with distance of the expectation of $V^{hd}$ observed in \cref{fig:wilson_time_averaged}c signals a confining potential between the doublons and holons. This attraction is apparent from the signal that measures the expected charge difference between doublons and holons $\Delta_i=\langle n^d_i\rangle -\langle n_i^h\rangle $. Note that the sum over the lattice $\sum_i(\langle n^d_i\rangle -\langle n_i^h\rangle)$ is constant in time, due to total number conservation. In \cref{fig:holon-doublon} we observe the real-time dynamics of the signal  $\langle n^d_i\rangle -\langle n_i^h\rangle $ for $U/J=0$ (left) and $U/J=4$ (right). Note that the expected local charge difference $\Delta_i$ for $U/J=0$ seems to form charge density waves, the $U/J=4$ signal tends faster to the uniform state $\Delta_i=0$

\begin{figure*}[t]
    \centering
    \includegraphics[width=\textwidth]{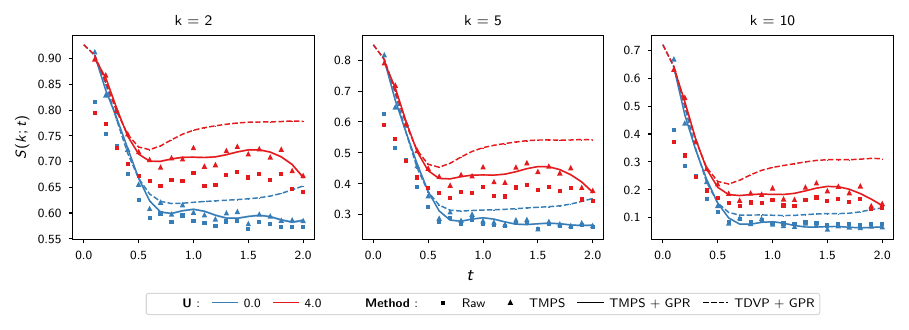}
    \caption{Moments of the operator $|\mathcal{L}|-N_{\rm doublons}$ that measures the absence of doublons. Each curve has been obtained by averaging over 1000 random subsets using      samples from the TDVP simulation (dashed lines) and all the experimental shots (the continuous line is a guide for the eye for the mitigated results). Tata data have been produced using the TMPS map obtained from TDVP data up to $t = 0.5$.}
    \label{fig:subsets}
\end{figure*}

Finally, to highlight the different states that the experiment and the TDVP algorithm produce, we study different moments of the (normalized) absence of doublons operator $X = 1-N_{\rm doublons}|\mathcal{L}|^{-1}$ given by
\begin{align}
    \langle X^k(t)\rangle = \sum_{\{j_m\}_{m=1}^k\in \mathcal{L}}\frac{1}{|\mathcal{L}|^k}\left\langle \prod_{m=1}^k(1- n_{j_m,\uparrow}(t)n_{j_m,\downarrow}(t))\right\rangle.
\end{align}
We estimate these moments through a Monte Carlo estimator by sampling random subsets of $k$ sites 1000 times. The results are shown in \cref{fig:subsets} for $k=2,5,10$. 
We observe divergence between the results of the tensor networks, mitigated experimental data and the results obtained by sampling the TDVP state after $t\sim 0.7$. This divergence is essentially the one observed for the number of doublons operator $N_{\rm doublons}$, as the moment $k$ of $X$ is very close to the mean $\mu=\langle X\rangle$ to the power $k$ as 
\begin{align}
    \langle X^k\rangle  = \langle [\mu+[X-\mu]]^k\rangle = \sum_{m=0}^k \binom{k}{m} \mu^m \langle [X-\mu]^{k-m}\rangle,
\end{align}
and the centred moments $\langle [X- \mu]^k\rangle $ are very small. In order to distinguish the value of the centred moments from zero and from the moments produced by sampling the TDVP state, more shots would be required. 

\begin{figure*}[t]
    \centering
    \includegraphics[width=\textwidth]{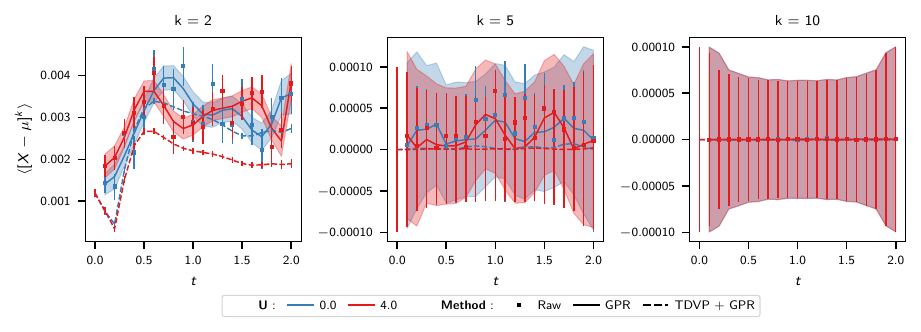}
    \caption{Central moments $\langle [X - 
    \mu]^k\rangle $ for different values of $k$.}
    \label{fig:subsets_central_moments}
\end{figure*}

{\section{Quantum Circuits}\label{sec:circuits}
\subsection{Fermion to Qubit Mapping} \label{sec:fermi-qubit}

We use the Jordan-Wigner (JW) mapping to represent the fermionic system on qubits. Every mode is assigned to a qubit and Fock states are mapped to computational basis states such that a mode is occupied (unoccupied) if its corresponding qubit is in the $\ket{1}$ $\left(\ket{0}\right)$ state. An ordering is chosen for the modes (correspondingly for the qubits) and creation/annihilation operators are mapped as 
\begin{equation}
c_j^\dagger = \sigma^+_j\prod_{i<j}Z_i,\quad c_j = \sigma^-_j\prod_{i<j}Z_i,\quad n_j=\frac{1}{2}(1-Z_j),
\end{equation}
where $\sigma^\pm_j=(X\mp iY)/2$. The $\sigma^{\pm}$ parts capture the creation/annihilation properties and the $Z$ strings ensure the correct anticommutation relations. This work uses the ``snake'' ordering for modes on a square lattice, where modes are ordered along rows from left to right and right to left in an alternating manner. Both spin-up and spin-down sectors are snake-ordered, with the spin-down sector coming after spin-up, as shown in \cref{fig:snake}.

\begin{figure}[htpb]
    \centering
        \centering
        \begin{tikzpicture}[scale=0.5] 

        \begin{scope}[shift={(.6,-.6)},black!30!white]
        \foreach \x in {0,1,2,3,4,5,6} {
            \draw[line width=3] (0,-2*\x)--(6,-2*\x);
        }
        \foreach \x in {0,1,2,3} {
            \draw (2*\x,0)--(2*\x,-12);
        }
        
        \draw[line width=3] (6,0)--(6,-2);
        \draw[line width=3] (0,-2)--(0,-4);
        \draw[line width=3] (6,-4)--(6,-6);
        \draw[line width=3] (0,-6)--(0,-8);
        \draw[line width=3] (6,-8)--(6,-10);
        \draw[line width=3] (0,-10)--(0,-12);
        \foreach \x in {0,1,2,3} {
            \foreach \y[evaluate={\xi=int((-1)^\y*\x + 1.5*(1-(-1)^\y))},evaluate={\l=int(4*\y+\x+28)}] in {0,1,2,3,4,5,6} {
            \node at (2*\xi,-2*\y) [circle,draw,fill=white,minimum size=5mm] {};
            \node at (2*\xi,-2*\y) {\l};
            }
        }    
        \end{scope}
    
        \foreach \x in {0,1,2,3} {
            \draw[white, line width=1.5] (2*\x,0)--(2*\x,-12);
            \draw (2*\x,0)--(2*\x,-12);
        }
        \foreach \x in {0,1,2,3,4,5,6} {
            \draw[white,line width=5] (0,-2*\x)--(6,-2*\x);
            \draw[line width=3] (0,-2*\x)--(6,-2*\x);
        }
        
        \draw[white,line width=5] (6,0)--(6,-2);
        \draw[white,line width=5] (0,-2)--(0,-4);
        \draw[white,line width=5] (6,-4)--(6,-6);
        \draw[white,line width=5] (0,-6)--(0,-8);
        \draw[white,line width=5] (6,-8)--(6,-10);
        \draw[white,line width=5] (0,-10)--(0,-12);
        \draw[line width=3] (6,0)--(6,-2);
        \draw[line width=3] (0,-2)--(0,-4);
        \draw[line width=3] (6,-4)--(6,-6);
        \draw[line width=3] (0,-6)--(0,-8);
        \draw[line width=3] (6,-8)--(6,-10);
        \draw[line width=3] (0,-10)--(0,-12);

        \draw[ibmRed,line width=3] (0,0)--(6,0)--(6,-2)--(0,-2);
        \draw[ibmBlue,line width=3] (2,-6)--(4,-6);
        
        \foreach \x in {0,1,2,3} {
            \foreach \y[evaluate={\xi=int((-1)^\y*\x + 1.5*(1-(-1)^\y))},evaluate={\l=int(4*\y+\x)}] in {0,1,2,3,4,5,6} {
            \node at (2*\xi,-2*\y) [circle,draw,fill=white,minimum size=5mm] {};
            \node at (2*\xi,-2*\y) {\l};
            }
        }
    
        \node at (0,0) [circle,draw=ibmRed,minimum size=5.1mm,fill=ibmRed!20!white] {};
        \node at (0,0) [ibmRed] {0};
        \node at (0,-2) [circle,draw=ibmRed,minimum size=5.1mm,fill=ibmRed!20!white] {};
        \node at (0,-2) [ibmRed] {7};
        \node at (2,0) [circle,draw=ibmRed,minimum size=5.1mm,fill=ibmRed!20!white] {};
        \node at (2,0) [ibmRed] {1};
        \node at (2,-2) [circle,draw=ibmRed,minimum size=5.1mm,fill=ibmRed!20!white] {};
        \node at (2,-2) [ibmRed] {6};
        \node at (4,0) [circle,draw=ibmRed,minimum size=5.1mm,fill=ibmRed!20!white] {};
        \node at (4,0) [ibmRed] {2};
        \node at (4,-2) [circle,draw=ibmRed,minimum size=5.1mm,fill=ibmRed!20!white] {};
        \node at (4,-2) [ibmRed] {5};
        \node at (6,0) [circle,draw=ibmRed,minimum size=5.1mm,fill=ibmRed!20!white] {};
        \node at (6,0) [ibmRed] {3};
        \node at (6,-2) [circle,draw=ibmRed,minimum size=5.1mm,fill=ibmRed!20!white] {};
        \node at (6,-2) [ibmRed] {4};
        
        \node at (2,-6) [circle,draw=ibmBlue,minimum size=5.1mm,fill=ibmBlue!20!white] {};
        \node at (2,-6) [ibmBlue] {14};
        \node at (4,-6) [circle,draw=ibmBlue,minimum size=5.1mm,fill=ibmBlue!20!white] {};
        \node at (4,-6) [ibmBlue] {13};
            
        \end{tikzpicture}
        \caption{}

\caption{Jordan-Wigner snake ordering on a spin-$\frac{1}{2}$, $4\times 7$ grid. The spin-up sector is in the foreground and the spin-down sector is in the background. In the qubit picture, the support of a hopping term is on the qubits corresponding to the hopping pair and all modes in between them in the ordering.  A horizontal hopping term is only supported on 2 qubits (blue) but a vertical hopping term can be supported on a longer chain covering the width of the lattice (red).
}
\label{fig:snake}

\end{figure}

\subsubsection{Encoded Interactions}
Undressed hopping interactions are represented as
\begin{equation}
c_i^\dagger c_j+c^\dagger_jc_i=\frac{1}{2}(X_iX_j+Y_iY_j)\prod_{i<k<j}Z_{k}.
\end{equation}
When a magnetic field is present these are represented as 
\begin{equation}
e^{i\phi_{ij}}c_i^\dagger c_j+e^{-i\phi_{ij}}c^\dagger_jc_i=\frac{1}{2}(\cos\phi_{ij}(X_iX_j+Y_iY_j)+\sin\phi_{ij}(Y_iX_j-X_iY_j))\prod_{i<k<j}Z_{k}.
\end{equation}

The onsite Coulomb terms of the Fermi-Hubbard Hamiltonian are represented on qubits as
\begin{equation}
n_{i,\uparrow}n_{i,\downarrow}=\frac{1}{4}(1-Z_{i,\uparrow}-Z_{i,\downarrow}+Z_{i,\uparrow}Z_{i,\downarrow}).
\end{equation}
The part proportional to 1 may be ignored as it will only contribute an unobservable global phase to the evolution. Furthermore, since this term appears for every spin pair, the Hamiltonian contains a part proportional to $\sum_i(Z_{i,\uparrow}+Z_{i,\downarrow})=\frac{1}{2}\sum_i(2-n_{i,\uparrow}-n_{i,\uparrow})$ which may also be ignored because it will only contribute a global phase to states of fixed particle number. For our purposes it then suffices to represent the Coulomb term as
\begin{equation}
n_{i,\uparrow}n_{i,\downarrow}\approx\frac{1}{4}Z_{i,\uparrow}Z_{i,\downarrow}.
\end{equation}

\subsubsection{Fixed Parity JW Loop}\label{sec:JW loop}

Since the Fermi-Hubbard model preserves parity in each spin sector, for states of fixed parity in spin sector $\sigma$, the parity operator $P_{\sigma}=\prod_i V_{i,\sigma}$ acts as a constant. In particular, for even (odd) parity states we have $P_{\sigma}=+1$ ($P_{\sigma}=-1$). This means that, for fixed parity states, the interactions can be multiplied by the $\pm$ parity operator with no effect on the physics. Specifically, one multiplies by $+P_\sigma$ when it is positive and $-P_\sigma$ when it is negative. This can be helpful for JW encoded systems as it allows the $Z$ string involved in hopping interaction to be ``flipped'' like so
\begin{equation}\label{eq: hopping with parity}
\pm P(c_i^\dagger c_j+c^\dagger_jc_i)=\mp\frac{1}{2}(X_iX_j+Y_iY_j)\prod_{\substack{0\leq k<i\\j<k\leq N}}Z_{k}.
\end{equation}
where $N$ is the number of modes in the sector and we have omitted the spin index. This flexibility of representation means that hopping terms between modes that are distant with respect to the ordering can be made much lower weight, providing alternative, more efficient avenues for their implementation.
In this work, we only apply this move to undressed hopping terms due to our choice of magnetic field, but the effect on dressed terms is analogous.

As a result of this, one can consider the mode ordering within a spin-sector to be a loop where the first and final spin-$\sigma$ are also adjacent to one another and hopping terms can be represented with a $Z$ string between the modes in either direction (so long as the appropriate sign is applied).
This is particularly useful when simulating lattices with periodic boundary conditions.

\subsection{Fermionic Swap Networks }\label{sec:swapnets}

Encoded hopping terms between horizontal pairs are two-qubit operators as they are adjacent in the ordering, so their evolution can be implemented with a simple quantum circuit involving two two-qubit gates (see \cref{sec:gate decomp}). The hopping terms between vertical pairs are not adjacent in this sense: they involve long strings of $Z$ operators and can be costly to implement (see \cref{fig:snake}). An efficient way to implement evolution under these terms is via fermionic swap networks~\cite{kivlichan18}. 

The fermionic swap gate (FSWAP) acts as
\begin{equation}
\textup{FSWAP}=\begin{pmatrix}1&0&0&0\\0&0&1&0\\0&1&0&0\\0&0&0&-1\end{pmatrix}=\textup{SWAP}\cdot\textup{CZ},
\end{equation}
and transforms Paulis under conjugation as
\begin{equation}
XI\leftrightarrow ZX,\quad YI\leftrightarrow ZY,\quad ZI\leftrightarrow IZ.
\end{equation}
Under the JW encoding, when acting on qubits $(j,j+1)$ adjacent in the ordering, it transforms the encoded fermionic operators as
\begin{equation}
c_j\leftrightarrow c_{j+1},\quad c^\dagger_j\leftrightarrow c^\dagger_{j+1},\quad n_j\leftrightarrow n_{j+1},
\end{equation}
effectively swapping the positions of the encoded modes, hence the name. FSWAP operations can be implemented on the Quantinuum hardware using only a single two-qubit gate (see \cref{sec:gate decomp} for further details). 

Networks of these operations can be used to rearrange the modes such that every interacting pair is adjacent at some point with respect to the ordering, at which time their interaction is implemented via the two-qubit hopping circuit (\cref{sec:gate decomp}).

\subsection{Second-Order Trotter Circuit}\label{sec:2nd order trot}

Our second-order Trotter step of time $t$ has the basic structure
\begin{equation}
    S_2(t) = \prod_{h\in \mathcal H}^\rightarrow e^{-ih\frac{t}{2}}\prod_{o\in\mathcal O}e^{-iot}\prod_{h\in\mathcal H}^\leftarrow e^{-ih\frac{t}{2}},
\end{equation}
where $\mathcal H$ is the set of hopping interactions in the Hamiltonian and $\mathcal O$ is the set of Coulomb interactions. The arrows indicate that the second round of hopping interactions is applied in the reverse order to the first, the Coulomb interactions all commute and are applied in parallel directly via $R_{zz}$ gates.
 The hopping parts are implemented via FSWAP networks, which specify the ordering of the terms. The second hopping part of the circuit is applied with the same gate schedule as the first, only in reverse order. Below, we present the instructions for applying the hops to a single spin sector, which are also applied to both the spin-up and spin-down sectors.

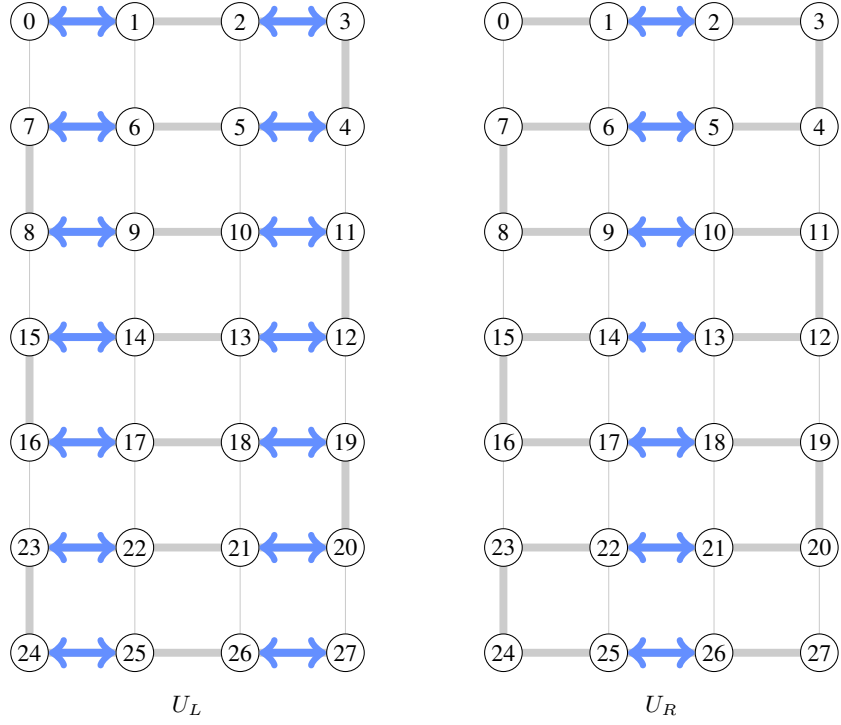
\begin{figure}
    \centering
\begin{tikzpicture}[scale=0.7]
\begin{scope}[black!20!white]
    \foreach \x in {0,1,2,3} {
        \draw[white, line width=1.5] (2*\x,0)--(2*\x,-12);
        \draw (2*\x,0)--(2*\x,-12);
    }
    \foreach \x in {0,1,2,3,4,5,6} {
        \draw[white,line width=5] (0,-2*\x)--(6,-2*\x);
        \draw[line width=3] (0,-2*\x)--(6,-2*\x);
    }
    
        \draw[line width=3] (6,0)--(6,-2);
        \draw[line width=3] (0,-2)--(0,-4);
        \draw[line width=3] (6,-4)--(6,-6);
        \draw[line width=3] (0,-6)--(0,-8);
        \draw[line width=3] (6,-8)--(6,-10);
        \draw[line width=3] (0,-10)--(0,-12);
\end{scope}
    \foreach \x in {0,1,2,3} {
        \foreach \y[evaluate={\xi=int((-1)^\y*\x + 1.5*(1-(-1)^\y))},evaluate={\l=int(4*\y+\x)}] in {0,1,2,3,4,5,6} {
        \node (\l) at (2*\xi,-2*\y) [circle,draw,fill=white,minimum size=5mm] {};
        \node at (2*\xi,-2*\y) {\l};
        }
    }

    \begin{scope}[<->,ibmBlue]
        \draw[line width=3] (0)--(1);
        \draw[line width=3] (7)--(6);
        \draw[line width=3] (8)--(9);
        \draw[line width=3] (15)--(14);
        \draw[line width=3] (16)--(17);
        \draw[line width=3] (23)--(22);
        \draw[line width=3] (24)--(25);
        
        \draw[line width=3] (2)--(3);
        \draw[line width=3] (5)--(4);
        \draw[line width=3] (10)--(11);
        \draw[line width=3] (13)--(12);
        \draw[line width=3] (18)--(19);
        \draw[line width=3] (21)--(20);
        \draw[line width=3] (26)--(27);
    \end{scope}

    \node at (3,-13) {$U_L$};

\begin{scope}[shift={(9,0)}]
\begin{scope}[black!20!white]
    \foreach \x in {0,1,2,3} {
        \draw[white, line width=1.5] (2*\x,0)--(2*\x,-12);
        \draw (2*\x,0)--(2*\x,-12);
    }
    \foreach \x in {0,1,2,3,4,5,6} {
        \draw[white,line width=5] (0,-2*\x)--(6,-2*\x);
        \draw[line width=3] (0,-2*\x)--(6,-2*\x);
    }
    
        \draw[line width=3] (6,0)--(6,-2);
        \draw[line width=3] (0,-2)--(0,-4);
        \draw[line width=3] (6,-4)--(6,-6);
        \draw[line width=3] (0,-6)--(0,-8);
        \draw[line width=3] (6,-8)--(6,-10);
        \draw[line width=3] (0,-10)--(0,-12);
\end{scope}
    \foreach \x in {0,1,2,3} {
        \foreach \y[evaluate={\xi=int((-1)^\y*\x + 1.5*(1-(-1)^\y))},evaluate={\l=int(4*\y+\x)}] in {0,1,2,3,4,5,6} {
        \node (\l) at (2*\xi,-2*\y) [circle,draw,fill=white,minimum size=5mm] {};
        \node at (2*\xi,-2*\y) {\l};
        }
    }

    \begin{scope}[<->,ibmBlue]
        \draw[line width=3] (2)--(1);
        \draw[line width=3] (5)--(6);
        \draw[line width=3] (10)--(9);
        \draw[line width=3] (13)--(14);
        \draw[line width=3] (18)--(17);
        \draw[line width=3] (21)--(22);
        \draw[line width=3] (26)--(25);
    \end{scope}

    \node at (3,-13) {$U_R$};

\end{scope}
\end{tikzpicture}
    \caption{Illustration of the action of $U_L$ and $U_R$ on a $4\times7$ grid, the blue arrows denote FSWAPs.}
    \label{fig:ULUR}
\end{figure}
\subsubsection{Swap Network}

Let $L_x$ and $L_y$ be the horizontal and vertical lattice dimensions and $U_L$ ($U_R$) be the circuit which FSWAPs even (odd) columns with those to their right, in cases where the leftmost column is even; see \cref{fig:ULUR}. For some pairs, FSWAPs will be merged with a hopping interaction, which can be done with the same circuit cost as an FSWAP; see \cref{sec:gate decomp}.

The first round of hopping terms is implemented via one of the following swap network schedules. For a step size of $t$ all hopping terms are applied for time $t/2$. 

\textbf{For even $L_x$:}
\begin{enumerate}
\item Implement all vertical hopping terms between modes adjacent in the ordering. 
\item Apply $U_L$, 
replacing any FSWAP with a merged FSWAP and hopping interaction if the pair of modes being swapped is involved in an unimplemented hopping term.
\item Implement all vertical hopping terms between modes adjacent in the ordering.
\item Apply $U_R$,  

replacing any FSWAP with a merged FSWAP and hopping interaction if the pair of modes being swapped is involved in an unimplemented hopping term.
\item Repeat from 2. terminating as soon as all hopping terms have been implemented.
\end{enumerate}

\textbf{For odd $L_x$:}
\begin{enumerate}
\item Implement all hopping terms between pairs on the leftmost column that are adjacent in the ordering.
\item Implement all hopping terms between pairs on the rightmost column that are adjacent in the ordering.
\item Apply $U_L$, 
replacing any FSWAP with a merged FSWAP and hopping interaction if the pair of modes being swapped is involved in an unimplemented hopping term.
\item Implement all hopping terms between pairs on the leftmost column that are adjacent in the ordering.
\item Apply $U_R$,  
replacing any FSWAP with a merged FSWAP and hopping interaction if the pair of modes being swapped is involved in an unimplemented hopping term.
\item Repeat from 2. terminating as soon as all hopping terms have been implemented.
\end{enumerate}
Note that the resulting swap network is shorter than that required for a first-order Trotter step. This is because for the first order, we would need to return the modes to their original positions in preparation for the next step, whereas here we are content to leave them in a jumbled state because the original positioning will be restored after the second swap network.

\subsubsection{Periodic Boundary Conditions}\label{sec: periodic boundary}

In either case, the horizontal hops across the periodic boundary are implemented automatically by the swap networks above, as every pair of modes on a row is adjacent at some point in the schedule.

The vertical hops are only automatically implemented by the above networks in the case of an even \textit{vertical} dimension $L_y$.

This is because of how the Jordan-Wigner ordering lines up between the top and bottom rows. In the even case, every vertical pair is adjacent w.r.t.\ the JW loop at some point in the swap network, but in the odd case, this never happens for the pairs connecting the top and bottom rows. 
These hopping terms are implemented by a bespoke circuit after the main swap network is completed. This circuit is detailed in \cref{sec:gate decomp}. 

\begin{figure}
    \centering
\begin{tikzpicture}
\begin{scope}[scale=0.45,every node/.style={scale=0.6,font=\fontsize{14}{16.8}}]

\draw[dashed] (0,1.1)--(3*6.2+4,1.1);
\node[fill=white] at (1.5*6.2+2,1.1) {$\textup{Step 1 }(t_1)$};

\begin{scope}[black!20!white]
    \foreach \x in {0,1,2} {
        \draw[line width=2] (0,-2*\x)--(4,-2*\x);
        \draw (2*\x,0)--(2*\x,-4);
    }
    
    \draw[line width=2] (4,0)--(4,-2);
    \draw[line width=2] (0,-2)--(0,-4);
\end{scope}
    \foreach \x in {0,1,2} {
        \foreach \y[evaluate={\xi=int((-1)^\y*\x + (1-(-1)^\y))},evaluate={\l=int(3*\y+\x)}] in {0,1,2} {
        \node (\l) at (2*\xi,-2*\y) [circle,draw,fill=white,minimum size=5mm] {};
        \node at (2*\xi,-2*\y) [] {\l};
        }
    }

        \draw[ibmRed, snake it,line width=2] (2)--(3);
        \draw[ibmRed, snake it,line width=2] (5)--(6);

    \draw[line width=2,->,>=stealth] (4.6,-2)--(5.6,-2);
\begin{scope}[shift={(6.2,0)}]
    
\begin{scope}[black!20!white]
    \foreach \x in {0,1,2} {
        \draw[line width=2] (0,-2*\x)--(4,-2*\x);
        \draw (2*\x,0)--(2*\x,-4);
    }
    
    \draw[line width=2] (4,0)--(4,-2);
    \draw[line width=2] (0,-2)--(0,-4);
\end{scope}
    \foreach \x in {0,1,2} {
        \foreach \y[evaluate={\xi=int((-1)^\y*\x + (1-(-1)^\y))},evaluate={\l=int(3*\y+\x)}] in {0,1,2} {
        \node (\l) at (2*\xi,-2*\y) [circle,draw,fill=white,minimum size=5mm] {};
        \node at (2*\xi,-2*\y) [] {\l};
        }
    }

    \begin{scope}[<->,ibmBlue,line width=2,opacity=0.75]
        \draw[snake it, ibmRed] (0)--(1);
        \draw[snake it, ibmRed] (5)--(4);
        \draw[snake it, ibmRed] (7)--(6);
        \draw (0)--(1);
        \draw (5)--(4);
        \draw (7)--(6);
    \end{scope}

    \draw[line width=2,->,>=stealth] (4.6,-2)--(5.6,-2);
\end{scope}
        
\begin{scope}[shift={(2*6.2,0)}]
\begin{scope}[black!20!white]
    \foreach \x in {0,1,2} {
        \draw[line width=2] (0,-2*\x)--(4,-2*\x);
        \draw (2*\x,0)--(2*\x,-4);
    }
    
    \draw[line width=2] (4,0)--(4,-2);
    \draw[line width=2] (0,-2)--(0,-4);
\end{scope}
    \foreach \x in {0,1,2} {
        \foreach \y[evaluate={\l=int(3*\y+\x)}] in {0,1,2} {
        \node (\x\y) at (2*\x,-2*\y) [circle,draw,fill=white,minimum size=5mm] {};
        }
    }
    \node at (0,0) {1};
    \node at (2,0) {0};
    \node at (4,0) {2};
    \node at (0,-2) {4};
    \node at (2,-2) {5};
    \node at (4,-2) {3};
    \node at (0,-4) {7};
    \node at (2,-4) {6};
    \node at (4,-4) {8};

    \begin{scope}[<->,ibmBlue,line width=2]
        \draw (10)--(20);
        \draw (11)--(21);
        \draw (12)--(22);
    \end{scope}
        \draw[ibmRed, snake it,line width=2] (01)--(02);
    \draw[line width=2,->,>=stealth] (4.6,-2)--(5.6,-2);
\end{scope}

\begin{scope}[shift={(3*6.2,0)}]
\begin{scope}[black!20!white]
    \foreach \x in {0,1,2} {
        \draw[line width=2] (0,-2*\x)--(4,-2*\x);
        \draw (2*\x,0)--(2*\x,-4);
    }
    
    \draw[line width=2] (4,0)--(4,-2);
    \draw[line width=2] (0,-2)--(0,-4);
\end{scope}
    \foreach \x in {0,1,2} {
        \foreach \y[evaluate={\l=int(3*\y+\x)}] in {0,1,2} {
        \node (\x\y) at (2*\x,-2*\y) [circle,draw,fill=white,minimum size=5mm] {};
        }
    }
    \node at (0,0) {1};
    \node at (2,0) {2};
    \node at (4,0) {0};
    \node at (0,-2) {4};
    \node at (2,-2) {3};
    \node at (4,-2) {5};
    \node at (0,-4) {7};
    \node at (2,-4) {8};
    \node at (4,-4) {6};

    \begin{scope}[<->,ibmBlue,line width=2,opacity=0.75]
        \draw[snake it, ibmRed] (00)--(10);
        \draw[snake it, ibmRed] (01)--(11);
        \draw[snake it, ibmRed] (02)--(12);
        \draw (00)--(10);
        \draw (01)--(11);
        \draw (02)--(12);
    \end{scope}
        \draw[ibmRed, snake it,line width=2] (20)--(21);
    \draw[line width=2,dotted] (4.6,-2)--(5.5,-2);
\end{scope}

\begin{scope}[shift={(0,-6.2)}]
\begin{scope}[black!20!white]
    \foreach \x in {0,1,2} {
        \draw[line width=2] (0,-2*\x)--(4,-2*\x);
        \draw (2*\x,0)--(2*\x,-4);
    }
    
    \draw[line width=2] (4,0)--(4,-2);
    \draw[line width=2] (0,-2)--(0,-4);
\end{scope}
    \foreach \x in {0,1,2} {
        \foreach \y[evaluate={\l=int(3*\y+\x)}] in {0,1,2} {
        \node (\x\y) at (2*\x,-2*\y) [circle,draw,fill=white,minimum size=5mm] {};
        }
    }
    \node at (0,0) {2};
    \node at (2,0) {1};
    \node at (4,0) {0};
    \node at (0,-2) {3};
    \node at (2,-2) {4};
    \node at (4,-2) {5};
    \node at (0,-4) {8};
    \node at (2,-4) {7};
    \node at (4,-4) {6};

    \begin{scope}[<->,ibmBlue,line width=2]
        \draw (10)--(20);
        \draw (11)--(21);
        \draw (12)--(22);
    \end{scope}
        \draw[ibmRed, snake it,line width=2] (01)--(02);
    \draw[line width=2,->,>=stealth] (4.6,-2)--(5.6,-2);
    \draw[line width=2,->,>=stealth,dotted] (4.6-6.2,-2)--(5.6-6.2,-2);
\end{scope}

\begin{scope}[shift={(6.2,-6.2)}]
\begin{scope}[black!20!white]
    \foreach \x in {0,1,2} {
        \draw[line width=2] (0,-2*\x)--(4,-2*\x);
        \draw (2*\x,0)--(2*\x,-4);
    }
    
    \draw[line width=2] (4,0)--(4,-2);
    \draw[line width=2] (0,-2)--(0,-4);
\end{scope}
    \foreach \x in {0,1,2} {
        \foreach \y[evaluate={\l=int(3*\y+\x)}] in {0,1,2} {
        \node (\x\y) at (2*\x,-2*\y) [circle,draw,fill=white,minimum size=5mm] {};
        }
    }
    \node at (0,0) {2};
    \node at (2,0) {0};
    \node at (4,0) {1};
    \node at (0,-2) {3};
    \node at (2,-2) {5};
    \node at (4,-2) {4};
    \node at (0,-4) {8};
    \node at (2,-4) {6};
    \node at (4,-4) {7};

    \begin{scope}[<->,ibmBlue,line width=2,opacity=0.75]
        \draw[snake it, ibmRed] (00)--(10);
        \draw[snake it, ibmRed] (01)--(11);
        \draw[snake it, ibmRed] (02)--(12);
        \draw (00)--(10);
        \draw (01)--(11);
        \draw (02)--(12);
    \end{scope}
        \draw[ibmRed, snake it,line width=2] (20)--(21);
    \draw[line width=2,->,>=stealth] (4.6,-2)--(5.6,-2);
\end{scope}

\begin{scope}[shift={(2*6.2,-1*6.2)}]
\begin{scope}[black!20!white]
    \foreach \x in {0,1,2} {
        \draw[line width=2] (0,-2*\x)--(4,-2*\x);
        \draw (2*\x,0)--(2*\x,-4);
    }
    
    \draw[line width=2] (4,0)--(4,-2);
    \draw[line width=2] (0,-2)--(0,-4);
\end{scope}
    
    \draw[ibmOrange, snake it,line width=2] (0,0)--(0,1);
    \draw[ibmOrange, snake it,line width=2] (2,0)--(2,1);
    \draw[ibmOrange, snake it,line width=2] (4,0)--(4,1);
    
    \draw[ibmOrange, snake it,line width=2] (0,-4)--(0,-5);
    \draw[ibmOrange, snake it,line width=2] (2,-4)--(2,-5);
    \draw[ibmOrange, snake it,line width=2] (4,-4)--(4,-5);
    
    \foreach \x in {0,1,2} {
        \foreach \y[evaluate={\l=int(3*\y+\x)}] in {0,1,2} {
        \node (\x\y) at (2*\x,-2*\y) [circle,draw,fill=white,minimum size=5mm] {};
        }
    }
    \node at (0,0) {0};
    \node at (2,0) {2};
    \node at (4,0) {1};
    \node at (0,-2) {5};
    \node at (2,-2) {3};
    \node at (4,-2) {4};
    \node at (0,-4) {6};
    \node at (2,-4) {8};
    \node at (4,-4) {7};
    
    \draw[line width=2,->,>=stealth] (4.6,-2)--(5.4,-2);
\end{scope}

\begin{scope}[shift={(3*6.2,-6.2)}]
\begin{scope}[black!20!white]
    \foreach \x in {0,1,2} {
        \draw[line width=2] (0,-2*\x)--(4,-2*\x);
        \draw (2*\x,0)--(2*\x,-4);
    }
    
    \draw[line width=2] (4,0)--(4,-2);
    \draw[line width=2] (0,-2)--(0,-4);
\end{scope}
    \foreach \x in {0,1,2} {
        \foreach \y[evaluate={\l=int(3*\y+\x)}] in {0,1,2} {
        \node (\x\y) at (2*\x,-2*\y) [circle,draw,fill=white,minimum size=5mm] {};
        \node (\x\y) at (2*\x,-2*\y) [circle,draw=ibmMagenta,minimum size=9mm,line width=2] {};
        }
    }
    \node at (0,0) {0};
    \node at (2,0) {2};
    \node at (4,0) {1};
    \node at (0,-2) {5};
    \node at (2,-2) {3};
    \node at (4,-2) {4};
    \node at (0,-4) {6};
    \node at (2,-4) {8};
    \node at (4,-4) {7};

    \draw[line width=2,dotted] (4.8,-2)--(5.4,-2);
\end{scope}

\begin{scope}[shift={(0*6.2,-2*6.2)}]
\begin{scope}[black!20!white]
    \foreach \x in {0,1,2} {
        \draw[line width=2] (0,-2*\x)--(4,-2*\x);
        \draw (2*\x,0)--(2*\x,-4);
    }
    
    \draw[line width=2] (4,0)--(4,-2);
    \draw[line width=2] (0,-2)--(0,-4);
\end{scope}
    \draw[ibmOrange, snake it,line width=2] (0,0)--(0,1);
    \draw[ibmOrange, snake it,line width=2] (2,0)--(2,1);
    \draw[ibmOrange, snake it,line width=2] (4,0)--(4,1);
    
    \draw[ibmOrange, snake it,line width=2] (0,-4)--(0,-5);
    \draw[ibmOrange, snake it,line width=2] (2,-4)--(2,-5);
    \draw[ibmOrange, snake it,line width=2] (4,-4)--(4,-5);
    
    \foreach \x in {0,1,2} {
        \foreach \y[evaluate={\l=int(3*\y+\x)}] in {0,1,2} {
        \node (\x\y) at (2*\x,-2*\y) [circle,draw,fill=white,minimum size=5mm] {};
        }
    }
    \node at (0,0) {0};
    \node at (2,0) {2};
    \node at (4,0) {1};
    \node at (0,-2) {5};
    \node at (2,-2) {3};
    \node at (4,-2) {4};
    \node at (0,-4) {6};
    \node at (2,-4) {8};
    \node at (4,-4) {7};

    \draw[line width=2,->,>=stealth] (4.6,-2)--(5.6,-2);
    \draw[line width=2,->,>=stealth,dotted] (4.6-6.2,-2)--(5.6-6.2,-2);
\end{scope}

\begin{scope}[shift={(1*6.2,-2*6.2)}]
\begin{scope}[black!20!white]
    \foreach \x in {0,1,2} {
        \draw[line width=2] (0,-2*\x)--(4,-2*\x);
        \draw (2*\x,0)--(2*\x,-4);
    }
    
    \draw[line width=2] (4,0)--(4,-2);
    \draw[line width=2] (0,-2)--(0,-4);
\end{scope}
    \foreach \x in {0,1,2} {
        \foreach \y[evaluate={\l=int(3*\y+\x)}] in {0,1,2} {
        \node (\x\y) at (2*\x,-2*\y) [circle,draw,fill=white,minimum size=5mm] {};
        }
    }
    \node at (0,0) {0};
    \node at (2,0) {2};
    \node at (4,0) {1};
    \node at (0,-2) {5};
    \node at (2,-2) {3};
    \node at (4,-2) {4};
    \node at (0,-4) {6};
    \node at (2,-4) {8};
    \node at (4,-4) {7};

    \begin{scope}[<->,ibmBlue,line width=2,opacity=.75]
        \draw[snake it, ibmRed] (00)--(10);
        \draw[snake it, ibmRed] (01)--(11);
        \draw[snake it, ibmRed] (02)--(12);
        \draw (00)--(10);
        \draw (01)--(11);
        \draw (02)--(12);
    \end{scope}
        \draw[ibmRed, snake it,line width=2] (20)--(21);
    \draw[line width=2,->,>=stealth] (4.6,-2)--(5.6,-2);
\end{scope}

\begin{scope}[shift={(2*6.2,-2*6.2)}]
\begin{scope}[black!20!white]
    \foreach \x in {0,1,2} {
        \draw[line width=2] (0,-2*\x)--(4,-2*\x);
        \draw (2*\x,0)--(2*\x,-4);
    }
    
    \draw[line width=2] (4,0)--(4,-2);
    \draw[line width=2] (0,-2)--(0,-4);
\end{scope}
    \foreach \x in {0,1,2} {
        \foreach \y[evaluate={\l=int(3*\y+\x)}] in {0,1,2} {
        \node (\x\y) at (2*\x,-2*\y) [circle,draw,fill=white,minimum size=5mm] {};
        }
    }
    \node at (0,0) {2};
    \node at (2,0) {0};
    \node at (4,0) {1};
    \node at (0,-2) {3};
    \node at (2,-2) {5};
    \node at (4,-2) {4};
    \node at (0,-4) {8};
    \node at (2,-4) {6};
    \node at (4,-4) {7};

    \begin{scope}[<->,ibmBlue,line width=2]
        \draw (10)--(20);
        \draw (11)--(21);
        \draw (12)--(22);
    \end{scope}
        \draw[ibmRed, snake it,line width=2] (01)--(02);
    \draw[line width=2,->,>=stealth] (4.6,-2)--(5.6,-2);
\end{scope}
\begin{scope}[shift={(3*6.2,-2*6.2)}]
\begin{scope}[black!20!white]
    \foreach \x in {0,1,2} {
        \draw[line width=2] (0,-2*\x)--(4,-2*\x);
        \draw (2*\x,0)--(2*\x,-4);
    }
    
    \draw[line width=2] (4,0)--(4,-2);
    \draw[line width=2] (0,-2)--(0,-4);
\end{scope}
    \foreach \x in {0,1,2} {
        \foreach \y[evaluate={\l=int(3*\y+\x)}] in {0,1,2} {
        \node (\x\y) at (2*\x,-2*\y) [circle,draw,fill=white,minimum size=5mm] {};
        }
    }
    \node at (0,0) {2};
    \node at (2,0) {1};
    \node at (4,0) {0};
    \node at (0,-2) {3};
    \node at (2,-2) {4};
    \node at (4,-2) {5};
    \node at (0,-4) {8};
    \node at (2,-4) {7};
    \node at (4,-4) {6};

    \begin{scope}[<->,ibmBlue,line width=2,opacity=0.75]
        \draw[snake it, ibmRed] (00)--(10);
        \draw[snake it, ibmRed] (01)--(11);
        \draw[snake it, ibmRed] (02)--(12);
        \draw (00)--(10);
        \draw (01)--(11);
        \draw (02)--(12);
    \end{scope}
        \draw[ibmRed, snake it,line width=2] (20)--(21);
    \draw[line width=2,dotted] (4.6,-2)--(5.4,-2);
\end{scope}
\begin{scope}[shift={(0*6.2,-3*6.2)}]
\begin{scope}[black!20!white]
    \foreach \x in {0,1,2} {
        \draw[line width=2] (0,-2*\x)--(4,-2*\x);
        \draw (2*\x,0)--(2*\x,-4);
    }
    
    \draw[line width=2] (4,0)--(4,-2);
    \draw[line width=2] (0,-2)--(0,-4);
\end{scope}
    \foreach \x in {0,1,2} {
        \foreach \y[evaluate={\l=int(3*\y+\x)}] in {0,1,2} {
        \node (\x\y) at (2*\x,-2*\y) [circle,draw,fill=white,minimum size=5mm] {};
        }
    }
    \node at (0,0) {1};
    \node at (2,0) {2};
    \node at (4,0) {0};
    \node at (0,-2) {4};
    \node at (2,-2) {3};
    \node at (4,-2) {5};
    \node at (0,-4) {7};
    \node at (2,-4) {8};
    \node at (4,-4) {6};

    \begin{scope}[<->,ibmBlue,line width=2]
        \draw (10)--(20);
        \draw (11)--(21);
        \draw (12)--(22);
    \end{scope}
        \draw[ibmRed, snake it,line width=2] (01)--(02);
    \draw[line width=2,->,>=stealth] (4.6,-2)--(5.6,-2);
    \draw[line width=2,->,>=stealth,dotted] (4.6-6.2,-2)--(5.6-6.2,-2);
\end{scope}

\begin{scope}[shift={(1*6.2,-3*6.2)}]
\begin{scope}[black!20!white]
    \foreach \x in {0,1,2} {
        \draw[line width=2] (0,-2*\x)--(4,-2*\x);
        \draw (2*\x,0)--(2*\x,-4);
    }
    
    \draw[line width=2] (4,0)--(4,-2);
    \draw[line width=2] (0,-2)--(0,-4);
\end{scope}
    \foreach \x in {0,1,2} {
        \foreach \y[evaluate={\xi=int((-1)^\y*\x + (1-(-1)^\y))},evaluate={\l=int(3*\y+\x)}] in {0,1,2} {
        \node (\l) at (2*\xi,-2*\y) [circle,draw,fill=white,minimum size=5mm] {};
        \node at (2*\xi,-2*\y) [] {\l};
        }
    }

    \begin{scope}[<->,ibmBlue,line width=2,opacity=0.75]
        \draw[snake it, ibmRed] (0)--(1);
        \draw[snake it, ibmRed] (5)--(4);
        \draw[snake it, ibmRed] (7)--(6);
        \draw (0)--(1);
        \draw (5)--(4);
        \draw (7)--(6);
    \end{scope}

    \draw[line width=2,->,>=stealth] (4.6,-2)--(5.6,-2);
\end{scope}

\begin{scope}[shift={(0*6.2,-4*6.2)}]
\draw[dashed] (0,1.1)--(3*6.2+4,1.1);
\node[fill=white] at (1.5*6.2+2,1.1) {$\textup{Step 2 }(t_2)$};
\begin{scope}[black!20!white]
    \foreach \x in {0,1,2} {
        \draw[line width=2] (0,-2*\x)--(4,-2*\x);
        \draw (2*\x,0)--(2*\x,-4);
    }
    
    \draw[line width=2] (4,0)--(4,-2);
    \draw[line width=2] (0,-2)--(0,-4);
\end{scope}
    \foreach \x in {0,1,2} {
        \foreach \y[evaluate={\xi=int((-1)^\y*\x + (1-(-1)^\y))},evaluate={\l=int(3*\y+\x)}] in {0,1,2} {
        \node (\l) at (2*\xi,-2*\y) [circle,draw,fill=white,minimum size=5mm] {};
        \node at (2*\xi,-2*\y) [] {\l};
        }
    }

        \draw[ibmRed, snake it,line width=2] (2)--(3);
        \draw[ibmRed, snake it,line width=2] (5)--(6);

    \node at (1,.62) [ibmRed,font=\fontsize{12}{14}] {\textbf{merged hops}};

    \draw[line width=2,->,>=stealth] (4.6,-2)--(5.6,-2);
    \draw[line width=2,->,>=stealth,dotted] (4.6-6.2,-2)--(5.6-6.2,-2);
\end{scope}

\begin{scope}[shift={(6.2,-4*6.2)}]
    
\begin{scope}[black!20!white]
    \foreach \x in {0,1,2} {
        \draw[line width=2] (0,-2*\x)--(4,-2*\x);
        \draw (2*\x,0)--(2*\x,-4);
    }
    
    \draw[line width=2] (4,0)--(4,-2);
    \draw[line width=2] (0,-2)--(0,-4);
\end{scope}
    \foreach \x in {0,1,2} {
        \foreach \y[evaluate={\xi=int((-1)^\y*\x + (1-(-1)^\y))},evaluate={\l=int(3*\y+\x)}] in {0,1,2} {
        \node (\l) at (2*\xi,-2*\y) [circle,draw,fill=white,minimum size=5mm] {};
        \node at (2*\xi,-2*\y) [] {\l};
        }
    }

    \begin{scope}[<->,ibmBlue,line width=2,opacity=0.75]
        \draw[snake it, ibmRed] (0)--(1);
        \draw[snake it, ibmRed] (5)--(4);
        \draw[snake it, ibmRed] (7)--(6);
        \draw (0)--(1);
        \draw (5)--(4);
        \draw (7)--(6);
    \end{scope}

    \draw[line width=2,->,>=stealth] (4.6,-2)--(5.6,-2);
\end{scope}
        
\begin{scope}[shift={(2*6.2,-4*6.2)}]
\begin{scope}[black!20!white]
    \foreach \x in {0,1,2} {
        \draw[line width=2] (0,-2*\x)--(4,-2*\x);
        \draw (2*\x,0)--(2*\x,-4);
    }
    
    \draw[line width=2] (4,0)--(4,-2);
    \draw[line width=2] (0,-2)--(0,-4);
\end{scope}
    \foreach \x in {0,1,2} {
        \foreach \y[evaluate={\l=int(3*\y+\x)}] in {0,1,2} {
        \node (\x\y) at (2*\x,-2*\y) [circle,draw,fill=white,minimum size=5mm] {};
        }
    }
    \node at (0,0) {1};
    \node at (2,0) {0};
    \node at (4,0) {2};
    \node at (0,-2) {4};
    \node at (2,-2) {5};
    \node at (4,-2) {3};
    \node at (0,-4) {7};
    \node at (2,-4) {6};
    \node at (4,-4) {8};

    \begin{scope}[<->,ibmBlue,line width=2]
        \draw (10)--(20);
        \draw (11)--(21);
        \draw (12)--(22);
    \end{scope}
        \draw[ibmRed, snake it,line width=2] (01)--(02);
    \draw[line width=2,->,>=stealth] (4.6,-2)--(5.6,-2);
\end{scope}

\begin{scope}[shift={(3*6.2,-4*6.2)}]
\begin{scope}[black!20!white]
    \foreach \x in {0,1,2} {
        \draw[line width=2] (0,-2*\x)--(4,-2*\x);
        \draw (2*\x,0)--(2*\x,-4);
    }
    
    \draw[line width=2] (4,0)--(4,-2);
    \draw[line width=2] (0,-2)--(0,-4);
\end{scope}
    \foreach \x in {0,1,2} {
        \foreach \y[evaluate={\l=int(3*\y+\x)}] in {0,1,2} {
        \node (\x\y) at (2*\x,-2*\y) [circle,draw,fill=white,minimum size=5mm] {};
        }
    }
    \node at (0,0) {1};
    \node at (2,0) {2};
    \node at (4,0) {0};
    \node at (0,-2) {4};
    \node at (2,-2) {3};
    \node at (4,-2) {5};
    \node at (0,-4) {7};
    \node at (2,-4) {8};
    \node at (4,-4) {6};

    \begin{scope}[<->,ibmBlue,line width=2,opacity=0.75]
        \draw[snake it, ibmRed] (00)--(10);
        \draw[snake it, ibmRed] (01)--(11);
        \draw[snake it, ibmRed] (02)--(12);
        \draw (00)--(10);
        \draw (01)--(11);
        \draw (02)--(12);
    \end{scope}
        \draw[ibmRed, snake it,line width=2] (20)--(21);
    \draw[line width=2,dotted] (4.6,-2)--(5.4,-2);
\end{scope}

\end{scope}
\end{tikzpicture}
    \caption{Illustration of the first Trotter step and the start of the second for a $3\times 3$ lattice. Identical operations are made on both spin sectors, so only spin-up is shown. Blue arrows indicate FSWAPs, red wavy lines indicate hopping interactions and orange wavy lines denote the special circuit used to implement hopping terms missed by the swap network in the odd $L_x$ case.
    All hopping terms are for time $t_1/2$ ($t_2/2$) in step 1 (2), except in the case where merging is noted.
    Blue arrows and red wavy lines superimposed indicate a simultaneous hop and FSWAP and the magenta circles indicate Coulomb interactions for time $t_1$ ($t_2$) with the corresponding spin-down modes. Parallelizable actions are shown on the same grid and labels indicate the location of the encoded fermionic modes at each stage. The final round of hops of step 1 and the first round of step 2 are merged into hops of time $(t_1+t_2)/2$. 
   }
    \label{fig:swapnet}
\end{figure}
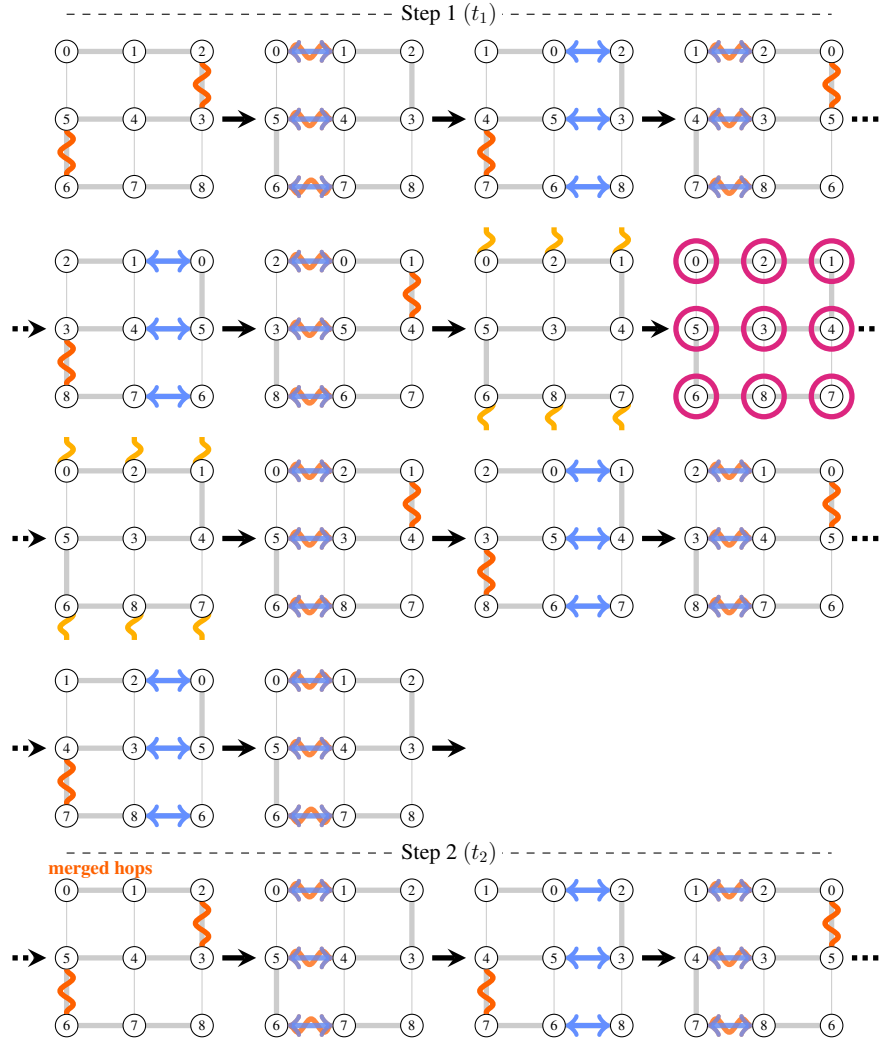

\subsubsection{Trotter Step Merging}\label{sec:step merging}

The symmetry of second-order Trotter circuits allows for further gate savings to be made.
In cases of even and odd $L_x$, the Trotter step begins and ends with the application of all vertical hopping terms available in the initial configuration (step 1. for even, steps 1. and 2. for odd). This means that the final operation in step $j$ can be merged with the first in step $j+1$ with the exception of the final Trotter step in the circuit. Specifically, for a circuit with $N$ Trotter steps numbered 1 to $N$ with step $j$ being for time $t_j$, steps $j<N$ have these hopping terms omitted from the end of their circuit and steps $k>1$ have the hopping terms at the start evolve for $(t_{k}+t_{k-1})/2$ instead of $t_k/2$. This reduces the number of gates required; the precise saving is discussed in \cref{sec:costs}.
A complete Trotter step for a $3\times 3$ lattice is illustrated in \cref{fig:swapnet}.

\subsubsection{Trotter Error}\label{sec:trotter-error}
We estimate the accuracy of the Trotterized dynamics using small-scale numerical simulations on the Fermi-Hubbard instance studied in this work. We simulate the error in the expectation values of all Pauli $Z$ observables up to weight-$3$, using $4$ Trotter steps and for times up to $T=2$. That is, for each observable $O$ and time $t$ we compute all 
\begin{equation}\label{eq:trotter-obs-error}
    |\bra{\psi(t)}O\ket{\psi(t)} - \bra{\psi'(t)}O\ket{\psi'(t)}|. 
\end{equation}
Where $\ket{\psi(t)} = e^{-i t H} \ket{\psi_0}$ and $\ket{\psi'(t)} = \prod_{j=1}^4 S_2(t/4)\ket{\psi_0}$. For $\ket{\psi_0}$, we choose the dimerized configuration used in the experiments. We then examine the average and maximum values of these errors.
We also compute infidelity between the exact time-evolved state $\ket{\psi(t)}$ and Trotterized time-evolved state $\ket{\psi'(t)}$, defined as
\begin{equation}\label{eq:state-error-trotter}
    2\sqrt{1 - |\langle\psi(t) | \psi' (t)\rangle|^{2}}
\end{equation}
in order to upper bound the Trotter error for all observables. The results of the simulation are shown in \cref{fig:trotter-error}. We observe that for late times and simulable sizes, although state infidelity increases with system size, the local observable error remains relatively stable. This error is significantly smaller than infidelity, with the average error being comparable to hardware noise.
For local observables and local Hamiltonians, there are theoretical grounds for assuming that Trotter error can be independent of system size, especially for fixed times \cite{Childs2021}.

However, we note that between times $1.5$ and $2$, Trotter error begins to play a more significant role, and our simulations begin interpolating between true time dynamics and Floquet dynamics. This is borne out in our comparisons to exact $U=0$ simulations, as can be seen in ~\cref{fig:doublons_and_triplets}.

We also repeat this analysis for observables whose weight scales with system size. Again we look at system sizes $L_{x} \times L_{y} =2 \times 2$, $2 \times 3$ and $3 \times 3$ but with observables with weight four, six and eight. The results of these calculations are shown in \cref{fig:trotter-error-weight}.

\begin{figure}[htpb]
    \centering
    \begin{minipage}[t]{0.45\textwidth}
        \centering
        \includegraphics[width=\linewidth]{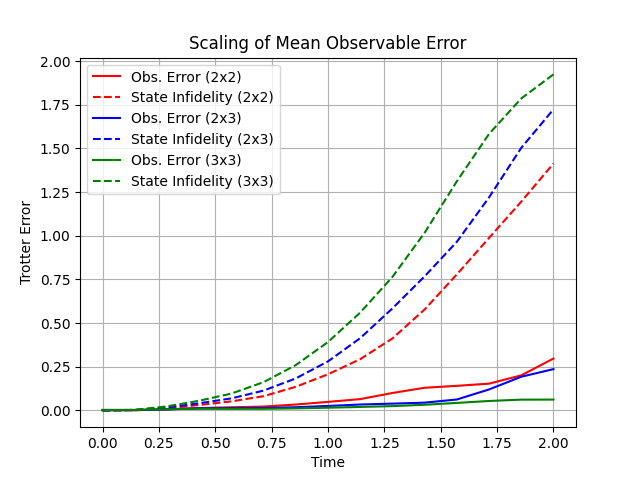}
        \\[-0.5em]
        {\small (a) Mean error over observables.}
    \end{minipage}
    \hspace{1cm}
    \begin{minipage}[t]{0.45\textwidth}
        \centering
        \includegraphics[width=\linewidth]{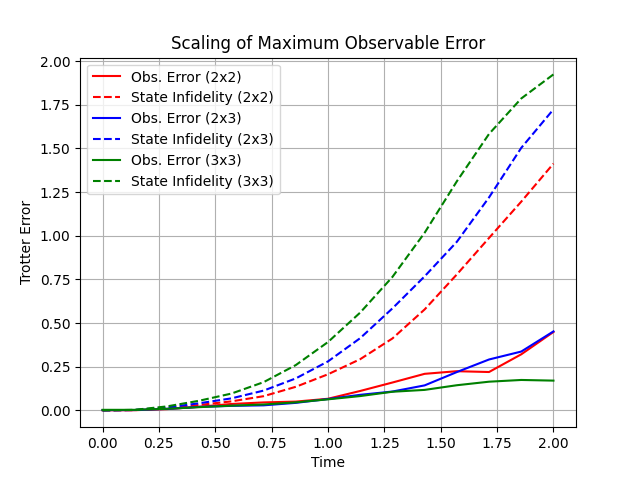}
        \\[-0.5em]
        {\small (b) Maximum error over observables.}
    \end{minipage}
    \caption{
    Trotter error on the expectation values (\cref{eq:trotter-obs-error}, solid lines) of all one, two and three qubit Pauli $Z$ observables for $4$ second-order Trotter steps over $15$ times between $t=0$ and $t=2$. As in our experiments, we take $U=4$, apply a magnetic flux in the smaller direction and use a dimerized configuration as input state. We then examine the system sizes $L_{x} \times L_{y} =2 \times 2$, $2 \times 3$ and $3 \times 3$ subject to double periodic boundary conditions. We look at the mean error over all observables at each time (a) and the maximum error over these observables at each time (b). We compute the state infidelity (\cref{eq:state-error-trotter}, dashed lines) between the exact time evolved state at each time $t$ and the state evolved using $4$ Trotter steps for each time $t$.
    }
    \label{fig:trotter-error}
\end{figure}

\begin{figure}[htpb]
    \centering
    \begin{minipage}[t]{0.45\textwidth}
        \centering
        \includegraphics[width=\linewidth]{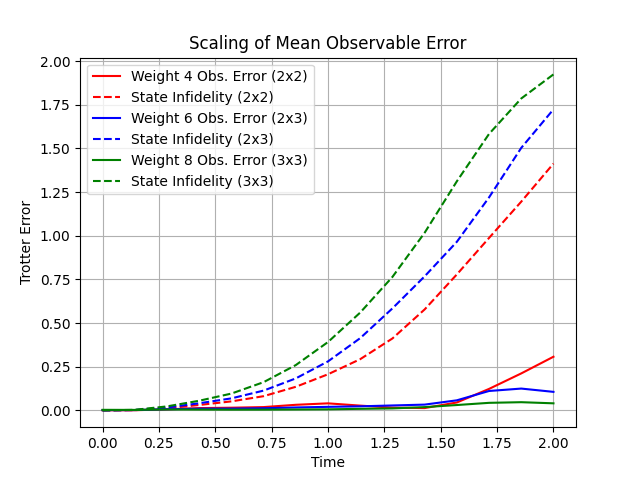}
        \\[-0.5em]
        {\small (a) Mean error over observables.}
    \end{minipage}
    \hspace{1cm}
    \begin{minipage}[t]{0.45\textwidth}
        \centering
        \includegraphics[width=\linewidth]{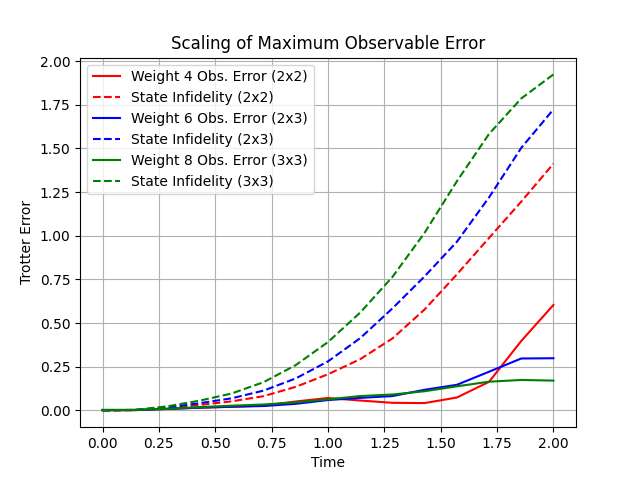}
        \\[-0.5em]
        {\small (b) Maximum error over observables.}
    \end{minipage}
    \caption{
    Trotter error on the expectation values (\cref{eq:trotter-obs-error}, solid lines) of all weight-four Pauli-$Z$ observables for $L_{x} \times L_{y} =2 \times 2$, all weight-six observables for $L_{x} \times L_{y} =2 \times 3$ and weight-eight observables for $L_{x} \times L_{y} =3 \times 3$. We use $4$ second-order Trotter steps over $15$ times between $t=0$ and $t=2$. As in our experiments, we take $U=4$, apply a magnetic flux in the smaller direction and use a dimerized configuration as the input state. We use double periodic boundary conditions. We look at the mean error over all observables at each time (a) and the maximum error over these observables at each time (b). We compute the state infidelity (\cref{eq:state-error-trotter}, dashed lines) between the exact time evolved state at each time $t$ and the state evolved using $4$ Trotter steps for each time $t$.
    }
    \label{fig:trotter-error-weight}
\end{figure}

\subsection{Gate Decompositions}\label{sec:gate decomp}
We use the notation for Pauli rotation
\begin{equation}
    R_{\sigma}(t)=e^{-\frac i 2 \sigma t}.
\end{equation}
The native two-qubit gate on the device is the arbitrary $R_{zz}$. 
Due to the all-to-all connectivity of the device, SWAP operations can be applied implicitly by relabeling qubits. With this, one can implement an FSWAP by applying a CZ and an implicit swap to a pair of qubits (in any order), where the CZ gate can be applied with the gates 
\begin{equation}
    \text{CZ}=R_{zz}\left(\frac \pi 2\right)\cdot S\otimes S.
\end{equation}

A dressed hopping interaction between modes $i$, $j$ adjacent in the JW ordering with $J=1$ 
for time $t$ 
is implemented with the circuit: 
$$
\Qcircuit @C=1.2em @R=1.5em {
\lstick{i} & \gate{R_z(\phi_{ij})} & \gate{H} & \multigate{1}{R_{zz}(-t)} & \gate{H} & \gate{S^\dagger} & \gate{H} & \multigate{1}{R_{zz}(-t)} & \gate{H} & \gate{S} & \gate{R_z(-\phi_{ij})} & \qw\\
\lstick{j} & \qw & \gate{H} & \ghost{R_{zz}(-t)} & \gate{H} & \gate{S^\dagger} & \gate{H} & \ghost{R_{zz}(-t)} & \gate{H} & \gate{S} & \qw & \qw
}
$$
If the pair of modes is adjacent across the loop discussed in \cref{sec:JW loop}, then the sign of the $R_{zz}$ rotations changes according to \cref{eq: hopping with parity}. 
In the cases where a hopping interaction is immediately followed by or preceded by an FSWAP, we can apply both at once using the circuit:
$$
\Qcircuit @C=1.2em @R=1.5em {
\lstick{i}   & \gate{R_z(\phi_{ij})}&  \gate{H} & \multigate{1}{R_{zz}\left(-t-\frac \pi 2\right)} & \gate{H} & \gate{S^\dagger} & \gate{H} & \multigate{1}{R_{zz}\left(-t-\frac \pi 2\right)} & \gate{H} & \qw & \qw\\
\lstick{j} & \qw &  \gate{H} & \ghost{R_{zz}\left(-t-\frac \pi 2\right)} & \gate{H} & \gate{S^\dagger} & \gate{H} & \ghost{R_{zz}\left(-t-\frac \pi 2\right)} & \gate{H}  & \gate{R_z(-\phi_{ij})}& \qw
}
$$

For a lattice with odd vertical dimension $L_y$, the vertical terms across the vertical boundary (i.e., hops between pairs $\{(0,N-L_x),(1,N-L_x+1),\dots,(L_x-1,N-1)\}$) are implemented by the following circuit
\begin{equation}
    \left(\prod_{i=0}^{L_x-2}\prod_{j=1}^{L_y-1-i}\text{CZ}(i,N-j)\right)
    \cdot\prod_{i=0}^{L_x-1} \text{HOP}(i,N-L_x+i,t) \cdot
    \left(\prod_{i=0}^{L_x-2}\prod_{j=1}^{L_y-1-i}\text{CZ}(i,N-j)\right)
\end{equation}
where HOP$(i,j,t)$ denotes the two-qubit hopping interaction circuit between qubits $i$ and $j$ for time $t$ specified above. The case of $L_x=4$ is illustrated below in circuit diagram form.
$$\Qcircuit @C=1.2em @R=1.5em {
 \lstick{N-4} & \qw & \qw & \qw & \qw & \qw & \qw & \multigate{7}{\text{Hops}} & \qw & \qw & \qw & \qw & \qw & \qw &  \qw\\ 
 \lstick{N-3} & \qw & \qw & \ctrl{3} & \qw & \qw & \qw & \ghost{\text{Hops}} & \qw & \qw & \ctrl{3} & \qw & \qw & \qw &  \qw\\ 
 \lstick{N-2} & \qw & \ctrl{2} & \qw & \qw & \ctrl{3} & \qw & \ghost{\text{Hops}} & \qw & \ctrl{2} & \qw & \qw & \ctrl{3} & \qw &  \qw\\ 
 \lstick{N-1} & \ctrl{1} & \qw & \qw & \ctrl{2} &\qw & \ctrl{3} & \ghost{\text{Hops}} & \ctrl{1} & \qw & \qw & \ctrl{2} &\qw & \ctrl{3} &  \qw\\ 
 \lstick{0} & \control \qw & \control \qw & \control \qw & \qw & \qw & \qw & \ghost{\text{Hops}} & \control \qw & \control \qw & \control \qw & \qw & \qw & \qw &  \qw\\ 
 \lstick{1} & \qw & \qw & \qw & \control \qw & \control \qw & \qw & \ghost{\text{Hops}} & \qw & \qw & \qw & \control \qw & \control \qw & \qw &  \qw\\ 
 \lstick{2} & \qw & \qw & \qw & \qw & \qw & \control \qw & \ghost{\text{Hops}} & \qw & \qw & \qw & \qw & \qw & \control \qw &  \qw\\ 
 \lstick{3} & \qw & \qw & \qw & \qw & \qw & \qw & \ghost{\text{Hops}} & \qw & \qw & \qw & \qw & \qw & \qw &  \qw\\ 
}
$$
The CZ gates in the circuit serve to cancel out the $Z$-strings in the encoded hopping terms. Consider the $L_x=4$ example. The hopping terms in question are (up to factors)
\begin{equation}\label{eq:boundary hops}
    \begin{split}
        h_{0,N-4}&=(X_0X_{N-4}+Y_0Y_{N-4})Z_{N-3}Z_{N-2}Z_{N-1},\\
        h_{1,N-3}&=(X_1X_{N-3}+Y_1Y_{N-3})Z_{N-2}Z_{N-1}Z_{0},\\
        h_{2,N-2}&=(X_2X_{N-2}+Y_2Y_{N-2})Z_{N-1}Z_{0}Z_{1},\\
        h_{3,N-1}&=(X_3X_{N-1}+Y_3Y_{N-1})Z_{0}Z_{1}Z_{2},
    \end{split}
\end{equation}
where we are using the JW loop to get a lower weight representation of the terms as discussed in \cref{sec:JW loop} and there is no magnetic field alteration. Defining the ``stringless'' versions of the hopping terms as
\begin{equation}
    \begin{split}
        \tilde h_{0,N-4}&=(X_0X_{N-4}+Y_0Y_{N-4}),\\ 
        \tilde h_{1,N-3}&=(X_1X_{N-3}+Y_1Y_{N-3}),\\
        \tilde h_{2,N-2}&=(X_2X_{N-2}+Y_2Y_{N-2}),\\
        \tilde h_{3,N-1}&=(X_3X_{N-1}+Y_3Y_{N-1}),
    \end{split}
\end{equation}
it is not hard to check that conjugating $\tilde h_{i,j}$ by the CZ circuit blocks above results in the corresponding $h_{i,j}$ for all pairs in question. Accordingly, evolution under the stringless hopping terms conjugated by these CZs is equivalent to evolution under the true, stringed terms.

The reader may recall that, as this is applied after the swap network, the modes will be shuffled from their original ordering. This does not affect this circuit, as at any point in the swap network (\cref{eq:boundary hops}) holds up to the resulting permutation of the labels on the LHS of the equations and since the circuit applies an evolution under all the Paulis, this relabelling is irrelevant to the outcome.

The total two-qubit gate cost of this circuit for the whole system is
\begin{equation}
    C_\text{b}=
        4\frac{(L_x-1)L_x}{2}+4L_x,
\end{equation}
where the first term is from the CZ gates, the second is from the hopping terms. As this circuit is always placed before and after the Coulomb interaction circuit, the set of CZ gates on either side of the Coulomb circuit can be commuted through to cancel each other out, making the effective cost of this circuit
\begin{equation}
    \widetilde{C}_\text{b}=
        2\frac{(L_x-1)L_x}{2}+4L_x=L_x^2+3L_x.
\end{equation}

\subsection{Trotter Circuit Gate Costs}\label{sec:costs}

The total two-qubit gate cost of a single forward pass of the swap network depends on the number of FSWAPs, vertical hopping terms and horizontal hopping terms implemented ($n_f$, $n_v$, $n_h$ respectively) and is given by
\begin{equation}
    C_\text{nw}=n_f+2n_v+n_h.
\end{equation}
Each vertical hopping term costs two two-qubit gates but horizontal hopping terms effectively cost only 1 because they are merged with an FSWAP for a total of 2 gates, this is counted as 1 gate due to the FSWAP and 1 due to the hopping. The numbers of each type of hopping term implemented in the swap network are given by
\begin{equation}
    n_h=2L_xL_y,\quad n_v=
    \begin{cases}
        2L_xL_y&:L_y\text{ even},\\
        2L_x(L_y-1)&:L_y\text{ odd}.\\
    \end{cases}
\end{equation}
This is due to the fact that the vertical hopping terms across the periodic boundary are not caught by the swap network as discussed in \cref{sec: periodic boundary}. 

The total number of swaps can be found by counting how many applications of $U_L$ and $U_R$ are needed for each mode to reach both ends of its horizontal row. For even $L_x$ this requires $L_x-1$ applications of $U_L$ and $L_x-2$ of $U_R$, with $U_L$ and $U_R$ containing ${L_xL_y}$ and ${(L_x-2)L_y}$ FSWAPs respectively (counting both spin sectors). For odd $L_x$ this requires $L_x-1$ of both $U_L$ and $U_R$ which both contain ${(L_x-1)L_y}$ FSWAPs. The usual swap network that returns the lattice to its original configuration on completion requires $L_x$ applications of $U_L$ and $U_R$ in both cases.

The total number of FSWAPs in our swap network is then:
\begin{equation}
    n_f =
    \begin{cases}
    (2L_x^2-5L_x+4)L_y&:L_x\text{ even},\\
    (2L_x^2-4L_x+2)L_y&:L_x\text{ odd}.\\
    \end{cases}
\end{equation}
Combining these, we get
\begin{equation}
    C_\text{nw}=
    \begin{cases}
        (2L_x^2+L_x+4)L_y&:L_x\text{ even, } L_y\text{ even},\\
        (2L_x^2+L_x+4)L_y-4L_x&:L_x\text{ even, } L_y\text{ odd},\\
        (2L_x^2+2L_x+2)L_y&:L_x\text{ odd, } L_y\text{ even},\\
        (2L_x^2+2L_x+2)L_y-4L_x&:L_x\text{ odd, } L_y\text{ odd}.
    \end{cases}
\end{equation}

As discussed in \cref{sec:step merging}, all but the last Trotter step have their final round of vertical hopping terms omitted as they are merged into the start of the next step. For each step, this amounts to a two-qubit gate saving of
\begin{equation}
    C_\text{merge}=
    \begin{cases}
        4L_y&:L_y\text{ even},\\
        4(L_y-1)&:L_y\text{ odd}.\\
    \end{cases}
\end{equation}
The merged hopping terms are all vertical hops adjacent in the ordering in the initial lattice configuration before any FSWAPs, $L_y$ for even $L_y$ and $L_y-1$ for odd, due to the misalignment of the JW ordering. The factor of 4 comes from the 2 spin sectors and the two-qubit gate cost for a hopping term.

As noted above, the cost of the circuit to implement vertical hops missed by the swap network is
\begin{equation}
    \widetilde{C}_\text{b}=
    \begin{cases}
        0&:L_y\text{ even},\\
        L_x^2+3L_x&:L_y\text{ odd},
    \end{cases}
\end{equation}
and the cost of the Coulomb interactions is
\begin{equation}
    C_\text{c}=L_xL_y.
\end{equation}

Putting these together, the total cost of an $N$ step second-order Trotter circuit is
\begin{equation}
    C_\text{tot}=N(2C_{\text{nw}} +2\widetilde C_\text{b} + C_\text{c} - C_\text{merge}) + C_\text{merge},
\end{equation}
where the final addition of $C_\text{merge}$ is due to the final step not having anything to merge its last hops with. In terms of lattice dimensions, this is
\begin{equation}
    C_\text{tot}=
    \begin{cases}
        N(4L_x^2+3L_x+5)L_y+4L_y&:L_x\text{ even, } L_y\text{ even},\\
        N((4L_x^2+3L_x+4)L_y+2L_x^2-2L_x+4)+4(L_y-1)&:L_x\text{ even, } L_y\text{ odd},\\
        N(4L_x^2+5L_x)L_y+4L_y&:L_x\text{ odd, } L_y\text{ even},\\
        N((4L_x^2+5L_x)L_y+2L_x^2-L_x+4)+4(L_y-1)&:L_x\text{ odd, } L_y\text{ odd}.
    \end{cases}
\end{equation}
\Cref{tab:gate cost comparison} shows a comparison between the gate cost of a second-order Trotter circuit using our optimisations vs a standard swap network~\cite{kivlichan18}.

\begin{table}
    \centering
    \begin{tabular}{|c|c|c|}
    \hline
        Lattice Size & 2q gates (our network) & 2q gates (standard network) \\
        \hline\hline
        $4\times 5$ & $428N+16$ & $484N+40$\\
        $4\times 6$ & $486N+24$ & $552N+48$\\
        $5\times 5$ & $674N+16$ & $717N+48$\\
        $4\times 7$ & $588N+24$ & $668N+56$ \\
    \hline
    \end{tabular}
    \caption{Comparison for various lattice sizes of two-qubit gate cost for an $N$ step second order Trotter circuit using the optimised circuit described in \cref{sec:2nd order trot} and the standard swap network used on JW lattices. Note that all lattices have periodic boundary conditions and the costs for the standard network also include the same circuit for vertical boundary terms and account for merged hops/swaps between Trotter steps.}
    \label{tab:gate cost comparison}
\end{table}

\subsection{Initial state preparation}\label{sec:init_state}
We prepare a series of triplets across our lattice. 
A triplet is prepared over two sites, $s_1, s_2$, given by 
\begin{equation}
  \ket{\psi_{\mathrm{triplet}}} = \frac{1}{\sqrt{2}}(
    c_{s_1, \uparrow}^\dagger c_{s_2, \downarrow}^\dagger 
    + c_{s_1, \downarrow}^\dagger c_{s_2, \uparrow}^\dagger) \ket{\Omega}
    = \frac{1}{\sqrt{2}} (
      \ket{1_{s_1, \uparrow} 0_{s_2, \uparrow} 0_{s_1, \downarrow} 1_{s_2, \downarrow}} 
      - \ket{0_{s_1, \uparrow} 1_{s_2, \uparrow} 1_{s_1, \downarrow} 0_{s_2, \downarrow}})
\end{equation}
This state is produced using the following circuit, which is a modified Greenberger–Horne–Zeilinger state creation circuit.
\begin{equation}
\Qcircuit @C=1.2em @R=1.5em {
\lstick{s_{1 \uparrow}} & \gate{H} & \multigate{1}{R_{zz}(\frac \pi 2)} & \gate{S^\dagger} & \qw & \qw & \qw & \qw & \qw & \qw & \qw & \qw \\
\lstick{s_{2 \uparrow}} & \gate{H} & \ghost{R_{zz}(\frac \pi 2)} & \gate{S^\dagger} & \gate{H} & \multigate{1}{R_{zz}(\frac \pi 2)} & \gate{S^\dagger} & \gate{X} & \qw & \qw & \qw & \qw \\
\lstick{s_{1 \downarrow}} & \gate{H} & \qw & \qw & \qw & \ghost{R_{zz}(\frac \pi 2)} & \gate{S^\dagger} & \gate{H} & \multigate{1}{R_{zz}(\frac \pi 2)} & \gate{S^\dagger} & \gate{Y} & \qw \\
\lstick{s_{2 \downarrow}} & \gate{H} & \qw & \qw & \qw & \qw & \qw & \qw & \ghost{R_{zz}(\frac \pi 2)} & \gate{S^\dagger} & \gate{H} & \qw
}
\label{eq:triplet_state_prep}
\end{equation}
This state preparation circuit incurs a cost of three two-qubit gates per triplet pair. 
In our system of 28 sites, \cref{fig:init_state_graph}, all sites are involved in a triplet, except for sites 0 and 13, which encode a holon and a doublon, respectively. 
Therefore, we have 13 triplet pairs, so state preparation accounts for 39 two-qubit gates in \cref{tab:gate_counts}. 

\begin{figure}
    \centering
    \includegraphics[width=0.7\linewidth]{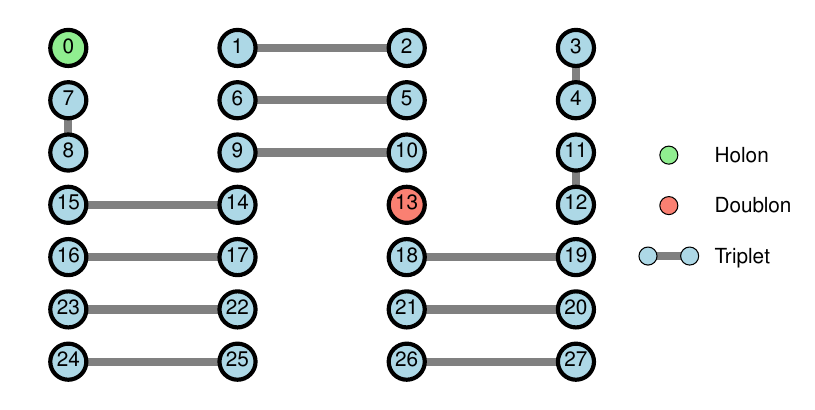}
    \caption{Initial state layout. A holon (green) and a doublon (red) are prepared, with all other sites prepared into triplet pairs (blue), 
    where each triplet connects sites adjacent in the JW encoding.
    }
    \label{fig:init_state_graph}
\end{figure}

\subsection{Compiler optimisations}\label{sec:compiler_optimisations}
We construct our circuits as described in \cref{sec:2nd order trot}, and then pass them through the compiler provided by \texttt{pytket} \cite{sivarajah2020t}. 
In particular, we apply \texttt{optimisation\_level = 2} of the compiler \cite{quantinuumPytketquantinuumDocumentation}, the main effect of which is captured by a pass of \texttt{FullPeepholeOptimise()}, which synthesises one-, two- and three-qubit subcircuits. 
The compiler also recognises and removes redundant gates and gate sequences, such as self-inverse gates and diagonal gates immediately before measurements, and performs some optimisations specific to Clifford circuits \cite{fagan2019optimising}. 
The main effect of this compilation on our circuits is to remove one-qubit gates: as described in \cref{sec:2nd order trot}, we optimise for two-qubit gate count such that there are no superfluous two-qubit gates, and therefore the compiler cannot find any further gates to remove. 
However, the subcircuits for each component, i.e., hopping, swap, and onsite terms, include numerous one-qubit gates that can be merged in between rounds of two-qubit gates, and so the compilation meaningfully reduces the one-qubit gate count.
\par 

In \cref{tab:gate_counts} we report the number of gates included in our circuits before and after compilation.
Note that our circuits include \texttt{SWAP} gates, but on the ion-trap platform, these are performed in software by relabelling qubits, i.e., no quantum operation occurs, so we omit these from gate counts. 
\Cref{tab:gate cost comparison} states that for $N$ Trotter steps, our swap network construction admits a cost of $588N + 24$; we use $N=4$ and \cref{sec:init_state} shows that the initial state preparation incurs a further 39 two-qubit gates, giving $(588\times4) + 24 + 39 = 2415$ two-qubit gates overall. 
\par 

\begin{table}[htpb]
    \centering
    \begin{tabular}{l@{\hskip 0.5in}c@{\hskip 0.5in}c}
     & one-qubit & two-qubit \\
    \hline
    Uncompiled & 11605 & 2415 \\
    Compiled & 4576 -- 4627 & 2410 -- 2415 \\
    \end{tabular}
    \caption{Number of gates in circuits before and after optimisation through pytket's compiler. There are minor variations due to the number of gates which can be compiled together following the pseudo-twirling routine.
    }
    \label{tab:gate_counts}
\end{table}

\subsection{Hardware specification}

\begin{table}[htbp]
    \centering
    \begin{tabular}{lr}
     & Error magnitude \\
    \hline
    Single-qubit gate error & 3E-05 \\
    Two-qubit gate error & 1E-03 \\
    Single-qubit leakage & 1E-05 \\
    Two-qubit leakage & 2E-04 \\
    Memory error & 2E-04 \\
    Measurement crosstalk error & 3E-06 \\
    SPAM error & 1E-03 \\
    \end{tabular}
    \caption{Hardware specifications of Quantinuum's H2-2 ion-trap device at the time of simulation.}
    \label{tab:hw_specs}
\end{table}

We ran our circuits on Quantinuum's \texttt{H2-2} ion-trap quantum processor in two batches: first between 19-23 June 2025, and second between 18-22 July 2025. 
The specified device performance at the time of running is listed in \cref{tab:hw_specs}, taken from \cite{h2_specs}.
We prepared time evolution circuits in intervals $\Delta t=0.1$, from $T=0.1$ to $T=2.0$, i.e., 20 circuits per timeseries. 
With two values of onsite term, $U=0,4$, and 8 pseudo-twirling instances per circuit (\cref{sec:pseudotwirling}), in total we ran $20 \times 2 \times 8 = 320$ distinct circuits, each for 160 shots. 
Each pseudo-twirling instance was executed in a random order. 
The list of pseudo-twirling instances was split into two halves and each half was executed on a different day. 
Given the depth of our circuits, and the gatespeed of ion-trap processors, each shot took $\sim5$ seconds, for a total runtime on the quantum device of $5 \times 20 \times 160  = 16000 \ \rm{seconds}\approx 4.5$ hours. 
Additionally, there were queues to access the device, as well as compilation time, so that the experiment took several days to run in practice.}
\section{Error suppression, error mitigation, and experimental analysis}
\label{sec:errormitigation}

Error suppression and mitigation are important to improve the accuracy of results obtained from a noisy quantum computer. The extremely restrictive constraints on the number of shots available in an experiment on ion-trap hardware mean that many techniques are not available or appropriate. Here we use a technique known as training with fermionic linear optics (TFLO)~\cite{tflo}, implemented using a variant which uses very few additional shots, together with Gaussian process regression (GPR) for further cleanup of the results. In addition, we implement a technique known as Pauli pseudotwirling to suppress systematic errors and facilitate error mitigation.
\par 

\begin{figure}[htpb]
    \includegraphics[width=\linewidth]{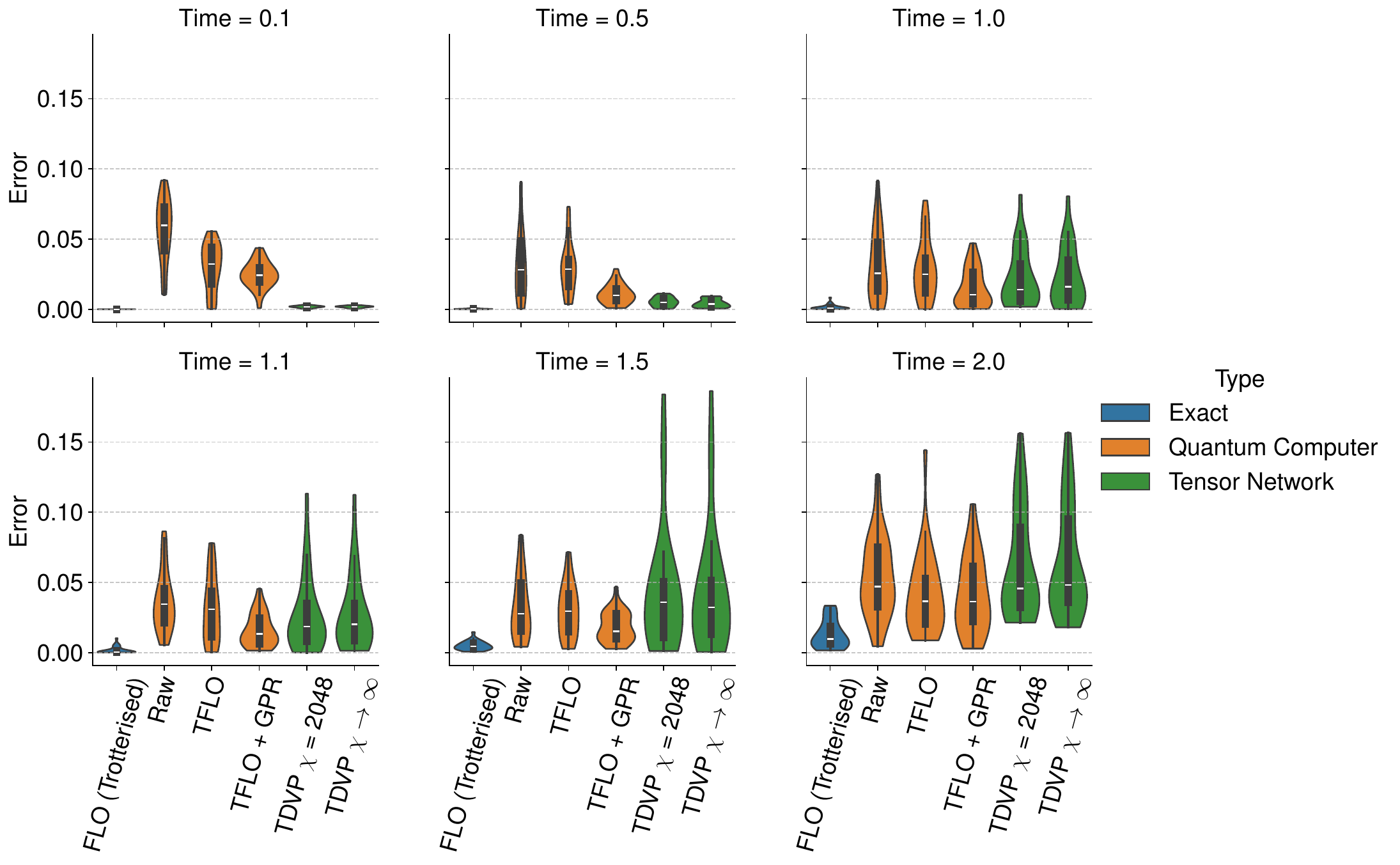}
    \caption{
    Error profile for the doublon observable compared with exact dynamics over time, computed through raw hardware data, mitigated hardware data (FLO, GPR), and TDVP simulation along with extrapolated TDVP results.
    }
    \label{fig:errors_violin_doublon}
\end{figure}

\begin{figure}
    \centering
    \includegraphics[width=0.85\linewidth]{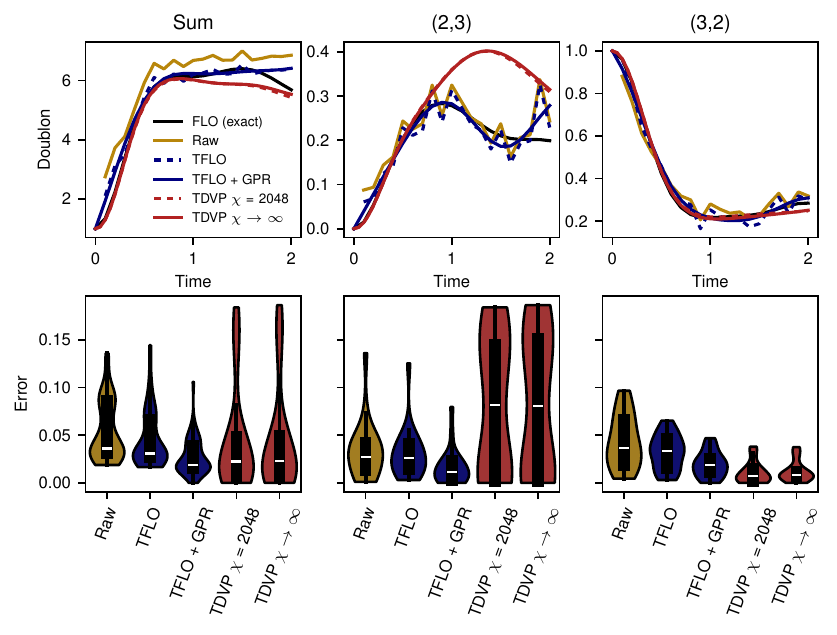}
    \caption{
        Dynamics of doublon observable for the raw, mitigated and TDVP-simulated data at $U=0$, along with corresponding errors with respect to exact FLO simulation.
        The left panel sums the doublon over the entire lattice.
        The central (right) panel highlights a particular site where we see the worst (best) performance for TDVP.
    }
    \label{fig:doublon_dynamics_errors}
\end{figure}

\begin{figure}
    \centering
    \includegraphics[width=0.85 \linewidth]{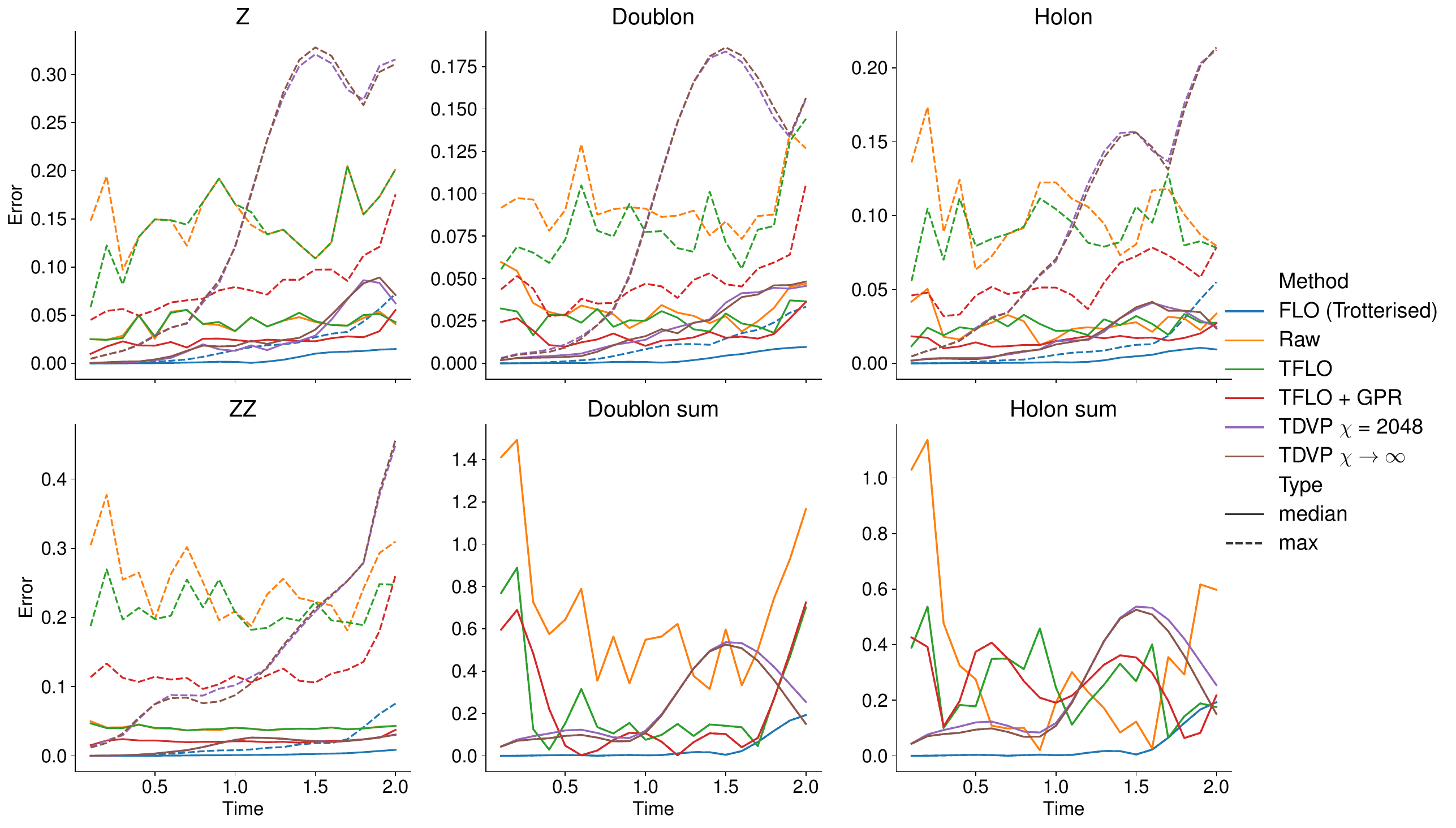}
        \caption{
        Median and maximum errors for all times for a range of types of observables. 
}
    \label{fig:errors_all_obs}
\end{figure}

We compare the performance of these methods with raw data in \cref{fig:errors_violin_doublon}, \cref{fig:doublon_dynamics_errors} and \cref{fig:errors_all_obs}. 
We see that usually the combination of TFLO and GPR significantly reduces the error, compared with the unmitigated results.
Further, we see that the error achieved by TDVP increases over time, as expected, and that this error cannot be explained by Trotter error alone. 
By contrast, the error achieved by our experimental results does not significantly increase with time, until the latest times (which we attribute to Trotter error). 
We expect that a significant portion of the error remaining in our results after error mitigation with TFLO can be explained by statistical noise, which is mitigated by GPR.
Next, we describe these techniques in detail.

\newpage
\subsection{Pauli pseudo-twirling}
\label{sec:pseudotwirling}

Pauli twirling~\cite{bennett96} is a well-known and effective method to suppress errors in quantum circuits~\cite{errormitigation}. 
The theoretical underpinning of this method is that in many realistic quantum hardware platforms, 2-qubit gates are the primary source of error, and the native 2-qubit gate is a Clifford gate (such as a CNOT or CZ gate). 
Each 2-qubit gate is sandwiched by a random choice of Pauli matrices picked from those for which, in principle, the ideal gate would be left unchanged. However, the approximate gate actually implemented by the hardware will not remain unchanged. Averaged over a number of instantiations, this tends to suppress systematic errors. 
Indeed, one can show that this process replaces coherent errors with incoherent errors, which facilitates certain error-mitigation procedures. 
Further, this procedure will have a dynamical decoupling effect and reduce other systematic errors occurring in the circuit.

Here, the native 2-qubit gate we execute is not a Clifford gate, but a $R_{zz}$ gate. Nevertheless, we can still carry out a more limited version of this procedure, known as Pauli pseudo-twirling \cite{pseudo-twirling}. Instead of wrapping the $R_{zz}$ gate between Pauli matrices that leave it unchanged, it is conjugated by random two-qubit Paulis and the sign of the rotation angle in $R_{zz}(t)=e^{-\frac i 2 ZZt}$ is either left unchanged or flipped, depending on whether the two-qubit Pauli commutes or anti-commutes with $ZZ$. Again, this procedure leaves the ideal circuit unchanged but helps shape the noise channel in the presence of noise. Similar procedures have been found to be effective on superconducting qubit~\cite{pseudo-twirling} and ion-trap~\cite{quantinuum} quantum hardware platforms.

To implement this in our experiment, we simply traversed each circuit, for each $R_{zz}(t)$ gate sampled a random two-qubit Pauli $P$, and replaced it with $P R_{zz}(t) P$ (or $P R_{zz}(-t) P$ if $ZZ$ and $P$ anti-commute). We sampled 16 such twirling instances for each logical circuit (corresponding to different values of $t$ and $U$). Note that the introduction of the additional two-qubit Paulis does not increase the gate count of the final compiled circuits, because runs of subsequent single-qubit gates get compiled into a single $U_3$ gate. 

\subsection{Symmetry averaging} \label{app:symmetry_averaging}
As explained in \cref{app:additional_results}, the initial state and Hamiltonian are invariant under the spin-reflection symmetry $R c_{i,\uparrow} R^\dagger = c_{i,\downarrow}$. This symmetry acts analogously on the $Z$ operators, i.e. $R Z_{i_1, \sigma_1} \cdots Z_{i_n, \sigma_n} R^\dagger = Z_{i_1, \overline\sigma_1} \cdots Z_{i_n, \overline\sigma_n}$. This allows us to increase the effective shot-count for the expectation values of all multi-qubit $Z$ strings and expectation values derived from them by averaging each multi-qubit $Z$ expectation value with its spin-reflected partner.

This is also the only obvious, simple symmetry of the initial state and Hamiltonian, because the irregular arrangement of triplets in the initial state breaks the translation and reflection symmetries of the lattice.

\subsection{Training with fermionic linear optics}
\label{app:subsec:tflo}
Training with fermionic linear optics (TFLO)~\cite{tflo} is a method which takes advantage of the well-known fact that quantum circuits that contain only fermionic linear optics (FLO) operations can be efficiently simulated classically~\cite{terhal02}. In some quantum algorithms, such as ours, many of the gates in the quantum circuits to be executed are FLO operations, so it is natural to expect that the behaviour of the ``true'' circuit in terms of errors should be similar to the quantum circuit that one would produce by removing all FLO operations from that circuit. Therefore, for any given observable, one can prepare a set of pairs of noisy and exact expectation values for that observable and use these to infer a map from noisy to exact expectation values. The hope is that applying this map to the corresponding quantum circuit, which is not FLO, will achieve similar results in terms of reducing the error in this observable.

Here, there is a particularly direct connection between the circuit that we wish to execute and an exactly simulable model. Every gate in the quantum circuit for simulating time-dynamics of the Fermi-Hubbard model for $U=0$ is FLO, and further, the structure of the circuit is exactly the same as the structure of the quantum circuit for $U\neq 0$, with the only difference being that for $U=0$, there is no time evolution by onsite terms.

This connection has previously been used in the context of VQE with the Hamiltonian variational ansatz to reduce errors in ground state simulation of the Fermi-Hubbard model~\cite{google-fhvqe}. In that work, training data was produced by choosing random parameters for the variational ansatz for many choices of parameters. Here, we do not have the shot budget to do this. So instead, we use the time-dynamics experimental results themselves as training data. We produce a set of samples for $U=0$ for the same set of time-points as for $U=4$, and aim to use this to infer a function for mapping noisy to exact observables. It is important that the quantum circuits for these two cases look as alike as possible, so that errors behave as similarly as possible. We therefore kept the same gate and fermionic swap network structure in the $U=0$ circuit as the $U=4$ case, by including the onsite terms with an evolution parameter of 0.

\begin{figure}
    \includegraphics{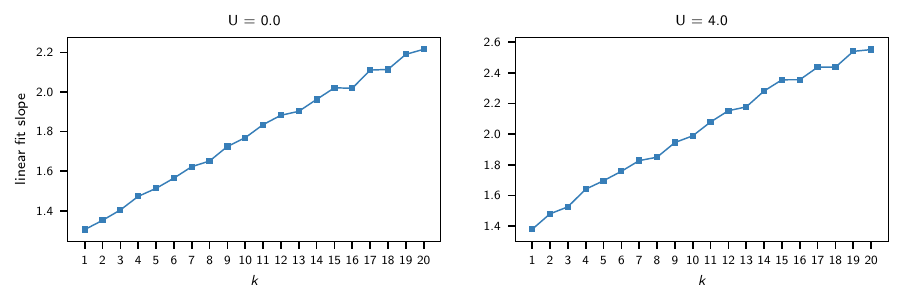} 
    \caption{Slope of TMPS fit for $S(k)$.}
    \label{fig:mad_tmps}
\end{figure}

\begin{figure*}[t]
    \centering
    \includegraphics{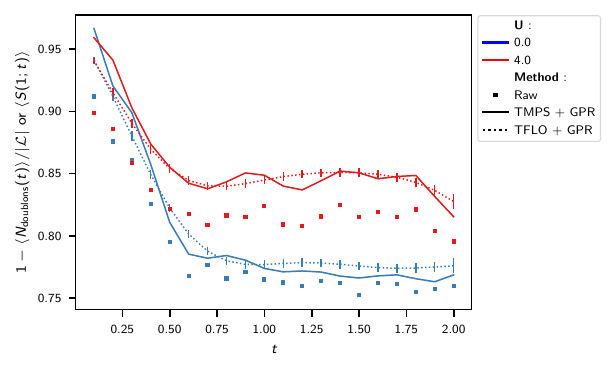}
    \caption{Comparison between $1-\langle N_{\mathrm{doublons}}(t)\rangle/|\mathcal{L}|$ and $\langle X(t)\rangle$. The agreement between the solid lines, obtained using TMPS, and the dotted lines, obtained using TFLO, shows that TMPS produces results compatible with TFLO.}
    \label{fig:subsets_vs_N_doublons}
\end{figure*}

An additional complication we face is that the full quantum circuit we execute for $U=0$ is not exactly simulable classically using standard FLO methods due to the initial spin triplet state preparation step. Thus, we refer to circuits of this form as ``NearFLO'' throughout.
Nevertheless, we can implement an efficient method for simulating NearFLO circuits classically in the case where one wishes to compute a low-weight observable (see \cref{sec:nearflo}).
To mitigate errors in high-weight observables, we use an alternative technique. Tensor network (TDVP) simulations are expected to achieve a high level of accuracy for short evolution times, and can be used to produce samples in the computational basis, and hence to estimate arbitrary observables. Therefore, we can use TDVP at short times to produce training data, on which the same fitting process as standard TFLO can be used. This is additionally justified by our observation that experimental errors for $U=0$ do not get significantly worse for long times (\cref{fig:errors_violin_doublon}). We called this technique training with MPS (TMPS).
For a given observable, we choose the time interval to be used to produce training data such that a good convergence in the bond dimension is observed.
We can use the slope of the fit to get a sense of the extent to which errors affect high-weight observables. We observe (see \cref{fig:mad_tmps}) that, as expected, the slope increases exponentially with the weight of the observable, but at a slow exponential rate. Moreover, \cref{fig:subsets_vs_N_doublons} shows that mitigating the same observable via TFLO and TMPS produces compatible results.

\begin{figure}
    \includegraphics{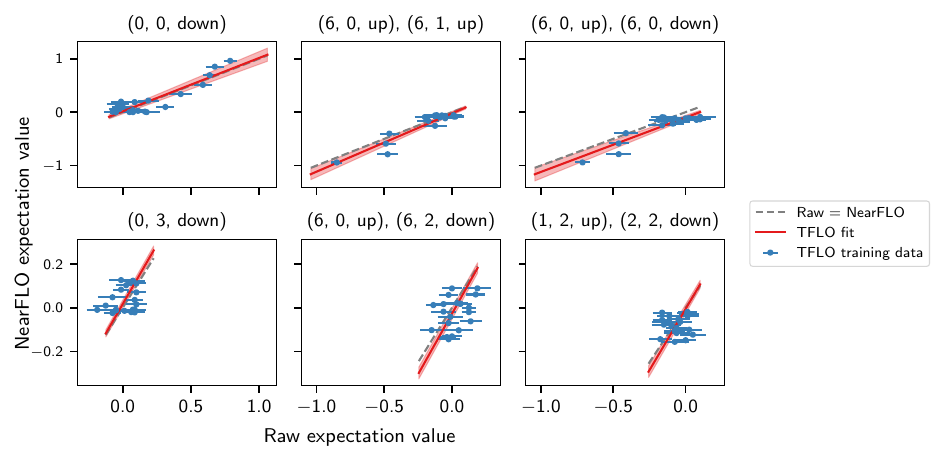} 
    \caption{A few examples of the linear fits performed for TFLO error mitigation. 
    The top row shows single- and two-qubit $Z$ expectation values between qubits that start out at $\pm 1$ and are thus easy to fit to, while the bottom row shows examples of single- and two-qubit $Z$ expectation values that start out at $0$ and thus require a regularization before performing a successful fit.}
    \label{fig:tflo}
\end{figure}

We find (see ~\cref{fig:tflo}) that a simple linear fit is often, though not always, a good model for the map between noisy and exact data. We use ordinary least-squares to perform this linear fit. We experimented with a more complex fitting function, which included dependence on the evolution time $t$, but we did not find that this significantly improved the quality of our results, as measured by the accuracy at $U=0$. We attribute this to the hardware error model not significantly depending on $t$. Given the risk of overfitting, we chose not to use this model.

Due to the triplets in the initial states, many observables vary little over time. Additionally, due to the relatively low shot count, the statistical uncertainty on all observables is relatively large (on average $\pm 0.08$). This means that for many observables, the signal-to-noise ratio (SNR) is too small to perform a well-controlled least-squares fit. To still be able to perform TFLO on these observables, we employed a two-stage fitting procedure. In the first stage, we identified all observables with a sufficiently large SNR and performed an ordinary least-squares fit to find the best TFLO rescaling parameters. In the second stage, we grouped all observables by the number of qubits they act on and whether they act on multiple Fermi-Hubbard sites. For each group, we computed a prior distribution over the optimal fit parameters. This prior distribution was then used (per group) to perform a linear least squares fit with regularisation to obtain well-behaved fit parameters even when the SNR of the time series was low.

\subsection{Postselection and using other occupation number sectors}

\begin{figure}
  \begin{center}
    \includegraphics{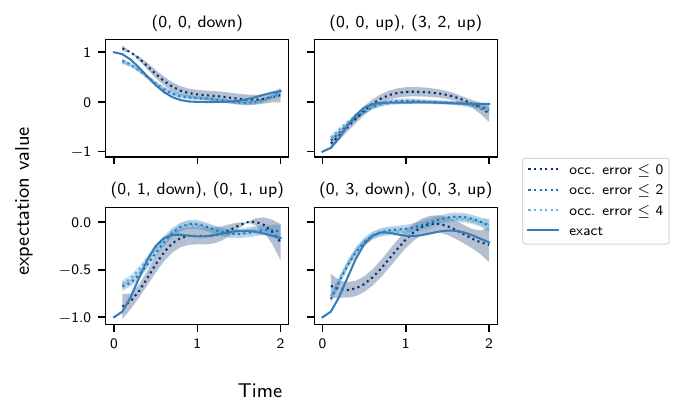}
  \end{center}
  \caption{The effect of post-selection on expectation values for a few examples of single and two-qubit $Z$ observables. This is done using the $U=0$ time series for exact comparison data and with GPR smoothing to decrease statistical noise. We observe that the most aggressive postselection on only the correct occupation numbers is the noisiest because of the small number of samples retained. When retaining samples with occupation error $\geq 2$, the time series don't change due to the fact that the vast majority of samples have 1 or 2 occupation errors.}
  \label{fig:post-selection-noise}
\end{figure}

\begin{figure}
  \centering
    \includegraphics{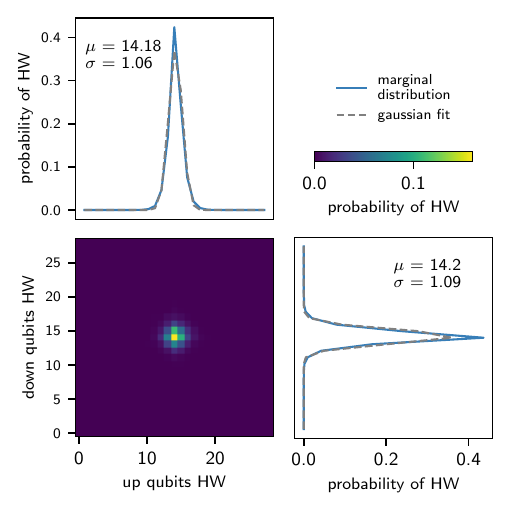} 
    \caption{The distribution of Hamming weights (HW) on the up and down qubits is well described by a Gaussian centered at $(14, 14)$, i.e., the correct Hamming weight in the absence of hardware errors. 19\% of shots have exactly Hamming weight $(14, 14)$.}
    \label{fig:hammingweights}
\end{figure}

Every gate in the quantum circuits that we execute preserves fermionic occupation number within each spin sector, which implies that we can detect errors which alter these quantities and, if we choose, postselect to only retain samples for which these quantities are correct~\cite{google-fhvqe}. However, in our experiments, the number of shots taken is heavily constrained, so it is not obvious whether postselection will be advantageous, because restricting the samples to a smaller subset tends to increase statistical error. Indeed, for $U=0$ we observe that postselection on occupation number makes errors worse due to the low number of shots retained and increases statistical noise (see \cref{fig:post-selection-noise}). On the other hand, we see that the Hamming weight in each spin sector tends to vary very little from the expected values (see \cref{fig:hammingweights}). This provides a simple metric for the quality of our results, and suggests that local observables should tend to vary only slightly from their expected values. For example, local densities ($Z$ operators) would be expected to vary by at most $\sim 1/56 \approx 0.018$ compared with their true values.

\subsection{Gaussian process regression}

As a final stage in our postprocessing of experimental data, we implement Gaussian process regression (GPR) smoothing for all time series. Briefly, a Gaussian process (GP)~\cite{Rasmussen_2004} is a probability distribution over functions. It is formulated as a marginal distribution over any finite set of function values such that the joint probability distribution for this finite set of function values is a multivariate Gaussian whose mean and covariance matrix only depend on the evaluation points and some hyperparameters encoded in what is called the kernel of a GP. This kernel can encode various things about the sampled functions, such as ($n$-times) differentiability, periodicity, or the correlation length $l$ on which it varies. A good overview of what can be encoded into the kernel and how is presented in~\cite[Chap. 4]{Rasmussen_2004}.

To use this probability distribution over functions for smoothing and fitting, a simple Bayesian procedure can be used. The kernel of the GP defines a prior distribution over functions, the noisy observations of the time series play the role of new evidence, and the application of Bayes' theorem then yields a posterior distribution over functions. The mean of this posterior distribution is then the ($n$-times) differentiable / periodic / function with correlation length $l$ that is most likely given the observed data. And the diagonal of the covariance matrix of the posterior distribution corresponds to the one-sigma confidence interval of the posterior time series. This Bayesian approach can also be used to add further prior information like the function value or derivative at special known points to the fitting procedure. Concretely, we know that the $Z$ expectation values of qubits that start out maximally polarised must be $\pm 1$ at $t = 0$ and that its derivative must be $0$ at $t = 0$. We included this prior information into the GPR fitting procedures whenever applicable. 

We used the Python implementation of GPR regression in \texttt{scikit-learn} \cite{scikit-learn} with a squared-exponential kernel with length scale $l = 0.4$ and variance $\sigma = 1.5$. These hyperparameters were determined by maximizing the marginal likelihood of $l$ and $\sigma$ given the observed time series for a variety of variables. We then found that the optimal hyperparameters were narrowly clustered around $l = 0.4$ and $\sigma = 1.5$ and thus fixed these values throughout all later analyses. Fixing the hyperparameters to the same values for all observables instead of determining them per observable also helped to avoid outliers due to sampling noise.

The error propagation procedure reported in \cref{sec:error-bars} gives error bars on the input data to the GPR smoothing procedure. But, by adding a Gaussian white noise kernel to the overall kernel, the average noise strength also becomes a parameter that is learnable through the hyperparameter optimisation procedure described in the previous paragraph. Fortunately, the error bars obtained through error propagation broadly agreed with those learned through hyperparameter optimisation in the GPR process. This gives further confidence that the approximations necessary for the error propagation are valid.

\subsection{Error bars}
\label{sec:error-bars}
The GPR error bars shown in the figures in the main text were
all computed using Gaussian error propagation and error bars of the initial raw values
computed from the shots were computed using
\begin{equation}
  \sigma_{\text{raw}}^2 = 
  \frac{1}{\ntwirling(\ntwirling - 1)} \sum_{i=1}^{\ntwirling} (\hat \mu_i - \mu)^2
  + \frac{1}{\ntwirling^2 \nshots} \sum_{i=1}^{\ntwirling} \sigma_i^2
  \label{eq:twirling-error-bar}
\end{equation}
where $\ntwirling$ is the number of twirling instances, $\nshots$ the number of shots
taken per twirling instance, $\mu_i$ and $\sigma_i^2$ are the per twirling instance
sample mean and sample variance and $\mu = \frac{1}{\ntwirling} \sum_{i=1}^{\ntwirling} \mu_i$ is the total
mean.

\subsubsection{Pseudo-twirling and error bars}
\label{sec:pseudo-twirling-and-error-bars}

When computing error bars in conjunction with (pseudo)-twirling, some care must be taken to account for the iterated expectations correctly. One might be tempted
to simply collect all $\ntwirling \times \nshots$ shots into one set, compute the expectation
of the observable $O$ as
\begin{equation}
  \hat\mu_O 
  = \frac{1}{\ntwirling \times \nshots} \sum_{i, j=1}^{\ntwirling, \nshots} O(x_{i,j})
\end{equation}
and the variance of $\hat \mu_O$ in the usual way as
\begin{equation}
  \hat \sigma^2_{\hat \mu_O} 
  = \frac{1}{\ntwirling \times \nshots (\ntwirling \times \nshots - 1)}
  \sum_{i, j=1}^{\ntwirling, \nshots} (O(x_{i,j}) - \hat \mu_O)^2.
  \label{eq:naive_twirling_error}
\end{equation}
But unless $\nshots = 1$ or all twirling instances are actually the same, this will
underestimate the error. An easy example to see this is to assume that we have very few twirling instances $\ntwirling$ from which we take many shots $\nshots$, and that the different twirling instances produce very different distributions.
Then the naive \cref{eq:naive_twirling_error} will be very small, even though we have only seen the data from very few twirling instances and they all give very different estimates for $\mu_O$.

To get an accurate estimate $\hat \sigma^2_{\hat \mu_O}$ for the variance of $\hat \mu_O$, we have to correctly account for the contribution to the variance coming from
sampling different twirling instances---the first term in \cref{eq:twirling-error-bar}---and
the contribution coming from shot noise for each circuit---the second term in \cref{eq:twirling-error-bar}. This can be done by employing the law of total variance. Carefully
going through the derivations then gives \cref{eq:twirling-error-bar}.

\subsubsection{Verification via bootstrapping}
For error propagation, we employed the \texttt{uncertainties} Python package \cite{uncertainties}, which implements automatic Gaussian error propagation for many basic mathematical functions. However, for Gaussian error propagation to be valid, we have to assume that (i) the input distributions are all Gaussian; (ii) all data transformation functions are (or at least well-approximated by) affine linear functions.
In addition to these constraints, the \texttt{uncertainties} package only tracks and propagates the individual uncertainty of each variable, but not correlations.
The most straightforward alternative to Gaussian error propagation is bootstrapping \cite[Chap. 8]{Wasserman2007}. Very briefly, bootstrapping is a procedure for estimating the distribution (or properties of the distribution) of an estimator (here, our mitigated and processed observable values) by replacing the full distribution of the in-going random variable (that we don't know) with the empirical distribution that we obtained from the samples we actually have.

\begin{figure}
  \begin{center}
    \includegraphics{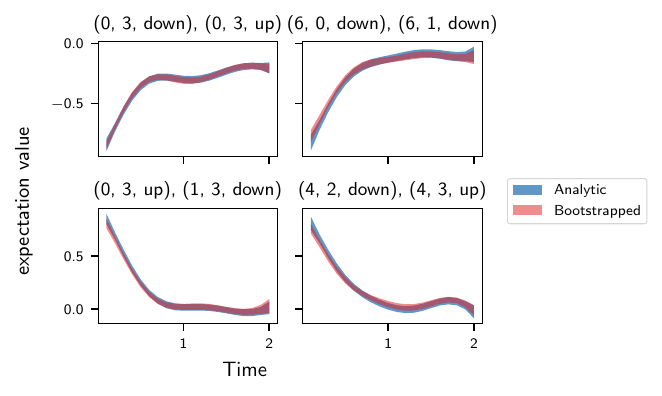}
  \end{center}
  \caption{Comparison of error bars computed via bootstrapping vs Gaussian error propagation for a random choice of 4 observables.}
  \label{fig:bootstrapping-vs-error-propagation}
\end{figure}

Similar to the discussion in \cref{sec:pseudo-twirling-and-error-bars}, one has to be careful when implementing the bootstrap method in conjunction with pseudo-twirling. The correct scheme is to first sample $\ntwirling$ circuits with replacement from the pool of twirling instances and then sample $\nshots$ shots with replacement from each of the sampled circuits. This dataset can then be piped through the whole error mitigation and data analysis pipeline. Repeating this procedure sufficiently many times allows us to estimate the standard deviations of the final mitigated observables. In \cref{fig:bootstrapping-vs-error-propagation} we show that the error bars obtained via bootstrapping agreed remarkably well with those obtained via the computationally much cheaper Gaussian error propagation. We thus decided to use Gaussian error propagation for all error bars reported here.

\subsection{Cross validation via particle filters}\label{subsec:ParticleFilter}

\begin{figure}
  \begin{center}
    \includegraphics{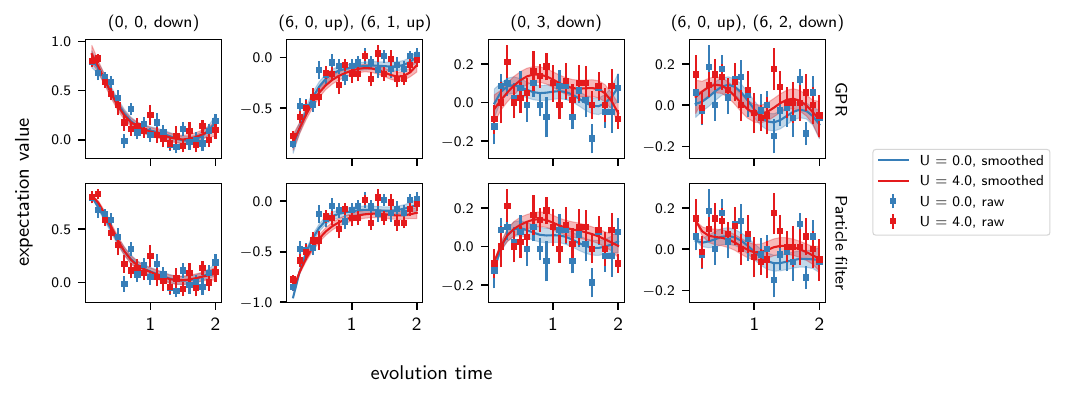}
  \end{center}
  \caption{Comparison of smoothing with Gaussian process regression (top) and particle filters (bottom) for a few single and two-qubit $Z$ timeseries. We find that the smoothed time series are very similar, even though the assumptions behind GPR (Gaussianity of input error bars) are not met.}
  \label{fig:gpr-vs-pf}
\end{figure}

In addition to GPR smoothing, we also experimented with a particle filter (PF)-based smoothing method to postprocess all time series. While GPR assumes Gaussian noise in the observations, the measurement noise in our setting is not strictly Gaussian. To account for this, we employ particle filtering as an additional validation method, which can accommodate non-Gaussian noise models. A particle filter~\cite{Gordon1993} is a Bayesian sequential Monte Carlo method that represents probability distributions over hidden states with a large ensemble of weighted samples, called particles. Each particle evolves forward in time under a stochastic state–space model and is reweighted by the likelihood of the observed data. Through repeated propagation and resampling, the PF generates full latent trajectories consistent with all observations, providing a smoothed estimate of the underlying latent signal. As shown in \cref{fig:gpr-vs-pf}, the smoothed time series obtained using PF are in close agreement with those obtained using GPR, despite the difference in the underlying noise assumptions. For a more in-depth review on particle filter with backward smoothing, refer to Sections~4 and~5 of Ref.~\cite{DoucetJohansen2011}.

We implemented the PF from scratch in Python, using two passes with ancestry-based backward smoothing. At each time step, the hidden state consists of the signal value and its rate of change (velocity), which evolve under a stochastic update with Gaussian noise and exponential velocity decay. Each particle is then reweighted according to the likelihood of the observed data at that time step, and then the resampling step systematically prunes unlikely particles and replicates likely ones, maintaining a representative ensemble. After this forward run, we perform backward smoothing using the particle ancestry. Each resampled particle at a given time step stores the index of its parent from the previous time step, which defines an ancestry tree. 
By tracing these links backward from a randomly chosen particle at the final time step, we generate full state trajectories consistent with all observations. Averaging over many such trajectories yields the smoothed time series, while the standard deviation across them quantifies the uncertainty.

Similar to GPR, we fixed the particle filter hyperparameters across all observables to maintain consistency and avoid overfitting to sampling noise in individual series. 
We used 6000 particles, while the remaining parameters (e.g., noise scales, initial uncertainties, velocity decay) were chosen based on test signals with known values. 
At the same time, they were kept sufficiently general to provide reliable smoothing across all observables.

\section{Classical simulation of non-Gaussian circuits}
\label{sec:nearflo}
The initial state considered in this work is a triplet covering alongside two lonely sites (doublon and holon). We therefore refer to it as a non-Gaussian or magic state~\cite{hebenstreit2019all, liu2022many,howard2017application,cudby2023gaussian,reardon2024improved,dias2024classical,hakkaku2022quantifying} because it cannot be generated by a FLO unitary (more details in \cref{app:exp_speedup_nearflo}), and thus in general has to be treated as a linear combination of the constituent Fock states, the number of which grow exponentially with the number of triplets present. 

Recall that fermionic Gaussian states are those of non-interacting fermions and hence correspond to the $U=0$ regime, which we will work in for the remainder of this section. 

\subsection{Non-Gaussian observables}
\label{app:exp_speedup_nearflo}
Recall that the initial state consists of lonely sites, which are either holes or doublons, and physical triplets across pairs of sites $(a,b)$. The triplet creation operator $\frac{c^{\dagger}_{a,\uparrow} c^{\dagger}_{b,\downarrow} + c^{\dagger}_{a,\downarrow} c^{\dagger}_{b,\uparrow}}{\sqrt{2}}$ is not a fermionic Gaussian state and cannot be generate by a FLO unitary~\cite{terhal02}. Thus, the simulation of observables in theory will scale exponentially with the number of triplets. However, we note that the initial state can be written as a product state across $N_{\text{triplets}}$ triplets and $N_{\text{lonely}}$ lonely sites according to \cref{eqn:triplet_init_state}
\begin{equation}
    \ket{\psi}_{\text{init}} = \bigotimes_{k=1}^{N_{\text{triplets}}} \ket{\text{triplet}}_{s_k} \bigotimes_{l=1}^{N_{\text{lonely}}} \ket{\text{lonely}}_{s_l} 
    \label{eqn:triplet_init_state}
\end{equation}
where $s_k$ denote the pair of sites participating in the $k^{th}$ triplet and $s_l$ is a single lonely site index. The observables required for TFLO are the single and two-point Pauli $Z$ correlators, $\langle Z_i \rangle$ and $\langle Z_i Z_j \rangle$ $\forall i,j$ fermionic mode indices (i.e,. site and spin indices are conglomerated). Here we work instead with the number operators $\langle \hat{n}_i \rangle$ and $\langle \hat{n}_i \hat{n}_j \rangle$  but there is a trivial conversion between $\hat{n} \leftrightarrow  Z$ via $\hat{n} = \frac{1-Z}{2}$. This set of densities and correlators is enough to perform TFLO corrections for all observables considered in the main text except for the Wilson lines. As we will discuss later in this section, arbitrary weight observables in theory would scale exponentially with $N_{\text{triplets}}$, leading to an algorithm which is cumbersome and hence not implemented for the Wilson loops. 

\subsubsection{Densities}
\label{app:density_near_flo}
The observables $\langle \hat{n}_i \rangle, \forall i$ where $i$ denotes a fermionic mode can be computed as follows
\begin{align}
    \bra{\psi_{\text{init}}(t)} c^{\dagger}_i c_i \ket{\psi_{\text{init}}(t)} &= \bra{\psi_{\text{init}}}e^{iHt} c^{\dagger}_i c_i e^{-iHt} \ket{\psi_{\text{init}}} \\
    &= \sum_{\alpha, \beta} R_{\alpha,i} R^*_{\beta,i} \bra{\psi_{\text{init}}} c^{\dagger}_{\alpha} c_{\beta} \ket{\psi_{\text{init}}} \\
    &= \sum_{\alpha} |R_{\alpha,i}|^2
\end{align}
where $H$ is the single band Hamiltonian defined in \cref{eqn:single_band_ham} and $R = e^{-iht}$ where $h$ is the adjacency matrix of the lattice $\mathcal{L}$. Due to the structure of the initial state, the coefficients $\alpha$ and $\beta$ must be equal for a non-vanishing value of $\bra{\psi_{\text{init}}} c^{\dagger}_{\alpha} c_{\beta} \ket{\psi_{\text{init}}}$, since flipping two bits within or across sites result in an orthogonal state. 

\subsubsection{Density-Density correlators}
The observable of interest is 
\begin{align}
    \bra{\psi_{\text{init}}(t)} c^{\dagger}_i c_i c^{\dagger}_j c_j \ket{\psi_{\text{init}}(t)} &= \bra{\psi_{\text{init}}}e^{iHt} c^{\dagger}_i c_i c^{\dagger}_j c_j e^{-iHt} \ket{\psi_{\text{init}}} \\
    &= \sum_{\alpha, \beta, \gamma, \eta} R_{\alpha,i} R^*_{\beta,i} R_{\gamma,j} R^*_{\eta, j} \bra{\psi_{\text{init}}} c^{\dagger}_{\alpha} c_{\beta} c^{\dagger}_{\gamma} c_{\eta} \ket{\psi_{\text{init}}}
\end{align}
Note that since $\psi_{\text{init}}$ is not a fermionic Gaussian state, one cannot simply compute this with a determinant/pfaffian. Hence, we need to consider the exponential $(2L)^P$ mode combinations in the sum. Here, $P=4$ is the fermionic operator weight. However, for the special case of $P=4$, the computational complexity can be reduced to scale as $O(L^2)$ due to the structure of the initial state.

The term $c^{\dagger}_{\alpha} c_{\beta} c^{\dagger}_{\gamma} c_{\eta}$ flips four qubits, turning $\ket{\psi_{\text{init}}} \rightarrow \ket{\psi'_{\text{init}}}$. The structure of a triplet state over sites $a$ and $b$ is given by
\begin{equation}
    \frac{1}{\sqrt{2}} (\ket{1_{i,\uparrow} 0_{i,\downarrow} 0_{j,\uparrow} 1_{j,\downarrow}} -\ket{0_{i,\uparrow} 1_{i,\downarrow} 1_{j,\uparrow} 0_{j,\downarrow}})
\end{equation}
and the only way to have a non-vanishing overlap with itself is to flip all four qubits. For a lonely site, the only non-vanishing overlap with itself is to flip nothing. Thus, out of all the possible combinations of $1,2,3$ and $4$ distinct/unique values of $\alpha, \beta, \gamma$ and $\eta$, for a non-trivial overlap one only needs to consider the cases where the indices land on a single or a pair of physical sites. 

\paragraph{Four distinct indices}
All non-trivial combinations of distinct $\alpha, \beta, \gamma$ and $\eta$ must be within a single triplet. 
The contribution to the total correlator becomes 
\begin{align}
    \langle n_i n_j \rangle_{\text{4 distinct}} (t) &= \sum_{(i, j)}^{N_{\text{triplets}}} \Bigg( \sum_{\alpha,\beta,\gamma,
    \eta \in \text{perm}[i\uparrow, i\downarrow, j\uparrow, j\downarrow]} R_{\alpha,i} R^*_{\beta,i} R_{\gamma,j} R^*_{\eta, j} \nonumber \\
    &\times \frac{1}{2} \bra{\Omega} (c^{\dagger}_{i,\uparrow} c^{\dagger}_{j, \downarrow} + c^{\dagger}_{i,\downarrow}c^{\dagger}_{j,\uparrow})^{\dagger} c^{\dagger}_{\alpha} c_{\beta} c^{\dagger}_{\gamma} c_{\eta} (c^{\dagger}_{i,\uparrow} c^{\dagger}_{j, \downarrow} + c^{\dagger}_{i,\downarrow}c^{\dagger}_{j,\uparrow}) \ket{\Omega}\Bigg)
\end{align}
where $\ket{\Omega}$ is the vacuum Fock state. Perm denotes the permutations of the list and $(i,j)$ denote the pairs of \textbf{site} indices forming the triplet covering. 
    
\paragraph{Two distinct indices} 
There are two unique cases here, $\alpha=\beta, \gamma=\eta$ or $\alpha=\eta, \beta=\gamma$, as only number operators are non-trivial. For both conditions, the two distinct mode indices can be 
\begin{align}
   & 1. \text{ Within the same triplet.} \nonumber \\
   & 2. \text{ Across two triplets.}\nonumber \\
   & 3. \text{ Within a lonely site.} \nonumber \\
   & 4. \text{ Across two lonely sites.}\nonumber \\
   & 5. \text{ Between a triplet and a lonely site.} \nonumber \\
\label{eqn:conditions_two_distinct}
\end{align}

The total correlator becomes
\begin{align}
    \langle n_i n_j \rangle_{\text{2 distinct}} (t) &= \sum_{(i,j)} \Bigg( \sum_{\alpha, \gamma \in \text{perm}[i\uparrow, i\downarrow, j\uparrow, j\downarrow]} |R_{\alpha,i}|^2 |R_{\gamma,j}|^2 \bra{\Omega} \mathcal{C}^{\dagger}_{i,j} n_{\alpha} n_{\gamma} \mathcal{C}_{i,j}\ket{\Omega}  \nonumber \\
    &+\sum_{\alpha, \beta \in \text{perm}[i\uparrow, i\downarrow, j\uparrow, j\downarrow]} R_{\alpha,i} R_{\beta,i}^* R_{\beta,j} R^*_{\alpha,j} \bra{\Omega} \mathcal{C}^{\dagger}_{i,j} n_{\alpha} (1-n_{\beta}) \mathcal{C}_{i,j}\ket{\Omega}  \Bigg)
\end{align}
where $(i,j)$ denote the pairs of \textbf{site} indices across all the conditions outlined in \cref{eqn:conditions_two_distinct}. $\mathcal{C}$ is the creation operator of the initial state type across sites $i$ and $j$.

\paragraph{One distinct index}
This is similar to the case of the number operator discussed in \cref{app:density_near_flo}, i.e. 
\begin{equation}
    \langle n_i n_j \rangle_{\text{1 distinct}}(t) = \sum_{\alpha} \sum_{i}^{2L} |R_{\alpha,i}|^2 |R_{\alpha,j}|^2\bra{\psi_{\text{init}}} n_i \ket{\psi_{\text{init}}}
\end{equation}
where $i$ now runs over all \textbf{modes}. 

The case of three distinct indices is trivially zero because in $c^{\dagger}_{\alpha} c_{\beta} c^{\dagger}_{\gamma} c_{\eta}$, there is either a double creation or annihilation acting on the same mode. Finally, 
\begin{equation}
     \langle n_i n_j \rangle (t) =  \langle n_i n_j \rangle_{\text{1 distinct}}(t) +  \langle n_i n_j \rangle_{\text{2 distinct}}(t) +  \langle n_i n_j \rangle_{\text{4 distinct}}(t)
\end{equation}
which scales as $O(L^2)$ from the combination of two sites. 

\subsubsection{Alternative observables not implemented: arbitrary weight Pauli Z observables}
\label{app:arb_weight_Z_nearflo}
The speedups discussed in \cref{app:exp_speedup_nearflo} allow for fast computations of $\langle Z\rangle(t)$ and $\langle ZZ\rangle(t)$ correlators that do not scale exponentially with system size. In general, it will be helpful to compute arbitrary weight $P$ observables of the form
\begin{equation}
    \langle \prod_k Z_k \rangle (t) = \bra{\psi_{\text{init}}} e^{iHt} \left(\prod_k^P Z_k\right) e^{-iHt} \ket{\psi_{\text{init}}},
\end{equation}
should they be needed for more error mitigation, such as with TFLO. Here, we leverage the fact that the Pauli $Z_k$ operator on mode $k$ is a FLO unitary, i.e.
\begin{equation}
    Z_k = -ie^{i\frac{\pi}{2}(1-2c^{\dagger}_k c_k)},
\end{equation}
and expand the initial state into its Fock state decomposition,
\begin{equation}
    \ket{\psi_{\text{init}}} = \sum_m \alpha_m \left(\prod_{i=1}^{W} c^{\dagger}_{m_i}\right) \ket{\Omega}
\end{equation}
where $m_i$ runs over the modes making up the Fock state $f$ with complex coefficient $\alpha_m$ and $\ket{\Omega}$ is the vacuum state. $W$ is the hamming weight of all the Fock states making up $\ket{\psi_{\text{init}}}$
\begin{align}
     \langle \prod_k Z_k \rangle (t)&= \sum_{m,m'} \alpha^*_{m'} \alpha_m \bra{\Omega} \Big(\prod_{i=1}^{W} c^{\dagger}_{m'_i} \Big)^{\dagger} e^{iHt} \Big(\prod_k^P Z_k \Big) e^{-iHt} \Big(\prod_{i=1}^{W} c^{\dagger}_{m_i} \Big) \ket{\Omega} \\
     &= \sum_{m,m'} \alpha^*_{m'} \alpha_m  \Bigg( \sum_{q_1, \dots, q_W} ( \prod_{i=W}^1 \mathcal{R}^{*}_{q_i, m'_i}) \Bigg) \bra{\Omega} c_{q_W} \dots c_{q_1} c^{\dagger}_{m_1} \dots c^{\dagger}_{m_W} \ket{\Omega} \\
     &= \sum_{m,m'} \alpha^*_{m'} \alpha_m  \Bigg( \sum_{q_1, \dots, q_W} ( \prod_{i=W}^1 \mathcal{R}^{*}_{q_i, m'_i}) \Bigg) \text{sgn} \Big(\sigma_{q1, \dots, q_W}^{m_1, \dots, m_W} \Big) \\
     &= \sum_{m,m'} \alpha^*_{m'} \alpha_m \text{det} 
\begin{pmatrix} 
    \textcolor{gray}{(m_1)} & \mathcal{R}^*_{m_1,m'_1} & \mathcal{R}^*_{m_1,m'_2} & \dots & \mathcal{R}^*_{m_1, m'_W} \\
   \textcolor{gray}{\vdots} & \dots & \dots & \vdots \\
   \textcolor{gray}{(m_W)} & \mathcal{R}^*_{m_W,m'_1} & \mathcal{R}^*_{m_W,m'_2} & \dots & \mathcal{R}^*_{m_W, m'_W}
\end{pmatrix}
\label{eqn:arbitrary_Z_near_FLO}
\end{align}
where $\mathcal{R} = e^{iht} \Big(\prod_k^P -i e^{i\frac{\pi}{2} \text{diag}(-1)^k} \Big) e^{-iht}$ is the $2L \times 2L$ matrix from the combination of the FLO time dynamics unitaries and Pauli $Z$ tensor products. The values $m_1$ and $m_W$ marked in grey denote the row indices; they are relative to each other and do not participate in the determinant. For example, if $m_2-m_1>1$ but $m_1+1$ does not exist as an index, then the $m_2$ row is directly below the $m_1$ row. Therefore, the absolute values of $m_1, \dots, m_W$ are not the actual row indices but give their relative position to each other so as not to wrongfully pick up an additional $-1$. This factor will not be a problem if the initial state was FLO, whereby the superposition is no longer necessary.  Note that if the fixed values of $m_1, \dots, m_W$ are monotonically increasing, this becomes the conventional definition of a determinant. $\text{sgn}(\sigma_{A}^B)$ is equal to $(-1)^{P_{AB}}$, where $P_{AB}$ is the number of swaps to permute the set of indices $A$ to match $B$. 

For the triplet covering considered in the main text, $|\alpha_m| = (\frac{1}{\sqrt{2}})^{N_{\text{triplets}}}$. Thus, as equation \cref{eqn:arbitrary_Z_near_FLO} suggests, the cost of simulating any weight $P$ Pauli $Z$ observable is bottlenecked by the summation over $(4^{N_{\text{triplets}}})$ combinations of Fock states from the initial state, and not $P$. The total cost of performing TFLO corrections for Wilson loops such as that in~\cref{fig:all_wilson_loops_fixed_perimeter} will practically scale as $O(4^{N_{\text{triplets}}}) \times N_T \times N_{\text{loops}}$ for the $N_T=20$ time steps taking in the experiment and there are $N_{\text{loops}} = 4182$ Wilson loops considered with Area = $12$. Not only is this numerically demanding, but TFLO is not guaranteed to be successful since each Wilson loop's time dynamics signal may vary too little for a good linear map fitting. As a result, we omit attempting TFLO on the Wilson observables discussed in the main text.

\subsection{Sampling from the triplet Magic state}
The ability to sample bitstrings from the triplet state can also be beneficial. For example, one can compare the sampled distribution against hardware to understand noise. Alternatively, it can be used to compute observables such as those discussed in \cref{app:arb_weight_Z_nearflo} in the non-Gaussian context, to be used as TFLO training data. 

Several existing algorithms would be applicable to sample from the magic state. To simulate measurements from a pure FLO system, one can employ the qubit-by-qubit sampling scheme proposed by Terhal and DiVincenzo~\cite{terhal02} with a naive time complexity upper bound of $O(N^4)$ for an $N$-mode measurement. This was subsequently improved upon by Bravyi and König~\cite{bravyi2011classical,bravyi2012disorder} to an upper bound of $O(N^3)$. Since the triplet covering initial state is a linear combination of FLO states but not FLO itself, naively applying the qubit-by-qubit sampling algorithm requires computations of overlaps of the form
\begin{equation}
    \bra{\psi_{\text{FLO}}(t)} \hat{O} \ket{\phi_{\text{FLO}}(t)} 
\end{equation}
where keeping track of phases is important, as $\psi_{\text{FLO}}(t)$ and $\phi_{\text{FLO}}(t)$ are different FLO states, and $O$ is some FLO operator/linear combinations of FLO operators. This can be done via the method developed by Bravyi and Gosset~\cite{bravyi2017complexity}, which generalizes Wick's theorem to compute matrix elements of exactly this type of Gaussian state overlaps. Although the theory exists, sampling from the magic state was not implemented again due to the exponential cost arising from the number of triplets in the initial state. Alternatively, a gate-by-gate sampling method proposed by Bravyi \textit{et al.}~\cite{bravyi2022simulate} or the sampling scheme proposed by Reardon-Smith \textit{et al.} ~\cite{reardon2024improved} are also viable but again hindered by the exponential cost of the triplet initial state. For these reasons, we omitted the implementation and analysis of magic state sampling.

\section{Majorana Propagation Simulations}
\label{app:sec:majorana_propagation_simulations}

\begin{figure}[h!]
    \centering
    \includegraphics{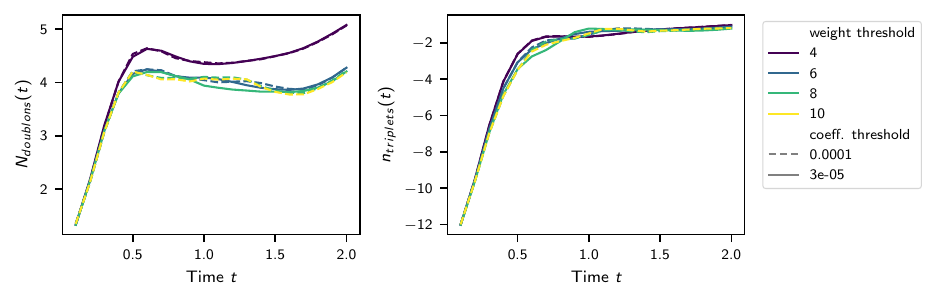}
    \caption{
      \mpsims time series for different lengths and coefficient truncation thresholds at $U=4$.
      The left panel shows the number of doublons shown in \cref{fig:densities_and_spin_correlation}
      of the main text, while the right panel shows the connected spin correlation function
      also shown in \cref{fig:densities_and_spin_correlation} of the main text.
    }
    \label{app:fig:mp_sim_convergence}
\end{figure}

\begin{figure}[h!]
    \centering
    \includegraphics{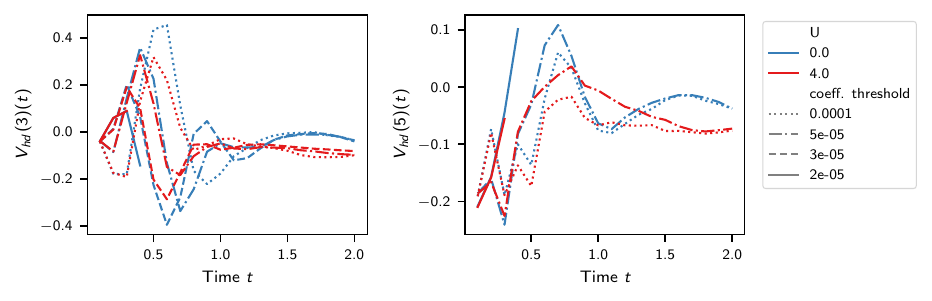}
    \caption{
      \mpsims time series of the open Wilson lines \cref{eq:open_wilson_lines} of length 3 and 5.
      The $U=0$ simulation was run at weight threshold 12 (the maximum weight of that observable) and the
      $U=4$ simulation at weight threshold 16. Even when lowering the coefficient
      threshold to $2\times 10^{-5}$ the results quickly diverge from the physically
      more sensible results reported in \cref{fig:wilson_time_averaged}c.
    }
    \label{app:fig:mp_sim_wilson_convergence}
\end{figure}

\begin{figure}[h!]
    \centering
    \begin{minipage}[t]{0.49\linewidth}
        \centering
        \includegraphics[width=\linewidth]{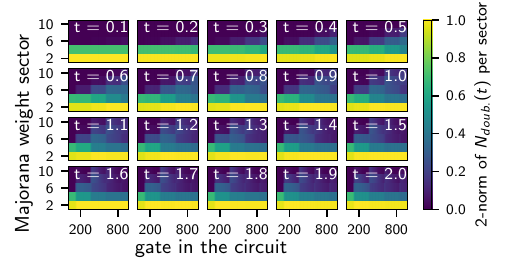}
        \\[-0.5em]
        {\small (a)}
    \end{minipage}
    \hfill
    \begin{minipage}[t]{0.49\linewidth}
        \centering
        \includegraphics[width=\linewidth]{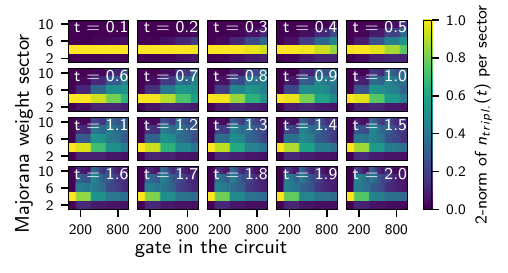}
        \\[-0.5em]
        {\small (b)}
    \end{minipage}
    \caption{
      Spreading of the observables over the Majorana weight sectors as they Heisenberg evolve through the circuit for number of doublons (left) and connected spin correlation function (right) at all simulated times for $U=4$. Note that for the MP simulations we wrote the circuit entirely in the fermionic language, i.e.\ as a sequence of hopping and onsite evolutions instead of a hardware native sequence of single qubit and $R_{zz}$ gates. This means the gate count for MP simulations is much lower than for the hardware circuits, even though they describe equivalent unitaries.
    }
    \label{app:fig:mp_sim_evolution}

\end{figure}

Besides the tensor network simulations discussed in \cref{app:sec:tensor_network_simulations}, we also implemented the Majorana Propagation (MP) algorithm first proposed in \cite{majorana_propagation}. It computes expectation values of observables by writing them as polynomials of Majorana operators, Heisenberg evolving these Majorana polynomials through the circuit and then computing the overlap of the Heisenberg evolved observable with the initial state. To keep the simulation feasible, the Majorana monomial is regularly truncated throughout the simulation by discarding terms whose coefficient is below a given coefficient threshold or whose Majorana weight is above a given weight threshold. For the latter truncation, it is important to note that the only gates in our circuits that increase the Majorana weight are evolution with the onsite terms, but that the majority of gates (FSWAP and hopping) leave the Majorana weight of each term unchanged. Furthermore, many observables of physical interest (in particular the triplet density \cref{eq:triplet_density} and the number of doublons \cref{eq:doublon_sum}) have only low Majorana weight. This makes the Majorana Propagation algorithm a natural candidate to simulate the time series shown in \cref{fig:densities_and_spin_correlation}.
Our implementation is largely inspired by the implementation of the related Pauli Propagation algorithm \cite{PauliPropagation1,PauliPropagation2} in Ref.~\cite{pauli_propagation_jl} and is built on top of the \texttt{Yao.jl} quantum simulation framework~\cite{YaoFramework2019}.

One part of the Majorana propagation simulation that is not entirely straightforward is handling the triplets in the initial state. Naively, we could simply absorb the state preparation circuit \cref{eq:triplet_state_prep} into the Heisenberg evolution and then compute the overlap with the computational zero state. But this would exponentially increase the number of terms in the Majorana polynomial (relative to the number of triplets) and would have prohibitive memory requirements. However, because we know that state preparation circuits are the last (in the Heisenberg picture) gates acting on the four qubits involved, we can discard all terms that are orthogonal to the zero-state on these four qubits. This implies that the presence of the triplets in the initial state does not increase the hardness of computing expectation values using MP.

To assess the accuracy of our \mpsims we ran each simulation at various coefficient and weight thresholds up to a length threshold of 10 and a weight threshold of $3 \times 10^{-5}$. The results of these various simulations are compared in \cref{app:fig:mp_sim_convergence,app:fig:mp_sim_wilson_convergence}. For $N_{\text{doublons}}$ and $n_{\text{triplets}}$, we observe that decreasing the coefficient threshold has relatively little effect. Additionally, once we include the Majorana weight 6 sector, the time series changes little. This suggests that the time series of $N_{\text{doublons}}$ and $n_{\text{triplets}}$ presented in the main text are quantitatively accurate. On the other hand, for the two open Wilson lines $V_{hd}(3)$ and $V_{hd}(5)$, the time series vary wildly as the truncation parameters are changed, even for $U=0$. This suggests that the MP simulations of $V_{hd}(3)$ and $V_{hd}(5)$ are not reflective of the exact values in the absence of truncation.

 In \cref{app:fig:mp_sim_evolution} we also visualise how the Majorana polynomials spread over the different Majorana weight sectors as they Heisenberg evolve through the circuit. The connected spin correlation function begins fully supported in the weight 4 sector and then spreads to the higher weight sectors, while the number of doublons is consistently supported in the weight 2 sector and spreads much less to the higher weight sectors. This indicates that the number of doublons is easier to simulate for MP than the connected spin correlation function. Note that for the observables in question, the simulations for $U=0$ are essentially exact because the Majorana weight of each term in the Majorana polynomial is never increased, and the weight 2 and 4 sectors can be simulated without any need for truncation.

\begin{table}
  \begin{tabular}{lr}
    Observable & Total runtime \\ \hline
    $n_{\text{triplets}}$ & 152 hours \\
    $N_{\text{doublons}}$ & 22 hours \\
    $n_{1, 0, \downarrow}$ & 9 hours \\
    $S^z_{0, 2} S^z_{1, 2}$ & 53 hours \\
    $V_{hd}(3)$ & 243 hours \\
    $V_{hd}(5)$ & 29 hours \\\hline
  \end{tabular}
  \caption{Total simulation cost for the Majorana propagation simulations reported. Note that the $29$
  hours reported for $V_{hd}(5)$ were stopped early when it became evident that these simulations would not converge to quantitatively accurate results.}
  \label{tab:mp-runtime}
\end{table}

\section{Tensor Network Simulations}
\label{app:sec:tensor_network_simulations}

We used a variety of tensor network methods to attempt to simulate the experiments we performed. These included the time-dependent variational principle (TDVP), direct circuit contraction, and other alternative time-evolution techniques for matrix product states.

\subsection{Time Dependent Variational Principle}\label{sec:tdvp}
A common algorithm for the simulation of the time evolution of quantum systems is the time-dependent variational principle (TDVP) \cite{haegeman2011time}. 
We found TDVP with two-site update to be the most effective tensor network method we used, so the figures throughout this article, used for comparison against results from hardware, report TDVP outputs. 
We use the implementation in Tenpy \cite{tenpy2024}, namely the \texttt{TwoSiteTDVPEngine}.
We vary bond dimension as seen in analysis throughout, but fix hyperparameters in all simulations: we iteratively evolve each system for $\Delta t=0.01$, and set the threshold for truncation in the Singular Value Decomposition (SVD) subroutine at $10^{-10}$. 

\begin{figure}[htpb]
    \centering
    \begin{minipage}[c]{0.42\linewidth}
        \centering
        \includegraphics[width=\linewidth]{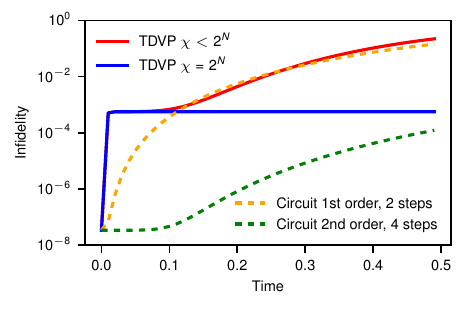}
        \\[-0.5em]
        {\small (a)}
    \end{minipage}%
    \hspace{0.03\linewidth} 
    \begin{minipage}[c]{0.5\linewidth}
        \centering
        \includegraphics[width=\linewidth]{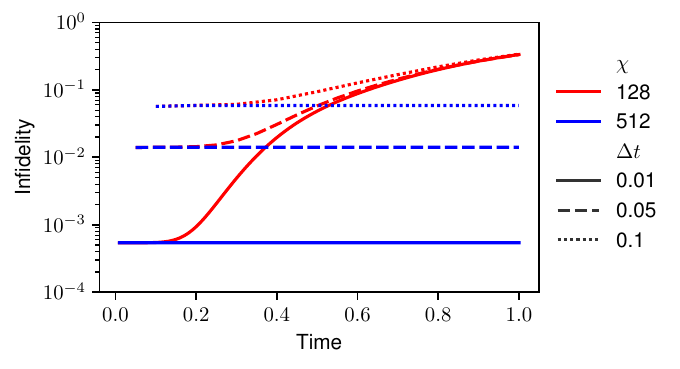}
        \\[-0.5em]
        {\small (b)}
    \end{minipage}
    \caption{
        Infidelity of TDVP for a $3\times3$ FH model with $U=0$. 
        \textbf{(a)} Comparison with simulation via quantum circuits. The solid (dashed) lines show simulation via TDVP (circuit emulation).
        The blue line shows $\chi=2^N=512$, while red line shows $\chi=2^{N-3}=64$.
        The green (orange) lines simulate the same configuration through an 18-qubit quantum circuit, following the same circuit structure as our experiment, i.e., four $2^{\rm{nd}}$ order Trotter steps (two $1^{\rm{st}}$ order steps). 
        \textbf{(b)} Infidelity of TDVP for varying $\chi$ and $\Delta t$. We observe a steep infidelity decrease as the time step size is decreased from 0.1 to 0.01, which explains the infidelity step-like shape curves in panel (a) for TDVP evolutions starting from low bond-dimension initial states.
    }
    \label{fig:tn_infidelity}
\end{figure}

To validate the correctness of our TDVP and experimental implementation, in \cref{fig:tn_infidelity} we compare the state infidelity achieved by TDVP against that of our circuits, for a $3\times3$ FH model. 
In all cases, statevector fidelity is with respect to exact diagonalisation through Quspin \cite{weinberg2017quspin}. 
We prepare a state of similar structure to that used in our experiment, i.e., a double-periodic lattice with two holons, one doublon, and triplet pairs between remaining sites. We set $U=0, \phi=0.5$ in \cref{eq:FH_Ham_main}.
Our system thus has 9 sites, with 8 occupied modes.
Tensor network representations of quantum systems are characterised by their number of sites, $N$, and physical dimension, $d$, i.e., each site corresponds to a $d$-dimensional Hilbert space.
An MPS simulation can completely capture the state of a target system given a bond dimension $\chi=d^{\frac{N}{2}}$, as stated in Ref.~\cite[Theorem 1]{perez2006matrix}. 
For fermionic systems such as the FH model under study, there are 4 available occupation levels for each site ($\ket{}, \ket{\uparrow}, \ket{\downarrow}, \ket{\uparrow\downarrow})$, so $d=4$ and hence we require $\chi=2^N$ to fully represent these systems, while an MPS with $\chi<2^N$ provides an approximate (inexact) model of the system.

We see the difference in accuracy in \cref{fig:tn_infidelity}(a): the blue (red) line represents an MPS of $\chi=2^N=512$ ($\chi=2^{N-3}=64)$. 
The constant infidelity of the $\chi=2^N$ line indicates the error due to the choice of $\Delta t$; decreasing $\Delta t$ improves the fidelity at the cost of a higher run time.
The circuit infidelities indicate a decrease in Trotter error as the circuit is given a higher Trotter order and more Trotter steps. 
In cases where the bond dimension is insufficient (for TDVP) or the number of Trotter steps is insufficient (for the quantum circuits), infidelity increases with time.
\par 

In \cref{fig:tn_infidelity}(b) we examine the effect of bond dimension, $\chi$, and time step, $\Delta t$, on the evolution through TDVP  as a function of time, by computing the state infidelity with respect to evolution of the same model with Quspin \cite{weinberg2017quspin}, which can exactly compute time evolution for this system size.
We study a $3\times3$ FH model of the same structure as in the main text, i.e. $J=1, U=4, \phi=\pi$ in \cref{eq:FH_Ham_main} and with periodic boundary conditions on a chain of triplet pairs, with one holon and one doublon.
We see that, where the bond dimension is sufficient to capture the entire state, i.e. $\chi=2^N=512$, and after a few time steps, at which point the bond dimension is reached, the error is determined solely by the time step. 
In the regime $\chi < 2^N$, however, the error is not fixed and grows with time.
The choice of time step directly determines the run time of the TDVP instance: we run $\frac{T_{\rm{max}}}{\Delta t}$ iterations of the algorithm, which take roughly the same time each (provided the MPS has reached the full bond dimension, which occurs at early times for large systems).
For the largest systems under study, we know that we cannot provide a large enough bond dimension to capture the state fully; therefore, we opt for a small $\Delta t$ to give the TDVP routine a fair opportunity to match or outperform the experiment.
Throughout this work, we have fixed $\Delta t=0.01$ to achieve a balance between accuracy and practical runtime.
We also fix the truncation threshold at $10^{-10}$ within the singular value decomposition subroutine.
\par

\begin{figure}
    \centering
    \includegraphics{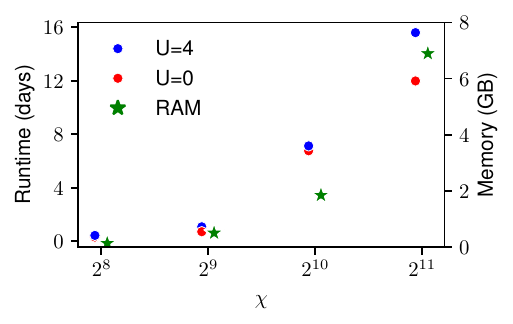}
    \caption{
    Time and memory requirements for the TDVP algorithm to simulate the $4\times7$ FH model, as a function of bond dimension $\chi$.
    These requirements refer to running on a 2.6 GHz Intel Cascade Lake processor with 8 CPUs.
    The time reported on the left-hand y-axis is to simulate the complete timeseries, i.e., up to $T=2$ in time steps $\Delta t = 0.01$, with an SVD truncation threshold of $10^{-10}.$ 
    The right-hand y-axis shows the RAM requirements for the same instances, which correspond to the green stars.
    }
    \label{fig:tn_timings}
\end{figure}

A practical consideration for the implementation of TDVP are the runtime requirements, as shown in \cref{fig:tn_timings}. 
We performed our simulations on the Google Cloud Platform with virtual machines of type \texttt{n2-highmem-32},i.e. on Intel Lake 2.6 GHz processors with 8 vCPUs and 12 GB RAM per TDVP instance \cite{gcp_n2_machine_types}.
Evidently, the gains in accuracy from higher bond dimension incur significant computational time and cost. 
We deem $\sim 2$ weeks for a single TDVP run to be feasible, but any meaningful boost to $\chi$ will yield calculations potentially lasting several months and tens of gigabytes of random access memory (RAM), without guaranteeing convergence in observables. In comparison, we saw in \cref{sec:compiler_optimisations} that the total run time on the quantum device was $4.5$ hours, i.e. $2.25$ hours per timeseries, albeit with additional overhead for compilation and queueing for access to the device.

\begin{figure}
    \centering
    \begin{minipage}[c]{0.55\linewidth}
        \centering
        \includegraphics[width=\linewidth]{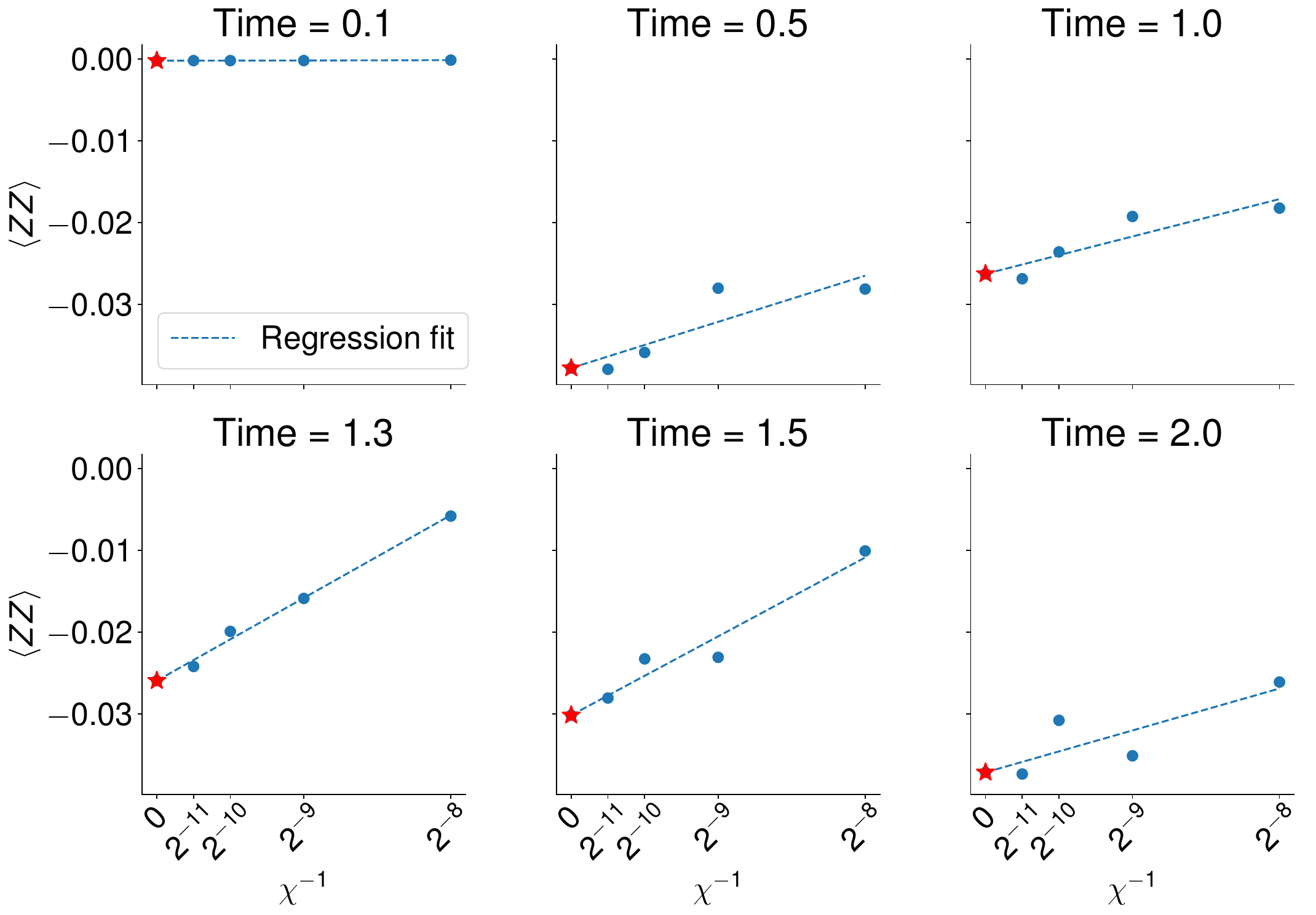}
        \\[-0.5em]
        {\small (a)}
        \label{fig:bond_dimension_regression}
    \end{minipage}%
    \hspace{0.03\linewidth} 
    \begin{minipage}[c]{0.39\linewidth}
        \centering
        \includegraphics[width=\linewidth]{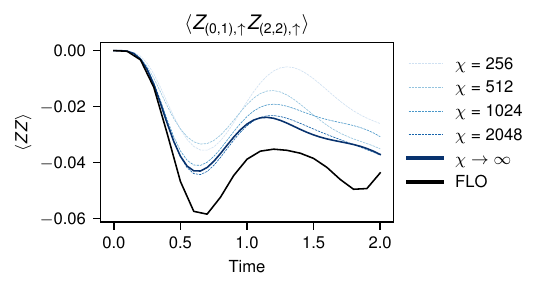}
        \\[-0.5em]
        {\small (b)}
        \label{fig:tn_extrapolation_dynamics}
    \end{minipage}

    \caption{
        \textbf{(a)} Extrapolation of observables to $\chi=\infty$.
        Blue dots are computed through TDVP with $\chi$, and a linear regression is fit through $\chi^{-1}$.
        The intercepts of the regression (red stars) are taken as estimates of the observable at $\chi^{-1} = 0$, i.e., to estimate the observable at $\chi\to\infty$.
        \textbf{(b)} Dynamics of $\langle ZZ \rangle$ on sites $(0,1), (1,2)$ for $\ket{\uparrow \uparrow}$.
        The true dynamics, computed through FLO, are shown in black.
    }
    \label{fig:tn_extrapolation}
\end{figure}

Following the procedure described in \cite{quantinuum}, we extrapolate our observables in an attempt to find more accurate approximations. 
Given access to observables for $\langle O\rangle \in\{\langle Z \rangle, \langle ZZ \rangle \}$ at all evolution times for $\chi\in\{2^8, 2^9, 2^{10}, 2^{11} \}$, we we wish to estimate the observable as computed by a tensor network of higher bond dimension.
We perform linear regression on the set $(\chi^{-1}, \langle O_{\chi} \rangle)$, and take the intercept of the regression as the estimate of the observable at $\chi^{-1} = 0$, i.e. $\chi \rightarrow \infty$.
\cref{fig:tn_extrapolation} illustrates this regression for a single observable, along with the resulting dynamics. 
\par 

We repeat this procedure on all $\langle O\rangle \in\{\langle Z \rangle, \langle ZZ \rangle \}$, for each combination of $(U, t)$ and report these results as $\chi \rightarrow \infty$ throughout. 
Overall, however, the outcome of this extrapolation does not systematically improve the accuracy of observables with respect to exact FLO calculations; see e.g. \cref{fig:errors_violin_doublon}. 
We therefore omit this extrapolation from the main text, and note that further efforts in this direction could yield more accurate TDVP observables without further simulations, e.g. by fitting only to the largest bond dimensions, or using polynomial fits.

\subsection{Direct circuit contraction with \texttt{quimb}}
\label{app:sec:quimb}

In this section, we quantify the space and time complexity for the exact contraction of the tensor network corresponding to the experimental quantum circuits. To do this, we use the rehearse feature of the \texttt{quimb} package~\cite{Gray_2018_quimb}, which allows us to estimate the complexity of contraction without the need to contract the tensor network explicitly. When contracting a tensor network, one has to consider both the contraction cost $C$, which is an estimate of the time complexity of the contraction, and the contraction width $W$, which represents the size of the largest intermediate tensor occurring in the contraction and which is directly connected to the memory required to store it.

\begin{figure}[htpb]
    \centering
    \includegraphics[width=\textwidth]{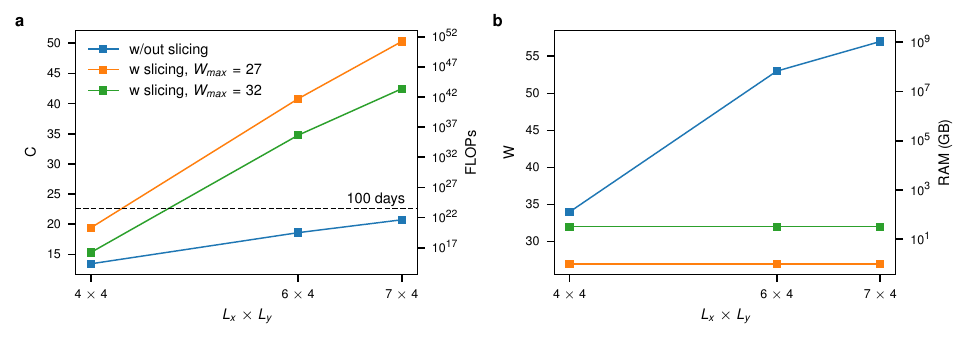} 
    \caption{Time and space complexity of the contraction of the quantum circuit with $U=4$ for different system sizes, $L_x$ and $L_y$. a) Contraction cost $C$ and b) contraction width $W$ of the largest intermediate tensor to compute a single amplitude via exact contraction of the tensor network with \texttt{quimb}. For a complex tensor with single precision, the contraction cost can be converted into FLOPs as $\mathrm{FLOPs} = 8^C$, while the memory required to store a tensor with width $W$ is $\text{RAM (GB)} = 2^{W}\times 8/20^{30}$. In a), the horizontal dashed line represents approximately the number of FLOPs corresponding to 100 days of computation on the supercomputer Summit~\cite{Gray_2021_hyperoptimized}. All the hyper-optimisers have run on four threads with the same time cutoff (six hours). }
    \label{app:fig:quimb_contraction}
\end{figure}

In ~\cref{app:fig:quimb_contraction} we show both the contraction cost $C$ and contraction width $W$ associated with the computation of a single amplitude~\footnote{The contraction cost and width associated with the computation of other quantities are similar and not shown here.} for different system sizes, $L_x$ and $L_y$. We used three high-quality \texttt{cotengra}~\cite{Gray_2021_hyperoptimized} hyper-optimisers with subtree reorganisation to find the optimal contraction path, namely, an unconstrained one targeting both cost and width, and two hyper-optimisers with dynamic index slicing with target sizes $W_{\mathrm{max}} = 27$ and  $W_{\mathrm{max}} = 32$, respectively~\cite{Gray_2021_hyperoptimized, Chen_2018_slicing, Villalonga_2019_flexible}. Recalling that storing a complex, single-precision tensor with width $W$
requires $2^W \times 8/20^{30}$ GB of memory, the two target widths $W_{\mathrm{max}}$ correspond to a memory of $1 \mathrm{GB}$ and $32 \mathrm{GB}$, respectively. The latter values have been chosen such that the contraction can be performed on a commercial GPU ($W_{\mathrm{max}} = 28$) or on an HPC GPU cluster ($W_{\mathrm{max}} = 32$). Indeed, note that this is the memory required to store the largest intermediate tensor, so for a whole computation, one usually has to take into account an additional buffer factor of $\approx 4$~\cite{Gray_2021_hyperoptimized}. As can be seen in ~\cref{app:fig:quimb_contraction}, dynamic slicing makes it possible to trade memory for time complexity. The latter is measured by the contraction cost $C$, which for a complex tensor with single precision translates to $8\times10^C$ real FLOPs. In panel c), we draw a horizontal line that approximately corresponds to a computational time of 100 days on the Summit supercomputer (281 petaFLOPs~\cite{Gray_2021_hyperoptimized, Villalonga_2020_establishing}). From ~\cref{app:fig:quimb_contraction}, it is therefore clear that the exact contraction of the tensor network corresponding to a $6 \times 4$ instance is already unrealistic (note that the memory required by unconstrained contraction is larger than $10^7$ GB).  

\newpage 
\subsection{Alternative matrix product state time evolution techniques}

We implement further algorithms for time dynamics simulation of the FH model through the MPS-FQE package \cite{provazza_fast_2024}.
We solve the time-dependent Schrödinger equation with two methods:  time-dependent version of the density matrix renormalisation group (TD-DMRG)and an explicit fourth-order Runge–Kutta (RK4) integrator. 
Both of these techniques use the Runge-Kutta formula.
TD-DMRG updates the MPS tensors in time while controlling bond dimensions via the standard multi-targeting approach \cite{feiguin_time-step_2005}.
The RK4 method corresponds to a global RK4 propagation via repeated MPO-MPS applications. 
\par 

For both of these methods, as well as TDVP, we assessed the accuracy, stability and algorithmic cost with respect to exact evolution, computed through Quspin \cite{weinberg2017quspin}.
The truncation error accumulates for long-time dynamics, as expected for the genuinely two-dimensional system under study, since the entanglement growth shows volume-law behavior. 
The TD-DMRG method suffers when targeting the basis, due to the level of truncation. 
On the other hand, the RK4 scheme outperforms TD-DMRG when small time steps are used, but loses fidelity due to the compression steps, which make the evolution non-unitary. 
The reduced density matrix used for optimising the basis targets multiple times within each step of time evolution to preserve the evolving subspace, thus avoiding a Trotter factorisation. 
We ran TD-DMRG++ (\cite{ronca_time-step_2017}), which improves the accuracy at fixed bond dimension by using independent MPS to represent states at different times. 
\par 

In contrast, the TDVP family of approaches offers variationally consistent flows and preserves unitarity better than TD-DMRG and RK4, which prove crucial when the matrix product operator contains long-range coupling from the 2D-to-1D mapping. 
This is achieved by enforcing a variationally consistent, norm-conserving evolution within the MPS manifold (strictly not-energy conserving within the two-site update variant), handling the MPO long-range terms without Trotterization. 
The TDVP method, therefore, is more suitable for the 2D problem of the main text, as for higher times, the aggressive truncation needed in the TD-DMRG or RK4 methods is penalised by effective long-range couplings from the Jordan-Wigner mapping. 
\par 

\begin{figure}[htpb]
    \centering
    \includegraphics[width=0.5\linewidth]{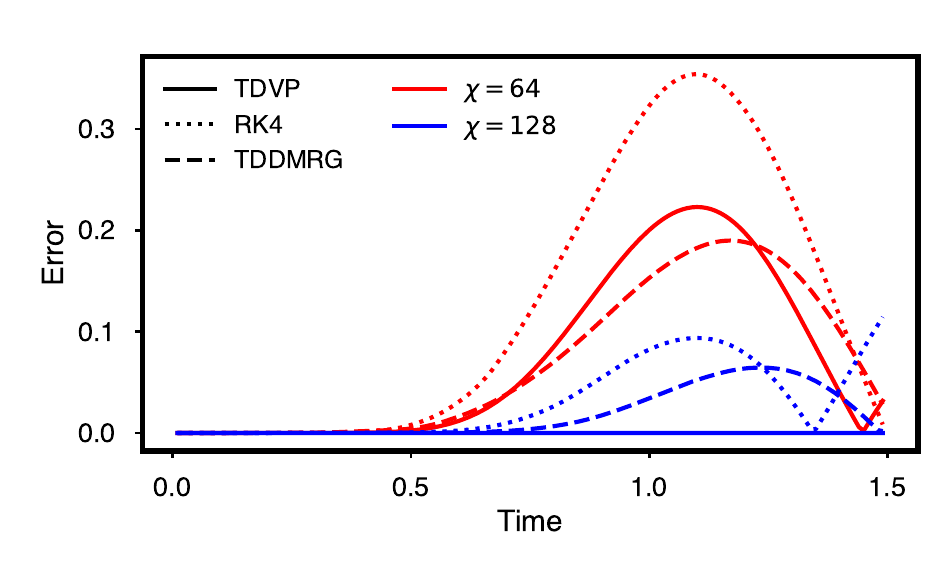}
    \caption{
    Error in expectation value of density-density correlation $n_0n_8$, i.e. the correlation between the furthest sites of a 3$\times$3 square Fermi-Hubbard model with open boundary conditions. 
    TDVP (solid), TK4 (dotted) and TD-DMRG (dashed) are shown for two values of bond dimension, $\chi=64$ (red) and $\chi=128$ (blue).
    }
    \label{fig:TDVP_v_fMPS_n0n8}
\end{figure}
We investigated non-local correlated observables for the $3 \times 3$ FH lattice with open boundary conditions. 
In \cref{fig:TDVP_v_fMPS_n0n8}, we see that, in the non-interacting limit ($U=0$), the long-time dynamics computed with restricted bond dimensions exhibit significant deviations from the exact solution, when we considered the density-density correlations between the first and last sites of the lattice. 
Among the methods considered, TDVP provides the best accuracy compared to both TD-DMRG and the RK4 scheme.
We hence conclude that TD-DMRG and RK4 techniques are less suited to the problem under study in our main text, and therefore include TDVP to represent tensor network simulations throughout.  

\newpage 
\subsection{Compressed MPS}

The final technique considered for the simulation of time dynamics is to group several lattice sites together within the MPS.
That is, instead of a lattice of size $L_x\times L_y$ sites, each with $d=4$, we group the sites along $L_x$ into a single \emph{super site}, such that we build a lattice of size $1\times 7$ with $d=4^{L_x}$. 
\par 

\begin{figure}[htpb]
    \centering
    \includegraphics{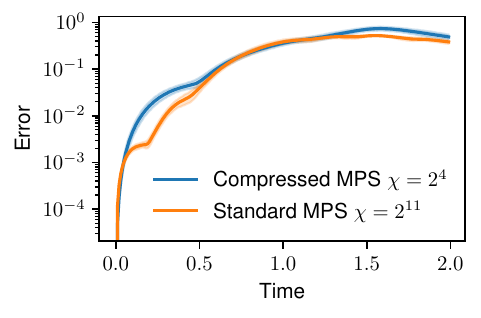}
    \caption{
        Errors in density observable computed through TDVP using standard MPS structure compared with compressed MPS structure. 
    }
    \label{fig:tn_compressed_comparison}
\end{figure}

We wish to determine whether this technique can outperform TDVP as described in \cref{sec:tdvp}.
For a fair comparison, we keep resources constant, i.e. we limit RAM to $\sim10$GiB, which allows $\chi = 2^4 = 16$ for the compressed method and $\chi = 2^{11} = 2048$ for the standard MPS.
\cref{fig:tn_compressed_comparison} shows how closely each method can recover the exact dynamics of a $4\times 7$ FH model with $U=0$, for which the ground truth can be computed through FLO.
We see that, while the compressed MPS technique can faithfully match the dynamics at low times, the standard MPS performs more reliably up to $T\approx0.5$, and both methods struggle similarly at higher times.
Ultimately, although the compressed technique is similarly viable for TDVP of this system, it does not clearly outperform the standard MPS, and we hence opt to use the more widely accepted standard MPS throughout. 

\subsection{Density Matrix Renormalization Group}
\label{app:subsec:dmrg}
\begin{figure}[htpb]
    \centering
    \includegraphics[]{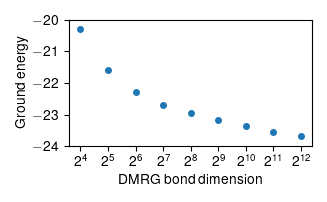}
    \caption{
    DMRG ground energies for varying maximum bond dimension, for a $4 \times 7$ Fermi-Hubbard model at half-filling with periodic boundary conditions and a flux of $\pi$.}
    \label{fig:4x7-dmrg-convergence}
\end{figure}

The ground state of the Hubbard model is estimated using the density matrix renormalisation group (DMRG) procedure~\cite{White1992}. 
We use the DMRG algorithm implemented in the ITensor Julia package~\cite{itensor, itensor-r0.3}. We start the DMRG algorithm with an initial state in the correct number sector and set the option to preserve quantum numbers during the DMRG procedure. Doing this reduces the amount of memory required and can speed up the code. For all runs of the algorithm, the cutoff is fixed at $10^{-7}$ and 50 sweeps are done, but in~\cref{fig:4x7-dmrg-convergence} we present results where the maximum bond dimension is varied. This figure demonstrates the convergence of DMRG as the bond dimension is increased and justifies our choice of bond dimension 4096 in~\cref{fig:wilson_panel_dmrg} as a reasonable approximation to the actual ground state.

\subsection{Competing tensor network methods}\label{app:subsec:competingTNS}
Here we comment on further tensor-network methods that may be able to efficiently simulate the fermionic dynamics studied in this work.
For low-weight observables and short to intermediate times
(or short/intermediate circuit depths), methods such as Heisenberg-MPO \cite{anand2023classical}
and BP-PEPS \cite{gray2025tensor} adapted to fermions \cite{mortier2025fermionic} could be competitive with the two-site TDVP approach described.
For open and/or closed Wilson line correlators, 
corresponding to large-weight observables, however,
these methods would naturally struggle, requiring large bond dimensions, especially at intermediate to large times.
Finally, quantum circuit simulations using the DMRG method introduced in \cite{ayral2023density} would be competitive for the explicitly Trotterized dynamics (see for instance \cite{haghshenas2025digital,granet2025superconducting}, where a large Trotter step was used).
We highlight that, although the system is Trotterized in our experiments, our numerical results align to continuous-time dynamics wherever it is possible to compare against exact classical simulations. 
Such comparisons are feasible at small scales for 
both $U=0$ and $U>0$ for MPS at the largest bond dimension, and at sufficiently small times and for the largest weight observables for larger systems.
As highlighted in the main text, in addition to local observable evaluation, we stress that any of these methods would need to pass the cross-entropy benchmark test (see Appendix~\ref{sec:xeb}) with a score higher than the MPS and experimental results obtained in this work.

\section{Cross-Entropy Benchmarking with FLO}
\label{sec:xeb}

The method of cross-entropy benchmarking (XEB)~\cite{arute2019quantum,morvan2024phase,zlokapa2023boundaries,cheng2025generalized,decross2025computational} is by now well-established in the context of random quantum circuits, as well as limited classes of circuits that allow efficient polynomial time computation of output amplitudes~\cite{chen2023linear,kaneda2025large} and also quantum time evolution of physical systems~\cite{cheng2025generalized,mark2023benchmarking}. In this work, we combine the latter two attributes into one protocol, leveraging the fact that amplitudes of the ideal $U = 0$ Near FLO time evolution can be exactly computed by practically efficient determinant methods described in \cref{sec:nearflo}.  The cross-entropy estimators we compute are
$$\textrm{XE}(\textrm{generator} || \textrm{FLO}) = \langle \log_2(1/p_\textrm{FLO})\rangle_\textrm{generator}$$
for the various choices of sample generators: experiment (exp), MPS, uniform (baseline) (see \cref{fig:XEB1}. Here we are using $p_\textrm{FLO}$ as the ideal model and testing the ``surprisal'' of applying this model to the different sample sets generated by the experiment, the highest bond dimension MPS ($\chi = 2048$), and a uniform sample generator.   Low scores are better in that they correspond to less surprise at the samples that are seen, but note that just like XEB for random circuits this is not meant to be a strict test of fidelity (it can be ``spoofed'' by low Shannon entropy distributions that output strings with a large value of $p_\textrm{FLO}$), rather it is meant to compare ``honest'' alternatives to the quantum circuit such as the large bond dimension MPS approximation we consider. 

Since the various sample generators occasionally produce bit strings $z$ with extremely small probability $p_\textrm{FLO}(z)$, a technical caveat is that we eliminate divergences in the $\log$ function by including some uniform noise in the ideal model
\begin{equation} p_\epsilon(z) = (1-\epsilon)p_\textrm{FLO}(z) + \epsilon/D \label{eq:pmixepsilon} \end{equation}
where $D=\binom{28}{14}^2$ is the dimension of the system and $\epsilon = 0.1$ is chosen in ~\cref{fig:XEB1}.  

\subsection{Experimental advantage in sampling.} 
When the difference in cross entropy $\textrm{XE}(\textrm{generator} || \textrm{FLO})$ between two sample generators is $r = \textrm{XE}(g_1 || \textrm{FLO}) - \textrm{XE}(g_2|| \textrm{FLO})$ bits, then the set of $M$ samples from $g_2$ is $2^{r M}$ times less surprising to see when one is expecting the samples to come from $p_{\textrm{FLO}}$. This means that a decrease in the LogXEB value increases the likelihood of the model exponentially in the number of samples (according to this particular test).  At evolution time $t = 2$, the experimental estimator improves on the MPS by 1.5 bits in expectation, a difference which is larger than the standard error (the MPS and uniform results are computed with 1000 samples per time point, and so the standard error is small on this scale). 

The results indicate a decisive advantage for the experiment over the $\chi = 2048$ bond dimension MPS as a sample generator.  The experiment achieves this advantage despite the fact that the ideal FLO model and the MPS are untrotterized models of the continuous time evolution.   Suppose one considers constraining to ``honest'' sample generators that reproduce local expectations as well as the MPS in \cref{app:sec:tensor_network_simulations}. In that case, the results support the conclusion that the experiment is the best sample generator we know of (since it is practical to compute Near FLO amplitudes, but not to sample from the output distribution).

\subsection{Modeling circuit fidelity with LogXEB}
\begin{figure}
    \centering
    \includegraphics[width=0.45\linewidth]{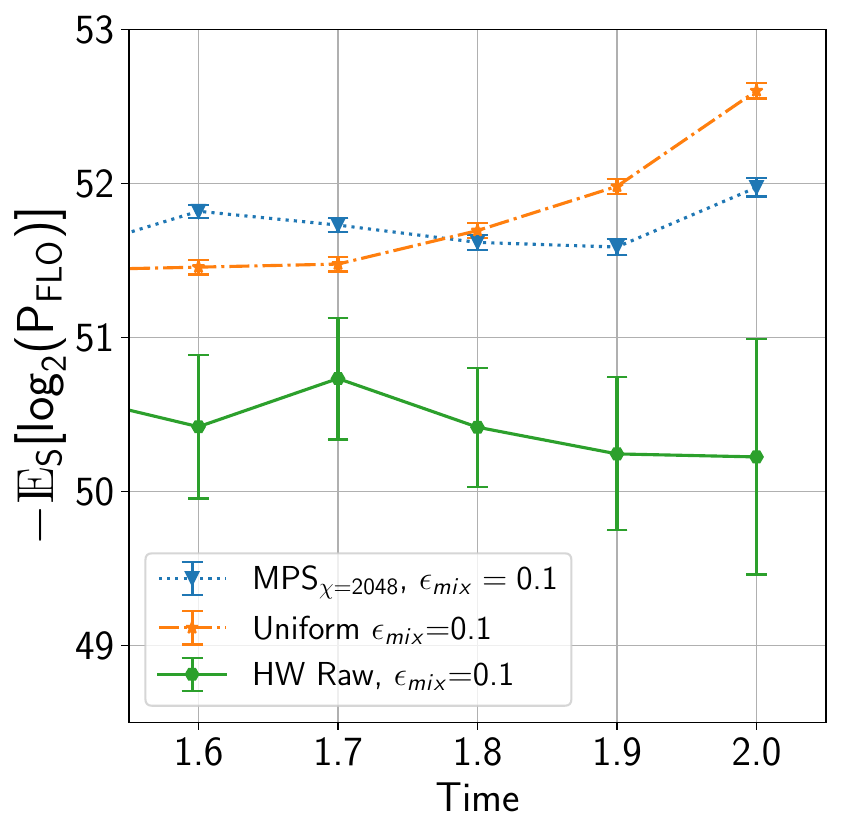}
    \includegraphics[width=0.45\linewidth]{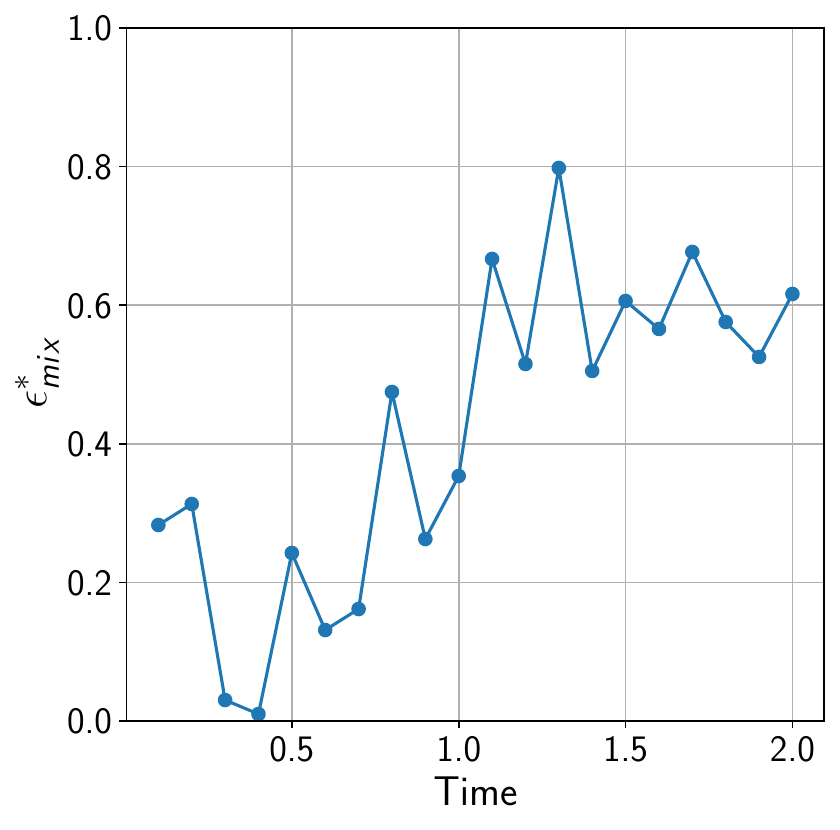}
    \caption{Left: A comparison of the ideal model $p_\textrm{FLO}$ against different sample generators: experiment, MPS with 
    $\chi = 2048$, and uniform (baseline) as a baseline.  A lower score means that the samples produced by the generator are less surprising according to the FLO likelihoods of those samples.  Right: Estimates of the circuit infidelity $\epsilon$ under the global depolarising noise model obtained by minimizing $\textrm{XE}(\textrm{exp}||\textrm{FLO},\epsilon)$ at each point in time.
    }
    \label{fig:XEB1}
\end{figure}

\begin{figure}
    \centering
    \includegraphics[width=\linewidth]{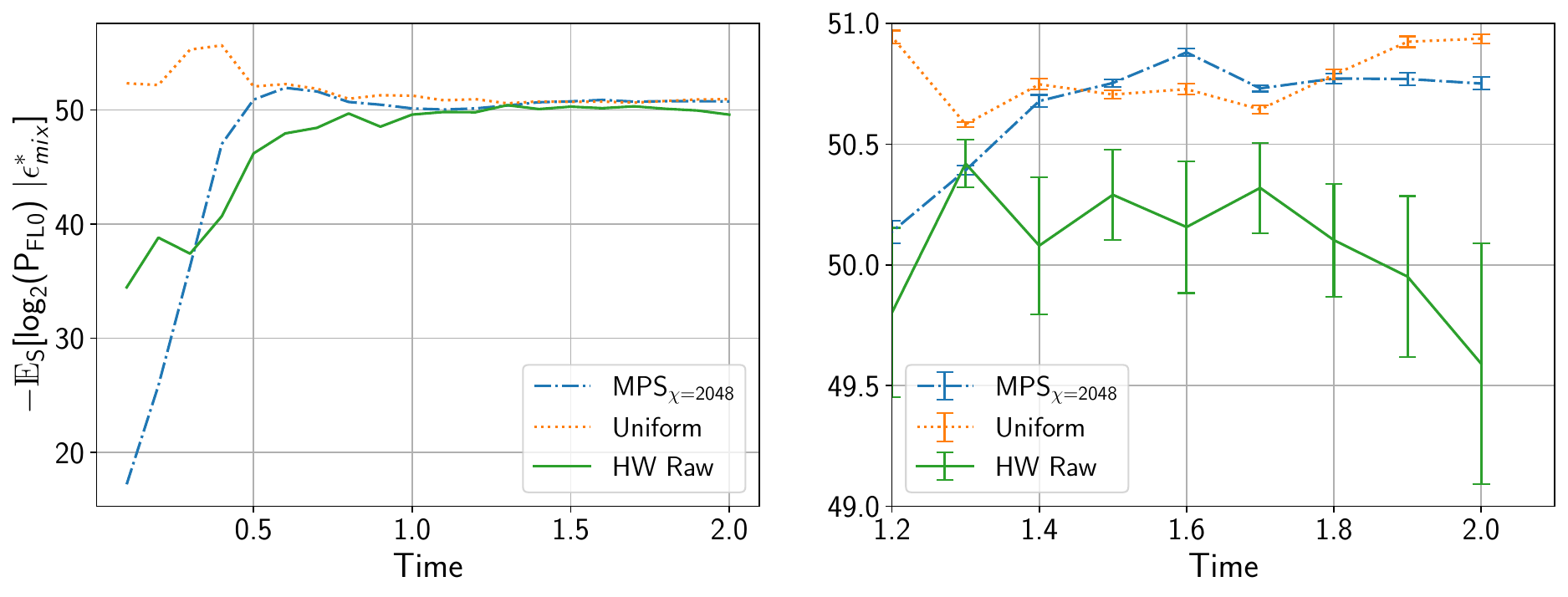}
    \caption{ Cross entropy estimators for the optimised choice of $\epsilon^*(t)$ for the FLO model mixed with depolarising noise in \eqref{eq:pmixepsilon}. The right-hand plot zooms in on the late-time behaviour in the left-hand plot.
    }
    \label{fig:XEB3}
\end{figure}

The term ``benchmarking'' in XEB refers to an estimate of circuit fidelity. Usually, this is done by computing the linear XEB estimators $\langle p_\textrm{FLO}\rangle_\textrm{generator}$ for the various sample generators, and then scaling and shifting them to relate to an estimate of the circuit fidelity $F$ under the global depolarizing model.    Instead, we devise an alternate approach to estimate the global circuit fidelity under the global depolarising model, by minimizing the LogXEB score $\textrm{XE}(\textrm{exp}||\textrm{FLO},\epsilon) = \langle \log(1/p_{\textrm{FLO},\epsilon})\rangle_\textrm{exp}$ for the distribution \eqref{eq:pmixepsilon} as a function of $\epsilon$. This noise estimation is a proper scoring rule with a fixed sample generator and a parameterized class of models: at each time point, we can apply it to extract the most likely $\epsilon$ among this set of models.   The value of the optimised circuit infidelity $\epsilon^*$ is shown as a function of time in ~\cref{fig:XEB1}, and the resulting cross entropy estimators in ~\cref{fig:XEB3} indicate an even larger advantage for quantum hardware over the MPS at late times.    

\bibliographystyle{apsrev4-2}
\bibliography{references}

\end{document}